\documentclass[aps,prd,twocolumn,superscriptaddress,showpacs,amsmath,amssymb]{revtex4}
\usepackage{graphicx,epsf}

\usepackage[]{latexsym}
\newcommand{\be}{\begin{eqnarray}}
\newcommand{\ee}{\end{eqnarray}}

\def\comment#1{}

\newcommand{\figref}[1]{Fig.\,\ref{#1}}
\newcommand{\eqeqref}[1]{Eq.\,\eqref{#1}}

\begin{document}

%
%
\title{Black Hole Remnants in the Early Universe}

\author{Fabio~Scardigli}
\email{fabio@phys.ntu.edu.tw}
\affiliation{Leung Center for Cosmology and Particle Astrophysics (LeCosPA),
National Taiwan University, Taipei 106, Taiwan.}

\affiliation{Department of Physics and Graduate Institute of Astrophysics,
National Taiwan University, Taipei 106, Taiwan.}

\author{Christine Gruber}

\email{chrisy_gruber@gmx.net}

\affiliation{Leung Center for Cosmology and Particle Astrophysics (LeCosPA),
National Taiwan University, Taipei 106, Taiwan.}

\author{Pisin~Chen}

\email{pisinchen@phys.ntu.edu.tw}

\affiliation{Leung Center for Cosmology and Particle Astrophysics (LeCosPA),
National Taiwan University, Taipei 106, Taiwan.}

\affiliation{Department of Physics and Graduate Institute of Astrophysics,
National Taiwan University, Taipei 106, Taiwan.}

\affiliation{Kavli Institute for Particle Astrophysics and
Cosmology, SLAC National Accelerator Laboratory, Stanford
University, Menlo Park, CA 94025, U.S.A.}
%
%

%
%
%
%
%
%
\begin{abstract}
We consider the production of primordial micro black holes (MBH) remnants in the
early universe. These objects induce the universe to be in a
matter-dominated era before the onset of inflation. Effects of such an
epoch on the CMB power spectrum are discussed and computed both
analytically and numerically.
By comparison with the latest observational data from the WMAP collaboration, we find that our
model is able to explain the quadrupole anomaly of the CMB power spectrum.
\end{abstract}
\pacs{xxx}
%
%
%
\maketitle

\section{Introduction}\label{SEc1}
%
Inflation is without doubt the best model to explain the observed spatially flat and homogeneous Universe.
Nevertheless, despite the great successes of the standard $\Lambda$CDM model in explaining almost all the data on CMB anisotropy
as most recently measured by WMAP observations, the suppression of the $l=2$ quadrupole mode still remains a puzzle
in the framework of the standard $\Lambda$CDM model (for a review on this subject, see e.g.~\cite{Inou2007}).

Recently, several authors \cite{NG, Powe2007} have been able to shed some light on this region of the CMB power spectrum,
by investigating the possibility of a pre-inflationary epoch, dominated by radiation, instead of the usual inflationary
vacuum. They found that a pre inflation radiation era can produce a suppression of the low $k$ modes of the
primordial power spectrum, and this in turn affects the low $l$ modes of CMB anisotropy power spectrum.
In fact, although inflation
has the effect of washing out the initial conditions of the Universe, it happens that, if the present Universe
is just comparable to the size of the inflated region, a pre-inflation era may leave imprints on the CMB power spectrum.

However, these early attempts suffered of an arbitrary initial condition in the pre-inflationary era.
Also, the space of the numerical parameters encoding the initial radiation density was merely explored,
without stating precise criteria for the choice of specific numerical values.

In the present paper we propose a pre-inflationary scenario that is based on the generic micro black hole (MBH)
production and a minimal set of first principles, namely the generalized uncertainty principle (GUP) and the
holographic principle (HP), that can give rise to the suppression of the CMB quadrupole self-consistently without
the need of arbitrary inputs.
Specifically, we consider the possibility of production of micro black holes in the early pre-inflationary Universe,
due to quantum fluctuations of the metric field \cite{gross, kapusta},
as the seeds for the suppression of the inflaton fluctuations. There are two salient features of this
MBH nucleation. One is that the production rate per unit volume of space and time is very high at the
Planck temperature. To prevent unphysical over-production of MBH, we invoke the holographic principle
(HP) to constrain the initial condition of MBH production. The other is that the rate of such MBH production
is a strong function of the background temperature. In particular, the rate is exponentially suppressed when
the temperature of the universe is sufficiently below the Planck temperature.
Inflation is in general assumed to start when the temperature of the
universe reaches the scale of the GUT energy, about $10^{15}-10^{16}\, GeV$.
Therefore one expects that the MBH production activity would cease long before the onset of the inflation,
and the MBH would have been totally evaporated and the universe would turn into radiation era before the
inflation begins. However,
when the Generalized Uncertainty Principle (GUP) is taken into consideration,
the complete decay of the nucleated MBH into radiation is prevented, and we have
massive, but inert black hole remnants \cite{ACSantiago} populating the pre-inflationary phase of the Universe.
Furthermore, the nucleation of MBH
is so efficient and fast that the Universe is put into a matter dominated era within a few Planck times,
just about $10^3\,\,t_p$ after the Big Bang (i.e., well before inflation) and there it stays until the onset of inflation.
Such a pre-inflation matter-dominated universe then suppresses the initial inflaton fluctuations at the onset of the inflation.

Accurate numerical simulations allowed us to single out almost unique numerical values for the relevant radiation and matter
parameters. We have computed the effects of a pre-inflationary matter epoch on the primordial power spectrum of the
quantum fluctuations of a scalar field, both analytically and numerically. Our analytical solution, also a new feature
of the present attempt with respect to the previous all-numerical investigations, has served as a guide for
the more precise numerical computations.
We have considered three alternative scenarios.
The main model presented in this paper attempts to explain the suppression of the quadrupole moment of the CMB with a
pre-inflationary matter era.
In order to isolate the cause of the CMB quadrupole anomaly, we further examine
two variations of this model, one without the GUP,
where the black holes decay into radiation completely, and one without any black hole
nucleated at all, both models resulting into a radiation-dominated era before inflation. \\
In all three cases the primordial power spectra have been fed to the CMBFAST code in order to obtain the CMB
power spectra, and then compared with each other and tested against the last WMAP observational data.
Boundary conditions have been set in the fully inflationary epoch, and in so doing we avoided any arbitrary
assumption on the state of radiation in the pre-inflationary era.
The pre-inflation matter model seems to be the only one, among those studied, which is able to describe the $l=2$ mode
suppression, although the radiation model still presents a better fitting of the data at high $l$ values.
This conclusion is widely discussed in the last section of the paper, where we also suggest avenues for future research.
We find it remarkable that, based on our {\it ab initio} model without arbitrary input parameters,
our resulting suppression of the CMB spectrum agrees well with observations.

The paper is organized as follows. In section II we will provide a brief overview of the black hole physics
under the Generalized Uncertainty Principle.
The GUP will lead to a new mass-temperature relation and define a minimum mass and maximum temperature for the black holes.
Section III will set up the basic equations governing the scenario, on the one hand the absorption and emission
processes which determine the black hole mass, and on the
other hand the evolution equations of the universe depending on its constitution. We will end up with a system of four
equations, containing
black hole mass, black hole density, radiation density and the scale factor as variables.
At the end of that section we derive a condition for a pre inflation matter era, and we present inflationary solutions.
In section IV we will numerically calculate the black hole production and fix the parameters in the evolution equations.
Section V will deal with the equations needed to be solved to obtain the primordial power spectrum of the quantum
fluctuations of a scalar field. First we state some approximate analytical solutions obtained
via the WKB method, and then we present the numerical result of the equations.
In section VI finally we present the CMB power spectrum of the temperature anisotropies obtained by our model,
and compare it to the two
alternative cases with a radiation-dominated era before inflation. The conclusions of our work are contained in section VII.

Throughout the paper the Planck length is defined as $\ell_p^2=G \hbar /c^3$,
the Planck energy as $\mathcal{E}_{p} \ell_{p} = \hbar c /2$, and the Planck mass as
$M_{p}=\mathcal{E}_{p}/c^2$.

\section{Black hole Physics}
\subsection{Generalized Uncertainty Principle}
\label{GUP}
As it is well known from the classical argument of the Heisenberg microscope \cite{H},
the size $\delta x$ of the smallest detail of an object, theoretically detectable with a beam of photons of energy $E$,
is roughly given by
\be
\delta x \simeq \frac{\hbar c}{2 E}\, ,
\label{HS}
\ee
since larger and larger energies are required to explore smaller and smaller details.

The research on viable generalizations of the Heisenberg uncertainty principle traces back to many
decades (see for early approaches \cite{GUPearly}; see for a review \cite{GUPreview},
and for more recent approaches \cite{MM, FS, Adler2}).
In the last 20 years, there have been important studies in string theory \cite{VenezETAL} suggesting
that, in gedanken experiments on high energy scattering with high momentum transfer, the uncertainty
relation should be written as
\be
\delta x \,\gtrsim \,\frac{\hbar}{2p} \,+\, 2\,\beta \, \ell_{p}^2 \,\frac{p}{\hbar}\, ,
\ee
where $\ell_{p}$ is the Planck length, and $\beta \ell_{p}^2 \sim \lambda_s^2$, where
$\lambda_s$ is the characteristic string length. Since in our high energy scattering
$E \simeq c p$, the stringy Generalized Uncertainty Principle (GUP) can be also written as
\be
\delta x \,\gtrsim \,\frac{\hbar c}{2 E} \,+\,  2\,\beta \,\ell_{p}^2 \,\frac{E}{\hbar c}\, ,
\label{STGUP}
\ee
where $E$ is the energy of the colliding beams.\\
A similar modification of the uncertainty principle has been proposed \cite{FS, SC},
on the ground of gedanken scattering experiments involving the formation of micro black holes
with a gravitational radius of $R_S \sim E$. It
reads
\be \delta x \gtrsim \left\{ \begin{array}{ll}
\frac{\hbar c}{2 E} \quad {\rm for} \quad E < \mathcal{E}_{p}\\ \\
\beta R_{S}(E) \quad {\rm for} \quad E \geq \mathcal{E}_{p} \, ,
\end{array}
\right.
\ee
where $R_{S}$ is the Schwarzschild radius associated with the energy $E$,
namely $R_S = \ell_p E / \mathcal{E}_{p}$.

Combining linearly the above inequalities we get
\be
\delta x \gtrsim \frac{\hbar c}{2 E} \,+\, \beta R_{S}(E)\, .
\ee
Thus, the GUP originating from micro black hole gedanken experiments (MBH GUP) can be written as
\be
\delta x \gtrsim \frac{\hbar c}{2 E} \,+\, \beta
\ell_{p} \frac{E}{\mathcal{E}_{p}}\, .
\label{MBH}
\ee
Also the stringy inspired GUP
(ST GUP,  eq.(\ref{STGUP})), using the relation $\mathcal{E}_{p} \ell_{p} = \hbar c /2$,  can be written as
\be
\delta x \gtrsim \frac{\hbar c}{2 E} \,+\, \beta
\ell_{p} \frac{E}{\mathcal{E}_{p}}\, .
\label{ST}
\ee
where $\beta$ is the deformation parameter, generally believed to be of $O(1)$. Thus, in 4
dimensions the two principles coincide. In $4+n$ dimensions, however, they lead to remarkably different
predictions (see \cite{Glimpses}).
%
%
%
%
%
%
\subsection{From the uncertainty principle to the mass-temperature relation}
\label{sec:masstemp}
Naturally, a modification of the uncertainty relation, i.e. of the basic commutators, has deep consequences
on the quantum mechanics, and on the quantum field theory built upon it. The general implementation of such
commutation rules, as regards Hilbert space representation, ultraviolet regularization, or modified dispersion
relations, has been discussed in a vast amount of literature (see~\cite{vari} for an incomplete list).
In the present section, we want to focus on the use of
(generalized) uncertainty relations to compute the basic feature of the Hawking effect, namely the formula
linking the temperature of the black hole to its mass $M$. The seminal results of Hawking and Unruh
\cite{Haw, Unr} are rigorously computed using QFT, based on Heisenberg uncertainty principle, on curved space-time.
However, it has been shown \cite{FS9506, ACSantiago, CDM03}
that the full calculation of QFT in
curved space-time (with standard commutators for the ordinary uncertainty principle, or with deformed
commutators for the GUP) can be safely replaced by a computation employing only the (generalized) uncertainty
relation and some basic physical considerations, in order to obtain the mass-temperature formula. \\

The GUP version of the standard Heisenberg formula (\ref{HS}) is
\be
\delta x \simeq \frac{\hbar c}{2 E} \,+\, \beta
\ell_{p} \frac{E}{\mathcal{E}_{p}}\, .
\label{He}
\ee
which links the (average) wavelength of a photon to its energy $E$.
Conversely, with the relation (\ref{He}) one can compute the energy
$E$ of a photon with a given (average) wavelength $\lambda \simeq
\delta x$.

Following loosely the arguments of
Refs.~\cite{FS9506,ACSantiago,CavagliaD,CDM03,nouicer,Glimpses},
we can consider an ensemble of unpolarized photons of Hawking radiation
just outside the event horizon. From a geometrical point of view,
it's easy to see that the position uncertainty of such photons is of
the order of the Schwarzschild radius $R_S$ of the hole. An
equivalent argument comes from considering the average wavelength of
the Hawking radiation, which is of the order of the geometrical size
of the hole. By recalling
that $R_S=\ell_p m$, where $m=M/M_p$ is the black hole mass in
Planck units ($M_p={\cal E}_p/c^2$), we can estimate the photon
positional uncertainty as
\be
\delta x \ \simeq \ 2\mu R_S \ =  \ 2\mu \ell_p m\, .
\label{VI.48a}
\ee
The proportionality constant  $\mu$ is of order unity and will be
fixed soon. With (\ref{VI.48a}) we can rephrase Eq.~(\ref{He}) as
\be
2\mu m \ \simeq \ \frac{{\cal E}_p}{ E} + \beta
\frac{E}{{\cal E}_p}\, .
\label{47}
\ee
According to the equipartition principle the average energy
$E$ of unpolarized photons of the Hawking radiation is
linked with their temperature $T$ as
\be
E \ = \ k_B T\, .
\ee
In order to fix $\mu$, we consider the semiclassical limit
$\beta \to 0$, and require that formula (\ref{47}) predicts the standard semiclassical
Hawking temperature:
%
\be
T_{H} \ = \ \frac{\hbar c^3}{8\pi G k_B M} \ = \ \frac{\hbar
c}{4 \pi k_B R_S}\, .
\label{Hw}
\ee
This fixes $\mu = \pi$.\\
Defining the Planck temperature $T_p$ so that ${\cal E}_p = k_B
T_p/2$ and measuring all temperatures in Planck units as $\Theta =
T/T_p$, we can finally cast formula (\ref{47}) in the form
\be
2m \ = \ \frac{1}{2\pi\Theta} \,+\, \zeta\, 2 \pi \Theta\, ,
\label{MT+}
\ee
where we have defined the {\em deformation} parameter $\zeta = \beta/\pi^2$.

As already mentioned, in the semiclassical limit both $\beta$ and $\zeta$
tend to zero and (\ref{He}) reduces to the ordinary Heisenberg
uncertainty principle. In this case Eq.~(\ref{MT+}) boils down to
\be
m \ =  \ \frac{1}{4 \pi \Theta}\, .
\label{H}
\ee
which is the dimensionless version of Hawking's formula (\ref{Hw}).

As we have seen, a computation of the mass-temperature relation for black holes
based on the GUP has resulted in a modification of the
Hawking formula for high temperatures. In the next subsection, we summarize as this leads also
to the remarkable prediction of black hole remnants (see \cite{ACSantiago}).

\subsection{Minimum masses, maximum temperatures}
\label{minmass}
The standard Hawking formula predicts a complete evaporation of a black hole, from an initial mass $M$
down to zero mass. As we have seen this is a direct consequence of the Heisenberg principle.
However, when the mass-temperature relation is derived from the GUP instead, the formulation immediately leads
to a minimum mass and a
maximum temperature for the
evaporating black hole. Precisely we have, for the GUP,
\begin{subequations}
\be
\Theta_{\rm max}&=&\frac{1}{2 \pi \sqrt{\zeta}}
\ee
\be
m_{\rm min}&=&\sqrt{\zeta}
\ee
\label{Thmax}
\end{subequations}
Note that, as expected, $\Theta_{\rm max} \to \infty$ and $m_{\rm min} \to 0$ in the Hawking limit
$\beta \to 0$.
Therefore the use of the GUP eliminates the problem of an infinite temperature at the
end of the evaporation process, which is clearly unphysical, and leads directly to the prediction of the existence of black hole
remnants (\cite{ACSantiago, CDM03, CavagliaD, ChenAd, Ghosh},
\footnote{It is interesting to note that analogous results have been obtained in a completely
different framework, via non commutative geometry inspired black holes. See \cite{Nicolini} and references
therein.}).
In references \cite{ACSantiago,Glimpses}, it has been shown that also the emission rate (erg/sec) is kept finite by the GUP
mass-temperature formula, in contrast with an infinite output predicted by the Hawking formula.

\section{Governing equations}\label{Sec2}
In this section we will write down the basic equations which govern a system of black holes and radiation in the early universe.
We will describe the evolution of a black hole mass as a balance of accretion and evaporation, as well as consider the
dynamical behavior of a universe constituted by black holes and radiation. \\
Then we shall derive a condition for a pre-inflation era dominated by matter, and the inflationary solutions for the
equations of motion of the scale factor $a(t)$, in both cases of pre-inflation matter, or radiation dominated eras.

\subsection{Emission rate equation}
In this subsection we will describe the evaporation behavior of a micro black hole (in 4 dimensions)
taking into account the GUP effects.

In the present model we consider only photons or gravitons,
nevertheless other kind of gauge or fermionic fields can be added in a straightforward way.\\
Before writing down the emission rate equation, we review some delicate issues about greybody
factors, emitted energy, and the Stefan-Boltzmann constant, in $4$ dimensions with the GUP.\\
The presence of a GUP, i.e. of a minimal length, forces us to take into account
the squeezing of the fundamental cell in momentum space (see
\cite{nouicer, LNChang, Kempf97_1, Kempf97_2, Niemeyer}).
The squeezing results in a deformation of the usual Stefan-Boltzmann law.
This deformation has to be considered,
at least in principle, since we deal with micro black holes close to their final evaporation phase, where
the predictions of the GUP are expected to differ noticeably from those of the Heisenberg principle.\\
Due to the deformation of the Heisenberg fundamental inequality,
\be
\Delta x \Delta p \geq \frac{\hbar}{2}\left(1 + \beta\, \frac{4\ell_p^2}{\hbar^2}\, \Delta p ^2\right)\, ,
\ee
the number of quantum states per momentum space volume (or the invariant phase space volume) is
\be
dn_x dn_y dn_z = \frac{V}{(2\pi\hbar)^3}
\frac{dp_x dp_y dp_z}{\left(1 + \beta\, \frac{4\ell_p^2}{\hbar^2}\, p^2\right)^3}
\ee
Since $p=\hbar k$, the number of quantum states (i.e. stationary waves) in the volume V, with wave vector
in $[k,\, k+dk]$ is
\be
dN=\int_\Omega dn_x dn_y dn_z = \frac{V}{(2\pi)^3}
\frac{4\pi k^2 dk}{\left(1 + \beta \, \frac{4\ell_p^2}{\hbar^2}\,(\hbar k)^2\right)^3}
\ee
Since $k=\omega/c$,
the number of photons (or gravitons) with frequency within
$\omega$ and $\omega + d \omega$ in a volume $V$ is given by
\be
d n_{\gamma}=\frac{V}{\pi^2 c^3}\,
\frac{\omega^2}{\left[1 + \beta\,\frac{4\ell_p^2}{\hbar^2}\,(\frac{\hbar\omega}{c})^2\right]^3}\,
\frac{\Gamma_{\gamma}(\omega)}{e^{\hbar\omega/k_BT}-1} \,\,d\omega\,.
\ee
In case of a perfect black body (perfect emitter) we have for the greybody factor
$\Gamma_{\gamma}(\omega) = 1$ for any $\omega$. The dependence of $\Gamma_{\gamma}(\omega)$ from
the frequency $\omega$
is in general very complicated. It has been studied in many papers (for 4 dimensional black holes see
\cite{Page},
for emission of gravitons in $4+n$ dimensions see \cite{CardCav}),
it is in some cases partially unknown, and in many cases can only be computed numerically.
In the present model,
we neglect the frequency dependence of $\Gamma_{\gamma}$, and therefore take the value
$\Gamma_{\gamma}:=\langle\Gamma_{\gamma}(\omega)\rangle$ averaged over all the frequencies.
Thus, for the number
of photons (or gravitons) in the interval ($\omega$, $\omega + d \omega$) in a volume $V$ we write
(in $4$ dimensions)
\be
d n_{\gamma}=\frac{V}{\pi^2 c^3}\,
\frac{\omega^2}{[1 + \beta\,\frac{4\ell_p^2}{\hbar^2}\,(\frac{\hbar\omega}{c})^2]^3}\,
\frac{\Gamma_{\gamma}}{e^{\hbar\omega/k_BT}-1} \,\,d\omega\,.
\ee
Obviously $\Gamma_\gamma < 1$ for a real non-ideal black body. \\
The total energy of photons contained in a volume $V$ (in $4$ dimensions) is then
\be
&&E^{\gamma}_{\rm TOT}(V) = \int_0^\infty \hbar \omega \,d n_{\gamma} \nonumber \\
&=&\Gamma_{\gamma}\frac{V (k_BT)^4}{\pi^2 c^3 \hbar^3}  \Gamma(4)\zeta(4)\,A(\beta, T),
\ee
where $\Gamma(s)$ is the Euler Gamma function, $\zeta(s)$ is the Riemann Zeta function, and
the function $A(\beta, T)$ accounts for the cells' squeezing in momentum space, due to GUP.
The function $A(\beta, T)$ can be formally written as
\be
A(\beta, T) = \frac{1}{\Gamma(4)\zeta(4)}
\int_0^\infty \frac{1}{\left[1+\beta(2\ell_p\, k_B T x \,/\, \hbar c)^2\right]^3}\cdot\frac{x^3}{e^x - 1}\,dx\nonumber\\
\,
\ee
and by this definition we have
\be
A(\beta, T) \to 1 \quad {\rm for} \quad \beta \to 0\,. \nonumber
\ee
Defining the Stefan-Boltzmann constant (in 4 dimensions) as
\be
\sigma_3 = \frac{c}{3}\, \frac{\Gamma(4)\zeta(4)}{\pi^2 c^3 \hbar^3} k_B^4,
\label{s3}
\ee
the total energy can be written as
\be
E^{\gamma}_{\rm TOT}(V) \,=\, \Gamma_{\gamma}\, \frac{3\, \sigma_3}{c}\, V \, T^4 \, A(\beta, T) .
\label{sb1}
\ee

The energy $dE$ radiated in photons (or gravitons) from the black hole, in a time $dt$,
measured by the far observer, can be written as
\be
d E = \Gamma_\gamma \frac{3\sigma_3}{c}\,\mathcal{V}_3\, T^4 \, A(\beta, T),
\ee
where ${\cal V}_3$ is the effective volume occupied by photons in the vicinity of the event horizon,
\be
{\cal V}_3 = 4\pi\, R_{S}^2\, c\, dt.
\ee

Thus, finally, the differential equation of the
emission rate is \cite{SMRD, Emparan, nouicer}
\be
-\frac{dE}{dt} = 12 \pi\, \Gamma_\gamma\, \sigma_3\,\, R_{S}^2\, T^4 \,A(\beta, T)\, .
\label{ereq}
\ee
where the minus sign indicates the loss of mass/energy.
With the explicit definitions of $\sigma_3$, $R_{S}$, and using Planck variables
$m=M/M_{p}=E/\mathcal{E}_{p}$, $\Theta=T/T_{p}$, $\tau= t/t_{p}$
(where $\mathcal{E}_{p}=\frac{1}{2} k_B T_{p}$ and $t_{p}=\ell_{p}/c$), we can rewrite
the emission rate equation as
\be
-\frac{dm}{d\tau} &=& \frac{8 \,\pi^3\,\Gamma_\gamma}{15}\,
m^2\,\Theta^4 \, {\mathcal A}(\beta, \Theta),
\label{ereqpl}
\ee
where we used $\Gamma(4)\zeta(4)=\pi^4/15$ and
\be
{\mathcal A}(\beta, \Theta) =
\frac{15}{\pi^4}
\int_0^\infty \frac{1}{\left[1 + 4\, \beta\, \Theta^2\, x^2 \right]^3}\cdot\frac{x^3}{e^x - 1}\,dx\nonumber\\
\,
\ee
In the applications presented in the following sections, the GUP will be implemented
by considering only the cutoff imposed on minimum masses and maximum temperatures. In other words, we mimic
the cutoff effects of the GUP by simply stating that the micro black hole evaporation stops when $T=T_{\rm max}$
or equivalently when $M=M_{\rm min}$, and otherwise using the "simpler" Hawking form of the mass-temperature relation.
This is tantamount to choose ${\mathcal A}(\beta, \Theta) \simeq 1$.
We adopt this choice in order not to render the calculation too tedious,
in particular for those involving the nucleation rate of black holes and the emission rate in the presence of absorption terms.

\subsection{Absorption terms in the evolution equation for micro black holes}

In this section we consider the absorption of radiation by black holes. Therefore, we extend the emission rate equation
(\ref{ereqpl}) in order to describe all the processes changing the mass of a black hole.
In principle, as our system consists of radiation and micro black holes, we should also take into account
the absorption of micro black holes by other micro black holes. However, we will see later that the black hole
density is low enough to neglect scattering processes among black holes themselves,
and thus we limit the analysis of absorption to the background radiation.
The calculations are particularly inspired by Refs. \cite{custodio, novikov, HawkingCarr, Barrow, Khlopov}.
Absorption terms will appear with a positive sign in Eq.(\ref{ereqpl}).
The general form of the absorption term will be
\be
\frac{dM}{dt} \,=\, \frac{\sigma}{c}\,\, \rho^{eff} \, \, ,
\ee
which contains the effective energy density $\rho^{eff}$ and the appropriate cross section $\sigma$ for the
gravitational capture of relativistic particles in the background by the black hole.
Since we want to consider relativistic background radiation,
the effective energy density can be defined as
\be
\rho^{eff} = \rho + 3 p(\rho)\,.
\ee
In the case of radiation with an equation of state parameter $w=\frac{1}{3}$, this results in an effective
energy density of
\be
\rho^{eff} = \rho_{rad} + 3 \frac{\rho_{rad}}{3} = 2 \rho_{rad}.
\ee

Thus, the absorption/accretion term for background radiation reads
\be
\frac{dM}{dt} \,=\, \frac{\sigma_{rad}}{c}\,\, 2 \rho_{rad}\, .
\ee
Since the environment is supposed to be isotropic and homogeneous, the cross section for the
absorption of relativistic particles is proportional to the square of the black hole mass
\cite{novikov},
\be
\sigma_{rad} = \sigma_{part} = 27 \pi \frac{G^2 M^2}{c^4}.
\ee
Note that a heuristic deduction of such cross section can be obtained
directly from the spherical geometry of the black hole
$dM = 4 \pi R_S^2\,\,\rho_{rad}\,dt/c = 16\, \pi\, G^2\, M^2 \rho_{rad} dt/c^5$.

In Planckian units the equation for accretion terms reads
\be
\frac{dm}{d\tau} = 27 \pi m^2 \frac{\rho^{eff}}{\rho_{pl}}
\ee
where $\rho_{pl}:={\mathcal E}_p/\ell_p^3$ is the Planck energy density.

Then, a more complete differential equation for the evolution of the mass of a micro black hole
can be given by
\be
\frac{dm}{d\tau} = -\frac{8 \,\pi^3\,\Gamma_\gamma}{15}\,\,
m^2\,\Theta^4 \, {\mathcal A}(\beta, \Theta) \,+\,
54 \,\pi \,m^2\, \frac{\rho_{rad}}{\rho_{pl}}
\label{bhevpl}
\ee
where we used the background equation of state through the effective energy density $\rho^{eff}=2\rho_{rad}$.\\
As stated before, for sake of simplicity we assume that the black hole evaporation evolves according to the standard Hawking
mass-temperature relation (\ref{Hw}), and thus we consider in Eq.(\ref{bhevpl}) the GUP correction function
${\mathcal A}(\beta,\Theta) \simeq 1$. We shall keep
in mind the cutoff on mass/temperature predicted by the GUP, and put it in by hand whenever needed.

Then, the differential equation for the evolution of micro black hole mass/energy $\varepsilon$
can be written as
\be
\frac{d\varepsilon}{dt} = - 12 \pi\, \Gamma_\gamma\, \sigma_3\,\, R_{S}^2\, T^4  +
\frac{54 \pi G^2 \varepsilon^2}{c^7} \rho_{rad},
\label{bhevpl1}
\ee
where $\varepsilon$ is the average energy content of a single black hole, $\varepsilon = M c^2$.
Using expression (\ref{s3}) for $\sigma_3$, and $R_S = 2 G \varepsilon / c^4$, we can write
\be
\frac{d\varepsilon}{dt} \ = \ - \frac{C}{\varepsilon^2} \ + \ D\, \varepsilon^2 \,\rho_{rad}
\label{bhev0}
\ee
with
\be
C=\frac{\Gamma_\gamma\, \hbar\, c^{10}}{3840\, \pi\, G^2}\,; \quad\quad
D=\frac{54 \pi G^2}{c^7}\,.
\ee

\subsection{Evolution equations}
\subsubsection{Cosmological equation}
Given the standard RW metric (with Weinberg conventions but $c\neq 1$)
\be
ds^2 = -c^2dt^2 + a^2(t)\left(\frac{1}{1-kr^2}dr^2 + r^2 d\Omega^2\right)
\label{RW}
\ee
where $d\Omega^2 = d\theta^2 + \sin^2\theta d\phi^2$,
and the energy-momentum tensor of a perfect fluid
\be
T_{\mu\nu} = (\rho + p) u_\mu u_\nu + p g_{\mu\nu}
\ee
where $\rho$ is energy density and $p$ is pressure,
the (00) component of the Einstein equation reads
\be
\left(\frac{\dot{a}}{a}\right)^2 \ + \ \frac{k\,c^2}{a^2} \ = \ \frac{8 \pi G}{3\,c^2} \rho,
\label{00}
\ee
while from the (ii) components we have
\be
2\frac{\ddot{a}}{a} + \left(\frac{\dot{a}}{a}\right)^2 + \frac{k\,c^2}{a^2} = -\frac{8 \pi G}{c^2} p.
\ee
In our model, the energy density has contributions of radiation and matter, and can thus be written as
$\rho = \rho_{rad} + \rho_{mbh}$.
For simplicity, and following Ref.\cite{NG}, we now consider a flat metric, i.e. $k=0$. The equation is then written as
\be
\left(\frac{\dot{a}}{a}\right)^2 = \frac{8 \pi G}{3\,c^2} (\rho_{rad} + \rho_{mbh}).
\label{00ii}
\ee
\subsubsection{Evolution equations for $\rho_{mbh}$ and $\rho_{rad}$}
We suppose our system to consist of a ``soup'' of micro black holes and radiation.
It is well known [see e.g. textbooks by Weinberg or Landau] that for the description of the evolution of the energy
densities $\rho_{rad}$ and $\rho_{mbh}$ under the cosmic evolution of the scale factor $a(t)$ usually
the (0) component of the continuity equation is considered,
\be
\nabla_\mu T^{\mu 0} = G^0\,.
\label{conteq}
\ee
Here $G^0$ is a source-sink term that can appear especially in the description of reciprocally interacting
subparts of the whole system, as we shall see in the next section. $T^{\mu\nu}$ is the energy-momentum tensor of
a perfect fluid.
Specializing this equation to the RW metric, we obtain, for the global energy density
$\rho_{rad} + \rho_{mbh}$,
\be
\dot{\rho}_{rad} + \dot{\rho}_{mbh} + 4 H \rho_{rad} + 3 H \rho_{mbh} = G^0
\label{56}
\ee
where $H=\dot{a}/a$ and $G^0$ is a possible source-sink term.
We shall now compute accurately the form of the continuity equation for both the
subsystems "radiation" and "black holes", in particular the form of the source-sink term.

The system we are investigating is a defined mixture of radiation and black holes, where
the Hubble radius $R_H$ of the universe contains a given fixed
total number $N$ of micro black holes, a given amount of radiation, and the only processes that can happen are exchanges
of energy between the black holes and the surrounding radiation. As already mentioned before,
\emph{in this phase no black hole merging, nor black hole nucleation, is supposed to happen}.

Let us first focus on the evolution equation for $\rho_{mbh}$. Micro black holes are a particular
type of dust: they can emit or absorb radiation. As a first step however, we suppose that micro black holes have a negligible
interaction with radiation (i.e. we treat them as standard dust).
Then the continuity equation, without any source term, can be written
\be
\dot{\rho}_{mbh} + 3 H \rho_{mbh} = 0\,; \quad \quad \quad H = \frac{\dot{a}}{a}
\label{eqcc}
\ee
This equation takes already into account the variations in mass/energy density due to the simple variation
in volume.
In fact, from (\ref{eqcc}) we have
\be
\frac{\dot{\rho}_{mbh}}{\rho_{mbh}} \ = \ -3 \left( \frac{\dot{a}}{a} \right)
\ee
and therefore
\be
\rho_{mbh} = \frac{A}{a^3}
\ee
where $A$ is an integration constant. Hence we see that at any time $t$ we should have
$\rho_{mbh}(t)\,a(t)^3 = A$, so the integration constant should be written as
\be
A = \rho_{mbh}(´t_c)\,a(t_c)^3,
\ee
where $t_c$ is the time point when the constant $A$ is determined. It is a characteristic time
for the onset of matter era, and will be investigated in section \ref{sec:bhnucl}.
Since $a(t)$ is adimensional, A should have the dimensions of an energy density. \\
Conversely, considering black holes as dust grains of constant mass $M$, then the link between mass/energy
density and volume can be immediately written as
\be
\rho_{mbh} = \frac{M c^2\,N}{R_H(t_c)^3} \cdot\frac{a(t_c)^3}{a^3}
\label{dens}
\ee
where $N$ is the {\em total} number of micro black hole in the volume $a(t)^3$ at any instant $t > t_c$.
As will be derived in section \ref{sec:bhnucl}, $N$ is
considered to be constant, since no creation, merging, or complete evaporation of micro black holes are
allowed after the time $t_c$. All the quantities $M$, $t_c$, $N$, $R_H(t_c)$, $a(t_c)$ will be
computed explicitly via numerical simulation in section \ref{sec:bhnucl}.
So we have
\be
\dot{\rho}_{mbh} = - 3 \ \frac{M c^2 N a(t_c)^3}{R_H(t_c)^3 a^4} \ \dot{a} = - 3 \left( \frac{\dot{a}}{a} \right) \rho_{mbh}\,.
\ee
which coincides with (\ref{eqcc}).\\
If now we suppose that also the mass/energy of the single black hole can change in time, then Eq.(\ref{dens})
reads
\be
\rho_{mbh} = \frac{\varepsilon(t)\,N}{R_H(t_c)^3} \cdot\frac{a(t_c)^3}{a(t)^3}
\ee
with $\varepsilon(t) = M(t) c^2$, and this expression immediately suggests by derivation the correct source
term in the continuity equation:
\be
\dot{\rho}_{mbh} \ + \ 3 \left( \frac{\dot{a}}{a} \right) \rho_{mbh} \ =
\  \frac{N\, a(t_c)^3}{R_H(t_c)^3\, a^3} \ \dot{\varepsilon}.
\label{eqcmbh}
\ee

Let us now focus on the equation for the radiation energy density $\rho_{rad}$. As first step, consider the
variation of $\rho_{rad}$ due to presence of emitting/absorbing micro black holes, when the system
radiation/black holes is contained in a
{\em box of fixed volume}. If $d\varepsilon$ is the variation in a time $dt$ of the energy content of a
single black hole, and the box contains $N$ black holes (all of the same mass), then the variation of the
energy of the radiation in the box is
\be
dE = - N d\varepsilon\,.
\ee
Since the volume of the box scales as $V(t)=R_H(t_c)^3\,a(t)^3/a(t_c)^3$, then the variation of the radiation energy density is
\be
d \rho_{rad} = \frac{dE}{V(t)} = - \frac{N d \varepsilon}{V(t)}
\ee
which means
\be
\dot{\rho}_{rad} \ = \ - \frac{N\,a(t_c)^3}{R_H(t_c)^3\,a^3} \ \dot{\varepsilon}
\ee
This relation is true in the hypothesis of a fixed box. If in particular the black hole were inert
(neither absorption nor emission) then $\dot{\varepsilon}=0$ and therefore $\dot{\rho}_{rad}=0$.

In an expanding box, containing only radiation or just a few inert black holes ($\equiv$ dust grains),
we know that from the continuity equation (\ref{conteq}) we can write for $\rho_{rad}$
\be
\dot{\rho}_{rad} + 4 H \rho_{rad} = 0 \quad \quad \Longleftrightarrow \quad \quad \rho_{rad} = \frac{B}{a^4}
\ee
where $B$ is an integration constant.\\
From the previous two steps, it is then clear that, considering both the expanding
box and the emitting/absorbing black holes, we can write globally for $\dot{\rho}_{rad}$
\be
\dot{\rho}_{rad} = - 4 \left(\frac{\dot{a}}{a}\right) \rho_{rad} - \frac{N\,a(t_c)^3}{R_H(t_c)^3\,a^3} \ \dot{\varepsilon}\,.
\label{eqcrad}
\ee
We see that in the equations (\ref{eqcmbh}), (\ref{eqcrad}) the term $N\,a(t_c)^3\,\dot{\varepsilon}/R_H(t_c)^3\,a^3$,
which accounts for the emitting/absorbing activity by black holes, appears with opposite signs, respectively.
This is physically very plausible since, if for example $\dot{\varepsilon} > 0$, then that
term contributes to the accretion of black holes' masses, while exactly the same amount
of energy is taken from the radiation surrounding the black holes. Coherently, we see that the global
continuity equation for black holes and radiation combined reads
\be
\dot{\rho}_{mbh} \ + \ 3 \left( \frac{\dot{a}}{a} \right) \rho_{mbh} +
\dot{\rho}_{rad} + 4 \left(\frac{\dot{a}}{a}\right) \rho_{rad} = 0\,,
\ee
that is, it {\em does not contain} any source term. This is reasonable, since our system contains
only black holes and radiation, and therefore the global energy content must be conserved (only
diluted by the cosmic expansion rate $H \neq 0$).
Systems of equations where one term appears as a source in one equation, and as a sink in another,
are quite common in physics and in cosmology. For example recent models dealing with the interaction between
dark matter and dark energy display such features \cite{jean}.
\subsubsection{Complete set of equations}

We are now able to write down a set of equations that should describe, hopefully in a complete way, the
primordial "soup" of radiation and micro black holes, in a temporal interval ranging from the end of
black hole production era ($t=t_c$) to the
starting of inflation ($t=t_{infl}$). Considering the equations (\ref{bhev0}),
(\ref{eqcmbh}), (\ref{eqcrad}), (\ref{00ii}), we can write the system ($t$ is, as before, the comoving time)
\be
&&\frac{d\varepsilon}{dt} \ = \ - \frac{C}{\varepsilon^2} \ + \ D\, \varepsilon^2 \,\rho_{rad} \nonumber \\
&&\dot{\rho}_{mbh} \ + \ 3 \left( \frac{\dot{a}}{a} \right) \rho_{mbh} \ =
\  \frac{N\, a(t_c)^3}{R_H(t_c)^3\, a^3} \ \dot{\varepsilon} \nonumber\\
&&\dot{\rho}_{rad} \ + \ 4 \left(\frac{\dot{a}}{a}\right) \rho_{rad} \ =
\ - \frac{N\, a(t_c)^3}{R_H(t_c)^3\, a^3} \ \dot{\varepsilon} \nonumber\\
&&\left(\frac{\dot{a}}{a}\right)^2 \ = \ \frac{8 \pi G}{3\,c^2}
\,(\rho_{rad} + \rho_{mbh})
\label{system}
\ee
As we see, this is a system of 4 equations for the 4 unknowns $\varepsilon(t)$,
$\rho_{mbh}(t)$, $\rho_{rad}(t)$, $a(t)$.
This is a good sign for the closure and solvability of the system. However
it is clear that this system is strongly coupled, and moreover nonlinear. Thus, to find an explicit solution
is surely hard and perhaps impossible. Nevertheless, the system can be studied in some physically
meaningful situations (as for example when the micro black holes are very weakly interacting with radiation,
with $\dot{\varepsilon} \simeq 0$, when they essentially behave like dust).
In these limits the equations can yield useful insights on the behavior of the scale factor $a(t)$, which can be
used (via a procedure similar to that of Ref.\cite{NG}) to compute the effects of this "soup" of
radiation and micro black holes on the successive inflation era, and possibly on particular features
of the power spectrum.
\subsection{Pre-inflation matter era}
We study here the regime just sketched at the end of the previous section,
when micro black holes are very weekly interacting with radiation ($\dot{\varepsilon} \simeq 0$).
Our primordial "soup" is therefore composed by radiation and dust.
In this approximation, the second and third equation of system (\ref{system}) can be immediately integrated
to give
\be
\rho_{mbh} = \frac{A}{a^3}\, , \quad \quad \quad \rho_{rad} = \frac{B}{a^4}\,.
\ee
where the integration constants $A$ and $B$ have dimensions as energy densities, and can be written as
\be
A \ = \ \rho_{mbh}(t_c)\, a(t_c)^3 \quad,\quad B \ = \ \rho_{rad}(t_r)\, a(t_r)^4\, .
\label{rho}
\ee
Here $t_c$ and $t_r$ are the characteristic times for the onset
of matter and radiation eras,
respectively. They will both be explicitly specified in the next sections.
Then the (00) equation of system (\ref{system}) reads
\be
\left(\frac{\dot{a}}{a}\right)^2 \ = \ \frac{8 \pi G}{3\,c^2} \,\left(\frac{A}{a^3} + \frac{B}{a^4}\right)\,.
\label{70}
\ee
Equation (\ref{70}) is separable, and can be integrated exactly. The solution obeying the boundary
condition
$a(0)=0$ for $t=0$ is
\be
\frac{2}{3\sqrt{\kappa}A^2}\left[(A a(t) -2 B)\sqrt{A a(t) + B}  +  2B\sqrt{B}\right] = t
\label{71}
\ee
where $\kappa = (8 \pi G)/(3\,c^2)$.
Using the binomial expansion
\be
( 1 + \epsilon)^{1/2} = 1 + \frac{1}{2}\epsilon - \frac{1}{8}\epsilon^2 + \dots
\ee
we find, in the limit $A \to 0$ or equivalently $a(t) \to 0$ for $t \to 0$,
\be
a^2 \ = \ 2 \sqrt{\kappa B}\,\, t
\ee
which is the well know behavior of pure radiation era.\\
In the other regime, when $a$ or $A$ are large, or equivalently $B \to 0$, we find
\be
a^{3/2} = \frac{3}{2}\sqrt{\kappa A}\,\, t
\label{matsol}
\ee
which is the the standard result for a matter dominated era.

We can now wonder how much matter (micro black holes =  dust) should be present in order to have a matter
dominated universe {\em before} the beginning of the inflation.
A condition for this can be easily derived by inspecting the exact solution (\ref{71})
and considering its expansion as
\be
&&\frac{2 a^{3/2}}{3 \sqrt{\kappa A}}
\left[1 - \frac{3}{2}\frac{B}{A a} - \frac{9}{8}\left(\frac{B}{A a}\right)^2   \right.\nonumber \\
&&+ \left. 2\left(\frac{B}{A a}\right)^{3/2}  + {\cal O} \left(\frac{B}{A a}\right)^3 \right] = t.
\ee
Clearly, the universe will be in a matter dominated era at the onset of inflation,
namely for $t = t_I \simeq 10^6 t_P$ (the time when the temperature of the universe corresponds to the GUT energy scale),
whenever the condition
\be
\frac{3}{2}\frac{B}{A\,a} \ll 1
\label{matera}
\ee
is satisfied. An even simpler derivation can be found by writing Eq.(\ref{70}) in the form
\be
\left(\frac{\dot{a}}{a}\right)^2 \ = \ \kappa \,\frac{A}{a^3} \left(1 + \frac{B}{A a}\right)
\ee
from which we read off that the evolution is matter dominated when $B/(Aa) \ll 1$.
In section \ref{sec:bhnucl} we shall compute explicitly via numerical simulation
every step of the black hole nucleation phase, and the associated evolution of the pre-inflationary radiation and matter eras.
We shall conclude that the above matter dominance condition is always met, even well before
the onset of inflation.

\subsection{Inflationary solutions}
\begin{figure}[ht]
\centerline{\epsfxsize=2.9truein\epsfysize=1.8truein\epsfbox{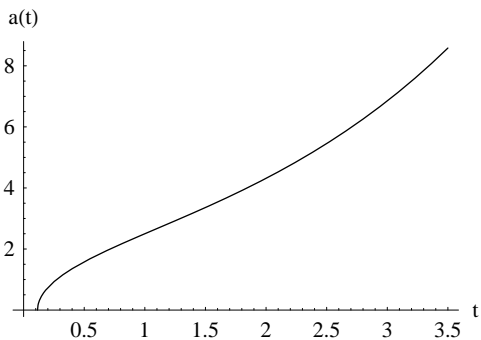}}
\caption[]{Diagram for $a(t)$ versus $t$, in a model assuming subsequent radiation and matter eras before inflation.}
\vspace{0.2cm} \hrule
\label{adotfig}
\end{figure}

In our equations we now also take into account a constant vacuum energy,
namely a cosmological constant.
In this way, we shall be able to generate inflationary exponential solutions.
Following Ref.\cite{NG}, the (00)
component of the Einstein equation in this case reads
\be
\left(\frac{\dot{a}}{a}\right)^2 \ = \ \kappa \,\left(\frac{A}{a^3} + \frac{B}{a^4} + C\right)\,.
\label{adot}
\ee
The constant $C$ in the Friedmann equation results from assuming a power law potential $V(\phi)$ for the inflaton field.
$C$ and the potential are connected by
\be
V(\phi) = \frac{3C}{2} \cdot \phi^2 + C\cdot c_1 \cdot \phi + c_2,
\ee
where $C$ is the quasi de Sitter parameter in the Friedmann equation, and $c_1$ and $c_2$ are constants to be fixed
from the inflationary model.
In other words, $C$ is mimicking the potential for the inflaton field.
The previous equation can be written as
\be
\left(\frac{\dot{a}}{a}\right)^2 \ = \ \kappa \,\frac{A}{a^3}
\left(1 + \frac{B}{A a} + \frac{C a^3}{A}\right)
\ee
and under the matter era condition, $B/(Aa)\ll 1$, it becomes
\be
\left(\frac{\dot{a}}{a}\right)^2 \ = \ \kappa \,\left(\frac{A}{a^3} + C\right)\,.
\label{adotm}
\ee
Again, this equation is easily separable, and the solution obeying the boundary condition $a(0)=0$
for $t=0$ is
\be
a(t) = \left(\frac{A}{C}\right)^{1/3}\left[\sinh\left(\frac{3}{2}\sqrt{\kappa C}\,\,t\right)\right]^{2/3}\,,
\label{atm}
\ee
which, for vanishing $C$, or small $t$, results in solution (\ref{matsol}),
\be
a(t) \simeq \left(\frac{3}{2}\right)^{2/3} \left(\kappa A\right)^{1/3}\, t^{2/3},
\ee
while for large $t$, it exhibits an
exponential (i.e. inflationary) behavior \cite{vilenkin},
\be
a(t) \simeq \left(\frac{A}{4 C}\right)^{1/3}\exp\left(\sqrt{\kappa C}\,\,t\right)\,.
\ee
We can grasp an idea of the overall solution $a(t)$ by numerically integrating Eq.(\ref{adot}).
The evolution of the scale factor is shown in Fig.\ref{adotfig}.

In case of radiation dominated pre-inflation era, i.e. no matter present ($A=0$), equation (\ref{adot}) reads
\be
\left(\frac{\dot{a}}{a}\right)^2 \ = \ \kappa \,\left(\frac{B}{a^4} + C\right)\, ,
\label{adotr}
\ee
and the solution obeying the boundary condition $a(0)=0$ for $t=0$ is
\be
a(t) = \left(\frac{B}{C}\right)^{1/4} \left[\sinh\left(2\sqrt{\kappa C}\,\,t\right)\right]^{1/2},
\label{atr}
\ee
which for small $t$ or vanishing $C$ is
\be
a(t) \ = \ \sqrt{2}\,\, (\kappa B)^{1/4}\,\, t^{1/2} \,.
\ee

It is also useful to derive a condition for the onset of inflation. From Eq.(\ref{adot}) we can obtain
the sign of $\ddot{a}$
\be
\left(\frac{\ddot{a}}{a}\right) \ = \ \kappa \,\left(-\frac{A}{2 a^3} - \frac{B}{a^4} + C\right),
\ee
and $\ddot{a}>0$ if
\be
C > \frac{A}{2 a^3}\left(1+ \frac{2 B}{Aa} \right) \gtrsim \frac{A}{2 a^3},
\label{inflcond}
\ee
where we used the pre-inflation matter era condition $B/(Aa)\ll 1$. Therefore we shall be in a full
inflationary era, $\ddot{a}>0$, when
\be
a \gtrsim \left(\frac{A}{2C}\right)^{1/3}\,.
\ee

\section{Black Hole Nucleation: Numerical Simulation}
\label{sec:bhnucl}
In this section we numerically simulate the nucleation of micro black holes in pre-inflation era,
and come up with a number of micro black hole remnants sufficient to make the universe pass
from a radiation dominated to a matter dominated pre-inflation era. We will express every quantity
in planckian units, meaning e.g. $\tau = t/t_p$, $\Theta = T/T_p$, $m = M/M_p$.
Thus every quantity is dimensionless. \\
In 1982 Gross, Perry and Yaffe~\cite{gross} investigated the stability of flat space and the arising
gravitational instabilities, which might lead to singularities. They used the formalism of path
integrals in a quantum version of Einstein's theory of gravity to analyze these gravitational
fluctuations. As a concrete example, they took a box filled with thermal radiation to derive an
expression for the probability for the spontaneous formation of black holes out of the gravitational
instabilities of spacetime. Two years later, Kapusta~\cite{kapusta} gave an alternative heuristic
derivation of the nucleation rate, using the classical theory of nucleation during a thermodynamical
phase transition. He reproduced the rate nearly completely with the classical approach
considering the change in free energy of the system during the nucleation of a black hole,
and completed the analogy by inserting by hand quantum corrections into his classically derived formula.
See Appendix 1 for the explicit steps of such derivation.\\
The nucleation rate for black holes reads (Eq.(\ref{nucformula}), Appendix 1)
\begin{equation}
    \Gamma_N(\Theta) = \frac{8\pi}{15\cdot 64\pi^3} \cdot \Theta^{-167/45}  e^{-1/16\pi \Theta^2},
    \label{eq:orignucl}
\end{equation}
where $\Theta$ is the temperature of the universe (the thermal bath), expressed in Planck units, and \emph{at the same time}
the temperature of the nucleated black holes, connected to their mass $m$ by
\begin{equation}
    \Theta = \frac{1}{4\pi m}.
    \label{eq:tempmass}
\end{equation}
At a given temperature $\Theta$, all the black holes created according to this nucleation rate
will have mass $m$. \\

As stated before, the pre-inflation era is supposed to take place from the Planck time $t_p\simeq 10^{-43}\,s$
to the onset of inflation, $t_{inf} \simeq 10^{-37}\,s$, when the temperature of the universe has
reached the GUT energy scale.
At very early times, when $\Theta > \Theta_* \simeq 1/(4\pi)$, the nucleation probability is very high,
but does not lead to black hole formation as it is forbidden by the GUP to create smaller than
the Planck mass black holes ($m_{\rm min} \simeq \sqrt{\zeta}\,\, M_p$ where $\zeta \sim 1$).
For temperatures above $\Theta_* \simeq \frac{1}{4\pi}$ production of small black holes is not possible.
So at least for this very early time, the universe is simply a chaotic hot sphere that we suppose to be
filled with primordial radiation, following the approach of Refs.\cite{Powe2007, NG}.
There might be regions with larger density than others, but the conditions are
too chaotic to allow formation of stable objects like black holes. The universe can thus be assumed
to be radiation-dominated at the beginning, and will migrate to being matter-dominated at a later time,
when black holes are starting to be formed. Considering an adiabatically expanding universe, we can write
$T(\tau)a(\tau) = T_p a(t_p)$, and since during the radiation era the scale factor evolves like
$a_r(\tau) \simeq a(t_p)\tau^{1/2}$, and we choose $a(t_p)=1$, we have
\begin{equation}
    \Theta_r = \frac{1}{\tau^{1/2}}
    \label{eq:temptimerad}
\end{equation}
during the time when the universe is dominated by radiation.
For a matter-dominated universe, the temperature evolution is analogously given by
\begin{equation}
    \Theta_m = \frac{1}{\tau^{2/3}}.
    \label{eq:temptimematter}
\end{equation}

\subsection{State of the universe}
\label{sec:stateofuniverse}

The parameters in the Friedmann equation containing matter and radiation energy densities
are defined as (Eq.(\ref{rho}))
\begin{subequations}
    \begin{align}
    A = \rho_m(\tau_c) \cdot a^3(\tau_c), \label{eq:ABA} \\
    B = \rho_r(\tau_p) \cdot a^4(\tau_p),  \label{eq:ABB}
    \end{align}
    \label{eq:AB}
\end{subequations}
and their dimension has to be energy per volume, as the scale factor is a dimensionless quantity. \\
At the Planck time, the universe is in a radiation-dominated stage, and it is reasonable to
assume that $\rho_r(\tau_p) = \rho_p$. Assuming $a(\tau_p)=1$ (which is an unconventional,
but convenient choice, and will be converted to the common notion
of $a(\tau_{today}) = 1$ later), the radiation parameter can
be fixed as $B=1$, expressed in Planck units. For the matter parameter, we have to choose a time $\tau_c$ when black holes
are starting to be nucleated, and calculate $\rho_m(\tau_c)$. This time and the parameter
$A$ will be derived in the next subsections via numerical simulations. \\
To calculate the time of the transition from radiation- to matter-dominance (which evidently takes place after
$\tau_c$), we consider the ratio
\begin{equation}
    R(\tau) = \frac{\rho_m(\tau)}{\rho_r(\tau)} = \frac{\rho_m(\tau)}{\rho_p} \cdot \tau^2.
    \label{eq:ratio}
\end{equation}
At the very beginning there is only radiation. So for a while $\rho_m(\tau)=0$, the evolution
of $\rho_r(\tau)$ is driven by radiation, $a(\tau) \sim \tau^{1/2}$, and therefore $\rho_r \sim \rho_p/\tau^2$.
This is correct at least before black holes are created, whereas after
the onset of nucleation, $a(\tau)$ evolves in a more complex manner dictated by equations (\ref{70}, \ref{71}).
As nucleation starts, $\rho_m(\tau)$ grows, and some time later $R(\tau)$ crosses 1.

$R(\tau)$ can only be used for qualitative statements at the beginning of nucleation, since,
in the way it is defined, it is rigorously valid only until the start of matter nucleation, and it
does not contain any information about the time-development of the scale factor according to the
full Friedmann equation, when matter and radiation are both present.
So it should only be applied during short time spans, just after the nucleation starts, when the
scale factor doesn't change significantly. The condition $3B/(2Aa)\ll 1$ is the only significant criterion
to fully determine the radiation to matter
transition of the universe at later times, when matter and radiation are both present.

\subsection{Nucleation Process}
\label{sec:nucleation}
\begin{figure}[ht]
    \begin{center}
    \includegraphics[width=0.4\textwidth]{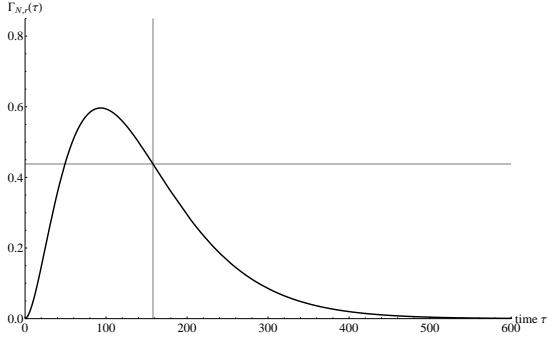}
    {\small \caption{Nucleation rate, $\Gamma_{N,r}(\tau)$ over time $\tau$.}
    \label{fig:rateradiation}}
    \end{center}
\end{figure}

Using~\eqeqref{eq:temptimerad}, in radiation era we can write the nucleation rate as a function of time as
\begin{equation}
    \Gamma_{N,r}(\tau) = \frac{1}{15\cdot 8\pi^2} \cdot \tau^{167/90}  e^{-\tau/16\pi}\,.
    \label{eq:nuclrate}
\end{equation}
The temperature $\Theta$ of the Universe is linked not only to time, but also to the mass of the
nucleated black hole, since a black hole, at the instant of its creation, is in thermal equilibrium with the
rest of the Universe.
In fact
\begin{equation}
    \Theta = \frac{1}{4\pi m} \quad {\rm and} \quad \Theta_r = \frac{1}{\tau^{1/2}}~\Rightarrow~\tau_r = 16\pi^2 m^2.
    \label{eq:masstime}
\end{equation}
It should be explicitly noted that the relation (\ref{eq:masstime}) does not express the time evolution
of the mass of one black hole, but the dependence of the initial mass of the nucleated black holes on time.
The evolution with time of the
black hole mass, due to evaporation/accretion
processes, is given by relation (\ref{bhevpl}), while equation (\ref{eq:masstime}) expresses the evolution
with time of the masses of the black holes \emph{at the instant of their creation}.
Therefore the cutoff from the Generalized Uncertainty Principle, which gives a minimum mass $m \sim 1$,
can be translated also in terms of time as $\tau_* = 16\pi^2 \simeq 158$.
This can be seen in~\figref{fig:rateradiation} - the curve is truncated at $\tau_*$
(vertical line), which corresponds to the cutoff at about one Planck mass.
This relation also implies that the black holes nucleated at later times have larger masses,
according to~\eqeqref{eq:masstime}, while the probability of their formation decreases.

From the onset of nucleation the number of black holes nucleated each Planck time per Planck volume
is given by the rate~\eqref{eq:nuclrate}, but we have to monitor closely the overall state of the
universe. When the ratio $R(\tau)$ crosses unity, then the universe changes to a
matter-dominated stage, in which case the nucleation rate is no longer given by~\eqeqref{eq:nuclrate}.
As soon as the phase transition happens, the nucleation rate has to be given in terms of the
temperature in a matter-dominated universe ($\Theta_m = \tau^{-2/3}$):
\begin{equation}
    \Gamma_{N,m}(\tau) = \frac{1}{15\cdot 8\pi^2} \cdot \tau^{334/135}
    \cdot e^{-\,\tau^{4/3}/\,16\pi}.
    \label{eq:nuclratematter}
\end{equation}
In order to get the number of black holes nucleated with time, we have to do a time integration
over the nucleation rate, multiplied with the volume of the universe in each time point.
Considering that the Hubble radius is defined as $R_H=c/H$, and that for power law expansion rates
($a(t) \sim t^{\alpha}$) we have $R_H \sim ct$, then, taking as initial condition $R_H(\tau_p)=\ell_p$,
we can write (in units of Planck length) $R_H(\tau) = \tau$. Therefore (since $\ell_p^3 = V_p = 1$)
\begin{equation}
    V_H(\tau) = \frac{4\pi}{3} \tau^3.
    \label{eq:Hubble}
\end{equation}
But before we go on to calculate the black hole number, we take into account another principle,
which will give a more stringent cutoff on the nucleation of black holes than the GUP does.

\subsection{A Cutoff from the Holographic Principle}

\label{sec:HolPrindetails}
\begin{figure}[ht]
    \begin{center}
    \includegraphics[width=0.4\textwidth]{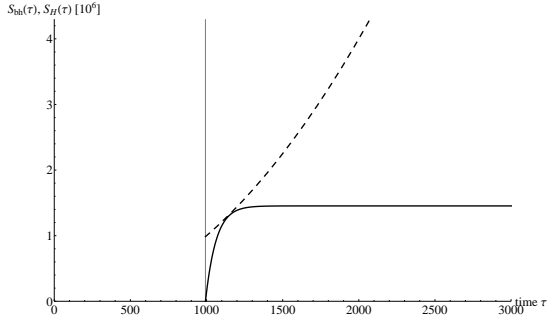}
    {\small \caption{The cutoff as defined by the holographic principle at $\tau=993$:
    the curves touch, but the entropy of the black holes (full line) never exceeds the
    maximum entropy that can be held by the Hubble sphere (dashed line).}
    \label{fig:goodcutoff}}
    \end{center}
\end{figure}

The holographic principle \cite{tHooftSusskindBousso} places a limit on the information content,
or entropy content, in a certain region of space-time. Quantitatively it states that (in units where $k_B=\ell_p=1$)
\begin{equation}
    S[L(B)] \ \leqslant \ \frac{A(B)}{4},
    \label{eq:covboundagain}
\end{equation}
where $L(B)$ is a so-called light sheet, which defines a certain region of space-time $B$,
and $A(B)$ is the codimension 1 boundary of that region. For our situation, applying the
holographic principle simply means that the total entropy contained in the Hubble sphere cannot exceed
the entropy of a black hole of size equal to the Hubble sphere, which represents the maximum
entropy that can be held in that spacetime region, as black holes are the most entropic objects.
According to Refs.~\cite{Beke1973,Wald2000}, the expression for the entropy of a black hole is
\begin{equation*}
    S_{bh} = \frac{A_{bh}}{4},
\end{equation*}
and so the entropy of a black hole of the size of the Hubble sphere (HS) is given by
\begin{equation}
    S_{HS}(\tau) = \frac{A_{HS}}{4} = \pi R_H^2(\tau).
    \label{eq:entropyHubble}
\end{equation}
This can be used to define a cutoff for the nucleation rate.
We demand that at no time point in the evolution of the universe the entropy of the black holes can exceed
the total entropy that Hubble sphere can maximally hold, and then we try to find a time point $\tau_c$,
from which this condition is fulfilled. If this condition is violated in
the course of black hole production, then \emph{it is simply not allowed to create black holes}.
The entropy of the black holes within the Hubble sphere is
\begin{equation}
    S_{bh}(\tau) = \int_{\tau_*}^{\tau}{\Gamma_{N,r}(\tau') \cdot \pi r_s^2(\tau')
    \cdot \Bigl( \frac{4\pi}{3} \tau'^3 \Bigr)\:d\tau'},
    \label{eq:entropybhs}
\end{equation}
where the Schwarzschild radius of the black holes is $r_s=m = \tau^{1/2}/(4\pi)$,
from Eq.(\ref{eq:masstime}),
and we require that at all times
\begin{equation*}
    S_{bh}(\tau) \leqslant S_{HS}(\tau).
\end{equation*}
Again, since the universe has started off in a radiation-dominated era, and the black hole
nucleation has not yet begun, the rate as a function of time is defined using $\Theta_r = \tau^{-1/2}$.
It is yet unknown when the production of black holes will change the state of the universe to
be matter-dominated, but to find out the time of nucleation start, we are for now bound to the
assumption of radiation dominance and thus will use $\Gamma_{N,r}(\tau)$ in the above expression
for the entropy.\\
The cutoff time turns out to be $\tau_c=993$, as can be seen in~\figref{fig:goodcutoff},
which shows the entropy of the Hubble sphere (dashed line) and that of the black holes (full line)
as a function of time.

\begin{figure}[ht]
    \begin{center}
    \includegraphics[width=0.4\textwidth]{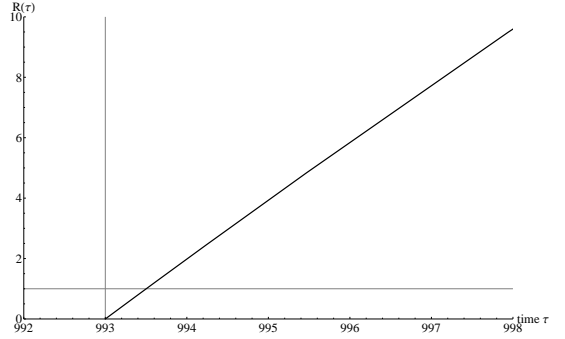}
    \small{ \caption{Ratio $R(\tau)$ of matter to radiation
    density in the early universe, after the nucleation starts at $\tau = 993$.}
    \label{fig:ratio}}
    \end{center}
\end{figure}

With $\tau_c$ being determined, we can calculate the number of black holes according to
\begin{equation}
    N_{bh}(\tau) = \int_{\tau_c}^{\tau}{\Gamma_{N,r}(\tau') \cdot
    \Bigl( \frac{4\pi}{3} \tau'^3 \Bigr) d\tau'},
\end{equation}
and the matter density as
\begin{equation}
    \rho_{m}(\tau) = \frac{1}{\tau^3} \int_{\tau_c}^{\tau}{\Gamma_{N,r}(\tau') \cdot
    m(\tau') \cdot \tau'^3 d\tau'}.
    \label{eq:massdensityrad}
\end{equation}
However, in order to correctly calculate the black hole number and the mass density,
we need to know when the universe will migrate from being radiation-dominated to being
matter-dominated. We know that black hole nucleation starts when the universe is in a
radiation-dominated state. For the very beginning of black hole nucleation, for a time span
of at least one Planck time, we have to use the rate $\Gamma_{N,r}$ to produce black holes.
After one Planck time, we have to check whether the universe is still in radiation- or already
in matter-dominated stage, and take the according production and expansion rates to follow
the black hole production.
A new parameter $d_r$ is introduced, which denotes the duration of this radiation-dominated
phase of black hole nucleation, until the universe reaches a matter-dominated state.
The matter density of the black holes can be expressed like
\be
    &\rho_{m}(\tau)& = \frac{1}{\tau^3} \int_{\tau_c}^{\tau_c+d_{r}}{\Gamma_{N,r}(\tau') \cdot
    m_{r}(\tau') \cdot \tau'^3 d\tau'} \nonumber \\
    &+& \frac{1}{\tau^3} \int_{\tau_c+d_{r}}^{\tau}{\Gamma_{N,m}(\tau') \cdot
    m_{m}(\tau') \cdot \tau'^3 d\tau'},
    \label{eq:massdensity}
\ee
where $m_r=\tau^{1/2}/(4\pi)$ and $m_m=\tau^{2/3}/(4\pi)$.
We have separated integrals for the two separate stages of black hole production.
We can now vary the parameter $d_{r}$ and investigate the ratio we are looking for, (here $\rho_p=1$)
\begin{equation}
    R = \rho_m(\tau) \cdot \tau^2,
\end{equation}
to determine at which point the dominant substance in the universe changes. \\
As can be seen from~\figref{fig:ratio}, it turns out that the universe very quickly reaches a matter-dominated stage
after the onset of
black hole production. After only a few Planck times, the ratio is clearly above unity,
and so we can safely assume that the universe is in a matter-dominated stage at about $3-4$ Planck
times after the black hole nucleation has begun.

With this assumption the estimated number of primordial black holes is given by
\begin{equation*}
    N(\tau_{inf}) \sim 10^4.
\end{equation*}
The masses of the black holes nucleated between $\tau_c=993$ and $\tau=996$, according to
the mass-time-relation~\eqeqref{eq:masstime}, are
\begin{equation*}
    m(\tau_c) \sim 2.5 .
\end{equation*}

\begin{figure}[ht]
    \begin{center}
    \includegraphics[width=0.4\textwidth]{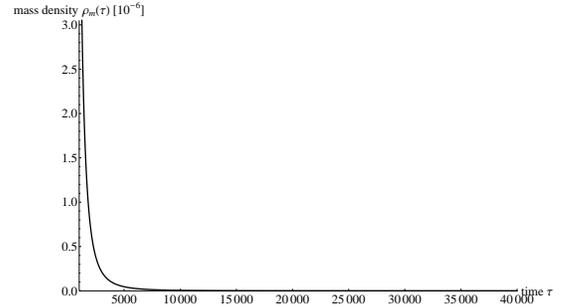}
    {\small \caption{The mass density of the universe as a function of time.}
    \label{fig:massdensity}}
    \end{center}
\end{figure}

We can now also calculate and plot the mass density,~\eqeqref{eq:massdensity} in evolution with time.
It is shown in~\figref{fig:massdensity}.

\subsection{Collision Rate and Black Hole Thermodynamics}

\label{sec:collisions}

\begin{figure}[ht]
    \begin{center}
    \includegraphics[width=0.4\textwidth]{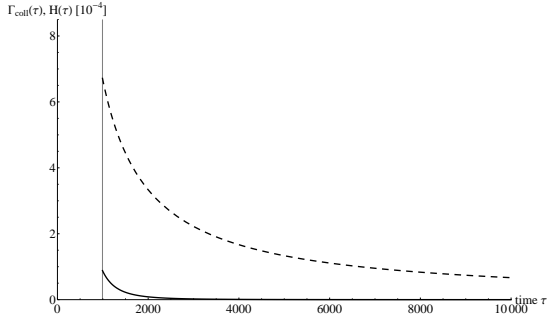}
    {\small \caption{The collision rate of black holes (full line) in development with time,
    together with the Hubble rate (dashed line).}
    \label{fig:collisionHubble}}
    \end{center}
\end{figure}

What is left to do to justify the assumption of a matter-dominated era due to the existence of
black holes is to analyze the results in the context of black hole collisions and merging.
The collision rate of black holes can be given by starting from a general definition of a
scattering rate. We consider black holes of mass $m(\tau_c)$, and their velocity to be determined by Brownian motion, therefore
$v = \sqrt{\frac{kT}{M}}$. Being $n$ the number of black holes per unit volume,
and $\sigma$ the scattering cross section, we arrive at
\begin{eqnarray}
    \Gamma_{coll} & = & n \sigma v \nonumber\\
    & = & n_{bh} \cdot 4\pi r_s^2 \cdot c\sqrt{\frac{\Theta}{m}}\nonumber\\
    & = & \frac{N_{bh}(\tau)}{V_H(\tau)} \cdot 4\pi m_c^{3/2} \cdot \tau^{-1/3}
\end{eqnarray}
where in the last line we used $r_s=m$ and $\Theta_m=\tau^{-2/3}$, with $c = \ell_p = 1$.
The rate can be seen in~\figref{fig:collisionHubble}, together with a dashed line denoting the
Hubble expansion rate. The collision processes are effective as long as the rate is higher than
the Hubble expansion rate - otherwise the reaction decouples, and collisions stop because the
expansion of the universe is faster than the time between collisions. As can be seen clearly
from the plot, the collision rate is at all times lower than the Hubble rate, and thus
no collisions or mergings of black holes take place.

\begin{figure}[ht]
\centerline{\epsfxsize=2.9truein\epsfysize=1.8truein\epsfbox{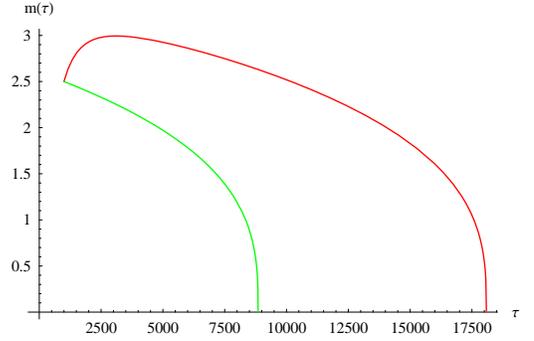}}
\caption[]{Black holes lifetime diagrams: the upper (red) for an evaporating/absorbing black hole;
the lower (green) for an evaporating only black hole.}
\vspace{0.2cm} \hrule
\label{BHfeed}
\end{figure}

We can also do some estimations of the black hole thermodynamics. As we have seen that the black
holes are not colliding with each other and don't have a chance to merge, we know that the black
hole masses can only change by accretion or evaporation.
This process is described by the differential equation (\ref{bhevpl}).
In our specific case we consider, as usual, the correction function ${\mathcal A}(\beta,\Theta) \simeq 1$ and
the greybody factor $\Gamma_\gamma=1$.\\
From $\tau_c \simeq 1000$ to the onset of inflation the Universe is matter dominated,
therefore we can write for the scale factor
\be
a(\tau) \simeq \left\{ \begin{array}{ll}
a(t_p)\,\tau^{1/2} \quad\quad\quad\quad\quad 1 < \tau \leq \tau_c \simeq 10^3\, t_p\\ \\
a(t_p)\,\tau_c^{-1/6}\, \tau^{2/3} \quad\quad \tau_c\leq \tau < \tau_{infl}\simeq 10^6 \, t_p
\end{array}
\right.
\label{scalef}
\ee
Since $a(t_p)=1$, we can write $a(\tau) \simeq 10^{-1/2}\,\tau^{2/3}$ and
\be
\rho_{rad}(\tau) \ = \ \frac{B}{a^4(\tau)}\ = \ \frac{\rho_r(t_p)a^4(t_p)}{a^4(\tau)}\
= \ \frac{100\,\rho_p}{\tau^{8/3}}
\ee
So the differential equation (\ref{bhevpl}) becomes
\be
\frac{dm}{d\tau} \ = \ -\frac{1}{480\, \pi\, m^2} \ + \  \frac{5400\, \pi\,m^2}{\tau^{8/3}}
\ee
which can be numerically integrated with the initial condition
\be
m(\tau=998) \simeq 2.5
\ee
The numerical integration confirms that, despite the absorption term, the black hole evaporates completely
in a time $T(m)\sim 17086\,\, t_p$. Actually, without the absorption term, the lifetime would be about
$50\%$ less, just $T(m)= 7854\,\,  t_p$. In any case, by a cosmic time of $\tau \sim 2 \cdot 10^4\,\, t_p$,
well before the inflation starts,
all the black holes have evaporated down to the Planck size remnants predicted by the GUP. In Fig.~\ref{BHfeed}
the two lifetime diagrams, one for evaporating/absorbing black holes, the other for
evaporating only black holes, are compared.

\subsection{Fixing the Friedmann Equation}
\label{sec:fixFriedmann1}
In this subsection we come back to the original goal of this part: to determine the numerical size of the parameters in
the Friedmann equation that correctly describe the universe developed in our model. We have already
settled $B=1$ by simply assuming that the density of radiation at the Planck time was equal to the
Planck density. For the matter component, we know that about $10^4$ black holes are nucleated during
a short period around $\tau \sim 10^3$ in the pre-inflation era. $A$ is given by~\eqeqref{eq:ABA}.
The matter density can be simply estimated by
\begin{equation*}
    \rho_m(\tau_c) \sim \frac{10^4~\mathrm{black~holes}}{R_H^3(\tau_c)} \sim
    \frac{10^4\,\sqrt{\zeta}\,\epsilon_p}{10^9\,V_p}\,,
\end{equation*}
where $m_{\rm min}=\sqrt{\zeta}$ is the minimum mass predicted by the GUP.
The scale factor is chosen as $a(t_p) = 1$, and then evolves in radiation-dominance until
$\tau_c \sim 10^3$; thus $a(\tau_c) \sim \tau_c^{1/2} \sim 10^{3/2}$. Putting these together,
the final result for the matter component is
\begin{equation}
    A = \frac{10^4\,\sqrt{\zeta}\,\epsilon_p}{10^9\,V_p} \cdot 10^{9/2}
    \sim 10^{-1/2} \,\sqrt{\zeta}\,\,\frac{\epsilon_p}{V_p}.
    \label{eq:Afinal}
\end{equation}
As stated before (see Sections \ref{GUP}, \ref{minmass}), we consider $\zeta \sim 1$.
With this number, we can confirm the actual validity of the matter-dominance condition
for the whole era $\tau_c < \tau < \tau_{infl}$. In fact in this era, according to Eq.~(\ref{scalef}),
the scale factor evolves over the values $10^{3/2} < a(\tau) < 10^{7/2}$,
and therefore the matter-dominance ratio spans the values
\begin{equation*}
    \frac{3}{2} \: \frac{B}{A\cdot a(\tau)} \simeq  \frac{1}{10^{-1/2}\cdot a(\tau)} \sim 10^{-1}-10^{-3} \ll 1\,,
\end{equation*}
which confirms that the Universe is in matter era from $\tau_c$  to the onset of inflation.\\
With all this information, we are fully equipped to commence the calculations of the primordial
power spectrum. It should be noted that $C$ is not yet determined - this will be taken care
of in the next sections.

\subsection{The Scenario without the GUP}
\label{nogup}
What if the nucleated black holes wouldn't be protected by the GUP from evaporating completely but would just transfer their
energy to radiation filling the universe? There would be no era of matter-dominance before inflation, but there would only
be radiation until the onset of inflation. Or would it? In the previous chapter, we have found out that the black hole
nucleation is very powerful and the universe is put into a matter-dominated stage nearly immediately. If the GUP is assumed,
the evaporation stops when the mass of the black hole reaches Planck mass, and the Universe stands in matter era until
the onset of the inflation. On the contrary, if the GUP is not valid, the black holes, once created, can evaporate
down to zero mass. The universe remains in a matter-dominated phase for the time of evaporation, but when all black
holes have vanished, radiation is the dominant species again, until inflation starts. That means, there are three
stages during the pre-inflationary era, and the evolution and growth behavior of the scale factor is changing at
three transition points in time. The parameters of the Friedmann equation have to be evaluated anew, and this leads
to a new differential equation for the field perturbation and a new $k(a)$-relation (see Sec.\ref{quintoA}). \\
We can summarize the time frame of this scenario as follows. The universe starts out radiation-dominated and remains
so until the start of black hole nucleation at time $\tau_c \sim 10^3$, where the scale factor has expanded to
about $a(\tau_c) = 10^{3/2}$. The subsequent matter phase is short and only lasts
from $\tau_c$ to $\tau_r \sim 2\cdot 10^4$, as we know from the previous calculations that the nucleated black holes
have an approximate life span of $2 \cdot 10^4 t_p$. During this time span the scale factor expands for
a factor of $10^{2/3}$. Again follows a period of radiation dominance, which lasts from $\tau_r$ to $\tau_{inf}$.
The temperature at the onset of inflation here again has to match the GUT energy scale, and for this reason
inflation is starting a little later, at $\tau_{inf}\sim 10^8$. From $\tau_r$ to $\tau_{inf}$ there are four
orders of magnitude in time, and so the scale factor expands for a factor of $10^2$. \\
This is the situation we are facing when the GUP is not included in the theory. However, there is a third
possibility that can be considered. \\

\subsection{The Scenario without any black hole}
After the previous subsection, it is also justified to ask for the case when there are no black holes at all,
and no black hole nucleation. What happens if we basically neglect all the black holes
nucleation processes in the early universe, and simply assume that the pre-inflation era was completely
radiation-dominated? The results from this scenario will then actually correspond to the investigations
done before (e.g.~\cite{Powe2007,NG}), completely cutting out the possibility of black hole formation.
In those previous works, a suppression of the CMB power spectrum for low multipole modes, resulting from
a suppression in the primordial power spectrum, was found. \\
Under these assumptions, the universe is in a radiation-dominated stage from the Planck time until the
temperature reached GUT energy scales, at about $\tau = 10^7 t_p$. During this time span, the scale factor
can expand from one Planck length to
\begin{equation}
    a_{inf} = 10^{7/2}.
\end{equation}
The power spectra resulting from the scenario introduced here are computed in section \ref{PS}.
%
%
%
\section{Scalar field fluctuations on an evolving background - The primordial power spectrum}
%
In this section we study the quantum fluctuations of a field living in a
universe whose background evolves with a given scale factor $a(t)$. To facilitate comparison
with the case of a pre-inflation radiation dominated universe already considered in Ref.\cite{NG},
we choose the same convention on the metric, namely
\be
ds^2 = g_{\mu\nu}dy^{\mu}dy^{\nu} = dt^2 - a(t)^2 d\vec{y}^2 \,.
\label{frw}
\ee
Here we set $c=1$, and we choose a flat metric (for simplicity, and since we deal with an
almost flat universe). We consider a zero mass scalar field $\Phi(t, \vec{y})$ and we perturb
the field around its classical expectation value,
$\Phi(t, \vec{y}) = \Phi_0(t, \vec{y}) + \varphi(t, \vec{y})$. The equation of motion for the scalar field
perturbation then reads
\be
\Box\varphi(t, \vec{y}) = 0,
\label{boxeq}
\ee
where
\be
\Box = -\frac{1}{\sqrt{-g}}\partial_\mu(\sqrt{-g} g^{\mu\nu}\partial_\nu)\,.
\ee
In the applications, the $a(t)$ of Eq.(\ref{atm}), or of Eq.(\ref{atr}), will be used in the metric $g_{\mu\nu}$.\\
Taking the Fourier transform for $\varphi$
\be
\varphi(t,\vec{y}) = \int \phi_k(t)\,\, e^{-i\, \vec{k}\cdot\vec{y}}\,\, d\vec{k},
\ee
we obtain from (\ref{boxeq}) the equation of motion for $\phi_k(t)$,
\be
\ddot{\phi}_k(t)\, +\, 3 \frac{\dot{a}}{a}\,\dot{\phi}_k(t)\,
+\, \left(\frac{k^2}{a^2}\right)\phi_k(t) \ = \ 0.
\label{eqphi}
\ee

In order to solve Eq.(\ref{eqphi}), different strategies are customarily used.\\
One possibility is to introduce the \emph{conformal time} $\eta$ defined by the relation
\be
d\eta = \frac{dt}{a(t)},
\ee
which implies
\be
\frac{d}{dt} = \frac{d\eta}{dt}\frac{d}{d\eta} = \frac{1}{a(\eta)}\frac{d}{d\eta}\,.
\ee
Then the equation for $\phi_k(\eta)$ becomes
\be
\phi_k''(\eta) \ +\  2\frac{a'}{a}\,\phi_k'(\eta) \ +\  k^2 \phi_k(\eta) \  = \ 0,
\ee
where a "prime" indicates a derivative with respect to $\eta$. Applying now a well known general procedure,
valid for any second order differential equation,
we can make the first derivative disappear. In fact, defining
\be
\phi_k(\eta) := v_k(\eta) p(\eta)
\ee
with
\be
p(\eta) = \exp\left[-\frac{1}{2}\int\frac{2a'}{a}\,d\eta\right],
\ee
we find
\be
p(\eta)=\frac{1}{a(\eta)},
\ee
and henceforth the equation for $v_k(\eta)$ reads
\be
v_k'' \ + \ \left(k^2 - \frac{a''}{a}\right)\,v_k \ = \ 0\,.
\label{eqv}
\ee
Another possible strategy to solve Eq.(\ref{eqphi}) is to change the independent variable from $t$ to $a$.
Then
\be
\frac{d}{dt} = \frac{da}{dt}\frac{d}{da} = \dot{a}\frac{d}{da}\,,
\quad\quad \dot{\phi}_k \ = \ \dot{a}\,\phi'_k,
\ee
and the equation for $\phi_k(a)$ reads
\be
\dot{a}^2 \phi''_k \ + \ \left(\ddot{a} + 3\frac{\dot{a}^2}{a}\right)\phi'_k \ + \
\left(\frac{k^2}{a^2}\right)\phi_k \ = \ 0,
\label{eqphia}
\ee
where of course $\dot{a}$ and $\ddot{a}$ have to be expressed as functions of $a$.
In the following, we shall use indifferently both procedures.

\subsection{The re-entering k-modes}
\label{quintoA}
\begin{figure}[ht]
\centerline{\epsfxsize=2.9truein\epsfysize=1.8truein\epsfbox{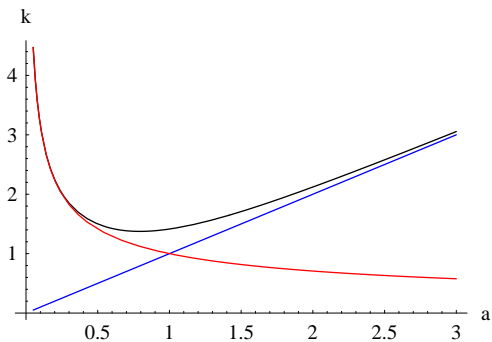}}
\caption[]{Diagram for $k$ versus $a$ (full line), in pre-inflation matter era}
\vspace{0.2cm} \hrule
\label{kafig}
\end{figure}

The k-modes which left the horizon at or shortly after the onset of inflation are those that are just now
re-entering our Hubble radius, and they represent the largest modes of fluctuations currently observable, having a size
comparable with that of the visible universe.
When we solve equation (\ref{eqphi}), or equivalently Eqs. (\ref{eqv}), (\ref{eqphia}),
we fix a particular value of the parameter $k$, i.e. a particular mode, and we obtain the solution for this mode, introducing
suitable boundary conditions.
We are concerned with modes that leave the horizon just at or shortly after the beginning of inflation, because
they correspond to the
largest scales observable today, and they could bring imprints of a possible pre inflation era.
A pre-inflationary era affecting these modes could thus
explain the anomaly of the
quadrupole moment of the CMB today. Simple geometric
considerations will help us to find a relation between $k$ and the scale factor. By comparing
the FRW metric (\ref{frw}), which is written in comoving coordinates, with the standard euclidean
physical metric $dt^2 - d\vec{Y}_{phys}^2$, we infer the relation between comoving and physical coordinates
\be
dY_{phys} = a(t) dy_{com}\,.
\ee
Recalling the Hubble law, $v = H d$, we get the physical Hubble radius as $c = H R_{H,\,p}$, and therefore
the comoving Hubble radius as
\be
R_{H,\,c} = \frac{R_{H,\,p}}{a} = \frac{c}{a H}.
\ee
Now, the wavelength of a perturbation crossing the horizon at a time $t_c$ (the largest visible perturbation)
should be given by the relation
\be
\lambda_c = 4 R_{H,\,p}(t_c),
\ee
which means, in comoving coordinates,
\be
\lambda_{c} = \frac{\lambda_{p}}{a} = 4\,\frac{c}{a_c H_c}.
\ee
Therefore a particular comoving $k$-mode at the time of its horizon crossing
obeys the relation
\be
k_{c} = \frac{2\pi}{\lambda_{c}} = \frac{\pi}{2}\frac{a_c H_c}{c}.
\ee
Since we chose units where $c=1$, and moreover $\pi/2 \simeq 1$, we can finally write
\be
k \ \simeq \ a\, H\,
\ee
as the horizon crossing condition.
Besides, reminding that $H = \dot{a}/a$, we have for a pre-inflation radiation era
\be
k \ = \ a \sqrt{\kappa} \left(\frac{B}{a^4} + C\right)^{1/2}
\label{kar}
\ee
while, for a pre-inflation matter era,
\be
k \ = \ a \sqrt{\kappa} \left(\frac{A}{a^3} + C\right)^{1/2}\,.
\label{kam}
\ee
In Fig.\ref{kafig}, we give a plot of the $(k,a)$ relation in the case of pre-inflation matter era,
with constants chosen arbitrarily as $A=C=1$.

\subsection{Scalar Field equations}
\label{scafieqs}
In this subsection we specialize Eq.(\ref{eqphia}) to particular $a(t)$ solutions.
For the pre inflation matter era case, using Eq. (\ref{adotm}) or (\ref{atm}) to compute $\dot{a},\,\ddot{a}$, we obtain
\be
\phi''_k \ + \ \frac{1}{a}\left(\frac{4Ca^3 + \frac{5}{2}A}{Ca^3 + A}\right)\phi'_k
+ \left(\frac{k^2}{\kappa a(Ca^3 + A)}\right)\phi_k \ = \ 0. \nonumber\\
\quad
\label{eqphiam}
\ee
In the case of pre inflation radiation era we have, using Eq. (\ref{atr}),
\be
\phi''_k \ + \ \frac{2}{a}\left(\frac{Ca^4}{Ca^4 + B} + 1\right)\phi'_k +
\left(\frac{k^2}{\kappa (Ca^4 + B)}\right)\phi_k \ = \ 0.\nonumber\\
\quad
\label{eqphiar}
\ee
The general plan of our work is to solve equations (\ref{eqphiam}), (\ref{eqphiar}), regarding $k$ as a
parameter, to obtain
$\phi(a,k)$. Alternatively, we can solve Eq.(\ref{eqv}) and get $\phi(\eta,k)$, and then,
since $a=a(\eta)$ and $\eta=\eta(a)$, make a substitution to $\phi(a,k)$. Once the solution $\phi(a,k)$ is available,
we can compute the power spectrum $P(k)$ of the quantum fluctuations of the field $\Phi$.
Since the power spectrum is usually given as a function of $k$ only, will be necessary to express $\phi$ as
a function of $k$ only, and this can be done through the relations (\ref{kar}), (\ref{kam}).
The final scope is to obtain the function
\be
P(k) = k^3 |\phi(a(k),k)|^2,
\ee
which represents the primordial power spectrum of the fluctuations (perturbations) of the field $\Phi$.
With $P(k)$ then we feed - as in our case - the CMBFAST code to generate the CMB anisotropy power spectra.

Unfortunately, analytical solutions of the full equations (\ref{eqphiam}),(\ref{eqphiar}),(\ref{eqv}),
when we deal with functions $a(t)$ given in (\ref{atm}), (\ref{atr}), are not easily expressible in
closed form. A power series solution is in general at hand, but still not much more comfortable than the
mere numerical one. However, luckily enough, it turns out that analytical solutions of the second order
approximations of equations (\ref{eqphiam}),(\ref{eqphiar}) can be obtained by means of the WKB method,
and these solutions are strongly corroborated by numerical insights.

Of course, when solving a differential equation, we need boundary conditions in order to fix the solution
explicitly. From the mathematical point of view, a boundary condition can be put anywhere in the realm of
definition of the solution. From the physical point of view, it is wise to put boundary conditions in regions
where the physics is reasonably well known. Since we are almost sure that inflation happened, while we know
little about a possible pre-inflation era, we choose to put our boundary conditions in the full
inflationary era.
This means that, for large $a$, or equivalently, for large $k$ (see equations (\ref{kar}), (\ref{kam})),
well after the beginning of inflation, the field $\phi$ must generate an \emph{almost} scale invariant, i.e. \emph{flat},
primordial power spectrum $P(k)$.
To be more precise, the latest observations have shown that the primordial power spectrum is not exactly flat,
but slightly tilted. The newest analysis of the WMAP data~\cite{Koma2010} indicate a form like
\be
P(k) \sim k^{n_s - 1}
\ee
with
\be
n_s \ = \ 0.963 \,\pm \,0.012 \,\,(68\% \,\,CL)\,.
\ee
Therefore the field $\phi$ must behave as
\be
|\phi(a(k),k)| \sim \frac{k^{\frac{1}{2}(n_s-1)}}{k^{3/2}}
\label{BCinf}
\ee
for large $k$.\\
This condition will allow us to fix properly the arbitrary constants in our solutions.
\subsection{Analytical investigations: First order approximation}

To begin with, let's figure out the solution of our differential equation for very large $a$, or $k$.
We use, as first example, the equation for $v_k(\eta)$, (\ref{eqv}). We need the expression of the
conformal time $\eta$ at large $a$, i.e. large cosmic time $t$. In fact, for $t \to \infty$, we have
from Eq.(\ref{atm})
\be
a_m(t) &=& \left(\frac{A}{C}\right)^{1/3}\left[\sinh\left(\frac{3}{2}\sqrt{\kappa C}\,\,t\right)\right]^{2/3}
\nonumber\\
&\simeq& \left(\frac{A}{4C}\right)^{1/3} \exp\left(\sqrt{\kappa C}\,t\right),
\ee
and for the radiation solution, Eq.(\ref{atr}),
\be
a_r(t) &=& \left(\frac{B}{C}\right)^{1/4} \left[\sinh\left(2\sqrt{\kappa C}\,\,t\right)\right]^{1/2}
\nonumber\\
&\simeq& \left(\frac{B}{4C}\right)^{1/4} \exp\left(\sqrt{\kappa C}\,t\right).
\ee
Note that for large $t$ the solutions are essentially the same, as it should be, since inflation
"washes out" everything (actually, \emph{almost} everything, as we shall see).
From $d\eta = dt/a(t)$, we can express $a(t)$ with the conformal time $\eta$,
\be
a(\eta) \ = \ \frac{1}{\sqrt{\kappa C}}\, \frac{1}{|\eta|} \ = \ -\frac{1}{\sqrt{\kappa C}\,\, \eta}
\ee
with $\eta < 0$, and this result holds, in the limit $t \to \infty$, for both solutions $a_r(t)$ or $a_m(t)$.
Thus, the equation for $v_k$ reads
\be
v''_k + \left(k^2 - \frac{2}{\eta^2}\right)v_k = 0\,.
\ee
The general solution in terms of Bessel functions is
\be
v_k(\eta) \ = \ \sqrt{|\eta|}\,\left[c_1(k)\,J_{3/2}\left(k|\eta|\right) \ +
\ c_2(k)\,J_{-3/2}\left(k|\eta|\right)\right],
\nonumber\\
\ee
where
\begin{eqnarray*}
J_{3/2}(x) &=& \sqrt{\frac{2}{\pi}}\frac{1}{\sqrt{x}}\left(\frac{\sin x}{x} - \cos x\right)\nonumber\\
J_{-3/2}(x) &=& \sqrt{\frac{2}{\pi}}\frac{1}{\sqrt{x}}\left(- \frac{\cos x}{x} - \sin x\right)
\end{eqnarray*}
with
\begin{equation*}
x = k|\eta| = \frac{k}{\sqrt{\kappa C}\,a}.
\end{equation*}
Then
\be
v_k(\eta)  &=&  \sqrt{\frac{2}{\pi}}\frac{1}{\sqrt{k}}\,
\left[c_1(k)\,\left(\frac{\sin x}{x} - \cos x\right)\right.\nonumber\\
&+&  \left.c_2(k)\,\left(\frac{\cos x}{x} + \sin x\right)\right].
\ee
Let's have a look at the behavior of the solution for $a\to\infty$, which means $|\eta| \to 0$, $x \to 0$.
Then
\be
v_k(\eta)  \simeq  \sqrt{\frac{2}{\pi\,k}}\,
\left[c_1(k)\,x^2  +  c_2(k)\left(\frac{1}{x} + \frac{x}{2} - \frac{x^3}{8}\right)\right].
\ee
Since $x\sim a^{-1}$,
we have, for any given $k$,
\be
&&\phi_k(a) = \\
&&\frac{v_k(a)}{a} \sim \sqrt{\frac{2}{\pi\,k}}
\left[c_1(k)\frac{1}{a^3}  +  c_2(k)\left(1 + \frac{1}{2a^2} - \frac{1}{8a^4}\right)\right]\,.\nonumber
\ee
If we require, as usual, $\phi_k(a) \to 0$ for $a \to \infty$, then this necessarily implies
$c_2(k) \equiv 0\,$. So finally the solution is
\be
v_k(\eta) \ = \ \sqrt{|\eta|}\,\left[c_1(k)\,J_{3/2}\left(k|\eta|\right)\right],
\ee
and
\be
\label{sol1}
&&\phi_k(a) = \frac{v_k(a)}{a} = \\
&&c_1(k)\sqrt{\frac{2}{\pi\,k}}
\left[\frac{\sqrt{\kappa C}}{k} \sin\left(\frac{k}{\sqrt{\kappa C} a}\right) -
\frac{1}{a} \cos\left(\frac{k}{\sqrt{\kappa C} a}\right)\right]. \nonumber
\ee
This is the solution which describes the field $\phi(a,k)$ in full inflationary era.
As we have seen, this solution holds for both cases of pre-inflation matter and pre-inflation radiation era.
The arbitrary integration constant $c_1(k)$ still has to be fixed, and this can be done by enforcing the
boundary condition in inflationary era, namely Eq.(\ref{BCinf}). Noticing that,
\emph{for large} $a$, from both equations (\ref{kar}),(\ref{kam}), we have
\be
k \ \simeq \ a\,\sqrt{\kappa C},
\ee
and therefore
\be
\phi(a(k),k) \ \simeq \ \sqrt{\frac{2\kappa C}{\pi}}\,\,\left[\sin(1)-\cos(1)\right]\,\,
\frac{c_1(k)}{k^{3/2}}\,,
\ee
then this means, from (\ref{BCinf}),
\be
c_1(k) \ = \ k^{\frac{1}{2}(n_s-1)}.
\ee
The solution (\ref{sol1}) will be employed as boundary condition in inflationary era for
the numerical integration of equations (\ref{eqphiam}), (\ref{eqphiar}).

Finally, we note that we can start from equations (\ref{eqphiam}), (\ref{eqphiar}) for $\phi_k(a)$,
instead of equation (\ref{eqv}) for $v_k(\eta)$. Taking the first order approximation for $a \to \infty$
in the coefficients, we get, from both equations (\ref{eqphiam}), (\ref{eqphiar})
\be
\phi''_k \ + \ \frac{4}{a} \ \phi'_k \ = \ 0\,.
\ee
This has the solution
\be
\phi(a,k) = \frac{c_1(k)}{a^3} + c_2(k),
\ee
which, for large $a$, when $k \ \simeq \ a\,\sqrt{\kappa C}$, can be matched with the
"almost flat spectrum" boundary condition (\ref{BCinf}) if
\be
c_1(k) \ \sim \ k^{\,\,\frac{3}{2} \ + \ \frac{1}{2}(n_s-1)}\,, \quad\quad\quad c_2(k) \equiv 0\,.
\ee

\subsection{Analytical investigations: Second order approximation}

Obviously, a solution like (\ref{sol1}) cannot be trusted when we look for information about the
behavior of $P(k)$
at low $k$-modes, since by construction it is built by taking into account the
first order approximation only, and requiring that it reproduces the almost flat spectrum for high $k$'s.
To have reliable analytical insights on the low $k$ behavior of $P(k)$, one should go to the second order
approximation. This can be done in principle by starting from equation (\ref{eqv}) for $v_k$, but
it would involve
a quite complicated construction of the conformal time $\eta$, and of the function $a(\eta)$,
in terms of Jacobian elliptic functions (see Appendix 2). A more straightforward path turns out to
be to start
directly from equations (\ref{eqphiam}), (\ref{eqphiar}) and develop the coefficients to the
second order in $a$
for $a \to \infty$.

For the pre-inflation matter era case, Eq.(\ref{eqphiam}), we have
\be
\phi''_k \ &+& \ \left(\frac{4}{a} \ - \ \frac{3A}{2C}\frac{1}{a^4} \ + \
O\left(\frac{1}{a^7}\right)\right)\phi'_k +\nonumber\\
&+& \ \left(\frac{k^2}{\kappa}\frac{1}{Ca^4} \ + \ O\left(\frac{1}{a^7}\right)\right)\phi_k \ = \ 0.
\label{eqphiam2}
\ee

\begin{figure}[ht]
\centerline{\epsfxsize=2.9truein\epsfysize=1.8truein\epsfbox{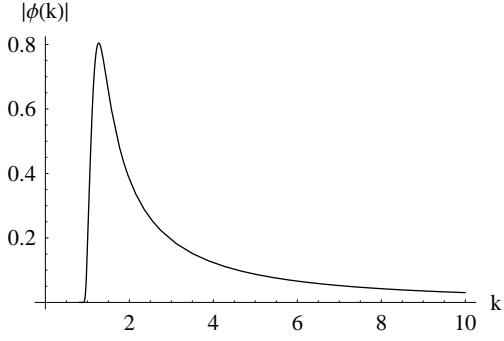}}
\caption[]{Field $|\phi(k)|$ versus $k$, in pre-inflation matter era (arbitrary units).}
\vspace{0.2cm} \hrule
\label{phim}
\end{figure}

\begin{figure}[ht]
\centerline{\epsfxsize=2.9truein\epsfysize=1.8truein\epsfbox{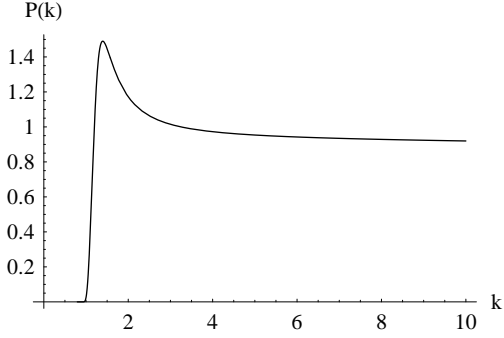}}
\caption[]{Primordial power spectrum $P(k)$ versus $k$, in pre-inflation matter era (arbitrary units).}
\vspace{0.2cm} \hrule
\label{Pphim}
\end{figure}

For the pre-inflation radiation era case, Eq.(\ref{eqphiar}), we have
\be
\phi''_k \ &+& \ \left(\frac{4}{a} \ - \ \frac{2B}{C}\frac{1}{a^5} \ + \
O\left(\frac{1}{a^9}\right)\right)\phi'_k +\nonumber\\
&+& \ \left(\frac{k^2}{\kappa}\frac{1}{Ca^4} \ + \ O\left(\frac{1}{a^8}\right)\right)\phi_k \ = \ 0.
\label{eqphiar2}
\ee
Following the usual well know method to get rid of the first derivative, we write
\be
\phi(a) = v(a) p(a),
\ee
with
\be
p(a) = \exp\left[-\frac{1}{2}\int f(a)da\right],
\ee
where $f(a)$ is the coefficient of $\phi'$. Then the equation for $v(a)$ reads
\be
v''(a) \ + \ \left(\frac{p''}{p} + f\cdot\frac{p'}{p} + g\right)v(a) \ = \ 0,
\label{eqv2}
\ee
where $g(a)$ is the coefficient of $\phi$.\\
To an equation of the form (\ref{eqv2}) we can apply the WKB method.
In what follows, we specify the main steps of the argument for the matter era
case, and report only the final
result for the radiation era case.\\
For Eq.(\ref{eqphiam2}), pre-inflation matter era, we find
\be
p(a) = \frac{1}{a^2} \exp\left(-\frac{A}{4C}\frac{1}{a^3}\right),
\ee
and thus
\be
v'' \ + \ \left(-\frac{2}{a^2} + \frac{k^2}{\kappa C a^4} -
\frac{9}{16}\frac{A^2}{C^2}\frac{1}{a^8}\right)v \ = \ 0
\ee
Setting
\be
F(a) = i \left[\frac{2}{a^2} - \frac{k^2}{\kappa C a^4} +
\frac{9}{16}\frac{A^2}{C^2}\frac{1}{a^8}\right]^{1/2}\,,
\ee
then the WKB ansatz suggests as solution for $v(a)$
\be
v(a) \ &=& \ \frac{1}{\sqrt{F(a)}}\left[c_+(k)\exp\left(i\int F(a)da\right)\right.\nonumber\\
&&\left. + \ c_-(k)\exp\left(-i\int F(a)da\right)\right]\,.
\ee
Writing
\be
G(a) \ = \ \int F(a) da
\ee
we see that, for example, for $a \to \infty$
\be
G(a) \ \simeq \ i \,\sqrt{2}\,\log a\,.
\ee
So, the WKB solution of Eq.(\ref{eqphiam2}) explicitly reads
\be
\label{WKBm}
&&\phi_k(a) = v_k(a)p(a) =\\
&&=\frac{2\,\sqrt[4]{\kappa C^2}\,\,\left[c_+(k)e^{iG(a)} + c_-(k)e^{-iG(a)}\right]\,e^{-i\pi/4}}
{\left[32\kappa C^2a^6 - 16k^2Ca^4 + 9\kappa A^2\right]^{1/4}\cdot \exp[A/(4Ca^3)]}. \nonumber
\ee
We can always find agreement with the boundary condition
(\ref{BCinf}), namely $|\phi|\sim k^{-3/2 + (n_s-1)/2}$ for large $a$ or $k$,
by defining the arbitrary constants $c_\pm(k)$ accordingly.
In fact for large $a$, $a \simeq k/\sqrt{\kappa C}$, and we can
choose $c_\pm(k)$ in a way that
\be
\left[c_+(k)e^{iG(a(k))} + c_-(k)e^{-iG(a(k))}\right]  \sim k^{(n_s-1)/2}
\label{bcc}
\ee
so that the condition (\ref{BCinf}) is fulfilled.
The expression of $\phi(a(k),k)$ when $a(k)\simeq k/\sqrt{\kappa C}$ (first approximation) is
\be
&&\phi(a(k),k) =\nonumber\\
&&\frac{2\,\sqrt[4]{\kappa C^2}\,\,\,\,k^{(n_s-1)/2}\,\,e^{-i\pi/4}}
{\left[16 k^6 /{\kappa^2 C} + 9\kappa A^2\right]^{1/4}\cdot \exp[A\sqrt{\kappa^3 C}/(4 k^3)]}\,.~~
\ee
We can also consider relation (\ref{kam}) to the second order of approximation, which reads
\be
a(k) \simeq \frac{k}{\sqrt{\kappa C}} - \frac{A \kappa}{2 k^2}\,.
\ee
Constants $c_\pm(k)$ should, and can, still be chosen as in (\ref{bcc}), in order to have
$|\phi|\sim k^{-3/2 + (n_s-1)/2}$ for large $k$.
Once this new expression of $a(k)$ is substituted in (\ref{WKBm}),
it is instructive to plot the quantities $|\phi(a(k),k)|$ and $P = k^{3}|\phi(k)|^2$.
In Figures (\ref{phim}), (\ref{Pphim}) we can see these qualitative plots,
where, for sake of simplicity, we arbitrarily set the parameters $A=B=C=\kappa=1$.

\begin{figure}[ht]
\centerline{\epsfxsize=2.9truein\epsfysize=1.8truein\epsfbox{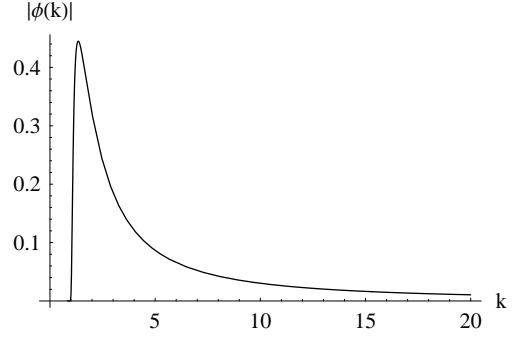}}
\caption[]{Field $|\phi(k)|$ versus $k$, in pre-inflation radiation era (arbitrary units).}
\vspace{0.2cm} \hrule
\label{phir}
\end{figure}

\begin{figure}[ht]
\centerline{\epsfxsize=2.9truein\epsfysize=1.8truein\epsfbox{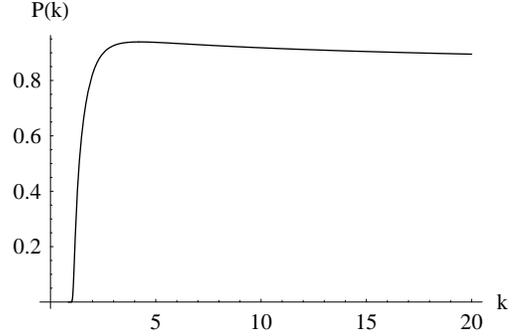}}
\caption[]{Primordial power spectrum $P(k)$ versus $k$, in pre-inflation radiation era (arbitrary units).}
\vspace{0.2cm} \hrule
\label{Pphir}
\end{figure}

For the pre-inflation radiation era, we start from Eq.(\ref{eqphiar2}). With analogous steps we find
\be
p(a) = \frac{1}{a^2} \exp\left(-\frac{B}{4C}\frac{1}{a^4}\right)
\ee
and
\be
v''(a) \ + \ F(a)^2\,\,v(a) \ = \ 0,
\ee
with
\be
F(a) = i \left[\frac{2}{a^2} - \frac{k^2}{\kappa C a^4} + \frac{9B}{C a^6} + \frac{B^2}{C^2 a^{10}}\right]^{1/2}\,.
\ee
The WKB solution of Eq.(\ref{eqphiar2}) explicitly reads
\be
&&\phi_k(a) = v_k(a)p(a) =\\
&&\frac{e^{-i\pi/4}\,\sqrt[4]{\kappa C^2}\,\,a^{1/2}\,\left[c_+(k)e^{iG(a)} + c_-(k)e^{-iG(a)}\right]}
{\left[2\kappa C^2a^8 - k^2 C a^6 + 9\kappa BC a^4 + B^2\kappa\right]^{1/4}\cdot \exp[B/(4Ca^4)]}. \nonumber
\label{wkbr}
\ee
Again the functions $c_\pm(k)$ should be chosen such that the square bracket's content in the numerator
goes as $\sim~k^{(n_s-1)/2}$,
so that $|\phi| \sim k^{-3/2 + (n_s-1)/2}$ for large $a\simeq~k/\sqrt{\kappa C}$.
Also here we may consider relation (\ref{kar}) to the second (or further) order in $k$,
\be
a(k) \simeq \frac{k}{\sqrt{\kappa C}} - \frac{\kappa B \sqrt{\kappa C}}{2 k^3}\,,
\ee
to get a better approximation. In Figures (\ref{phir}), (\ref{Pphir}) we see plots of the field $|\phi|$
and of the primordial power spectrum $P(k)$.

\begin{figure}[ht]
\centerline{\epsfxsize=2.9truein\epsfysize=1.8truein\epsfbox{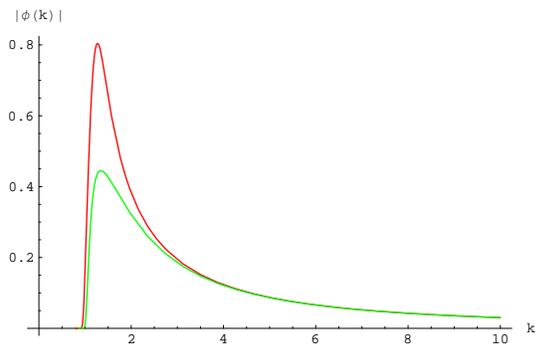}}
\caption[]{Field $|\phi(k)|$ versus $k$, for pre-inflation matter era (red/upper line),
and radiation era (green/lower line).}
\vspace{0.2cm} \hrule
\label{phimr}
\end{figure}

\begin{figure}[ht]
\centerline{\epsfxsize=2.9truein\epsfysize=1.8truein\epsfbox{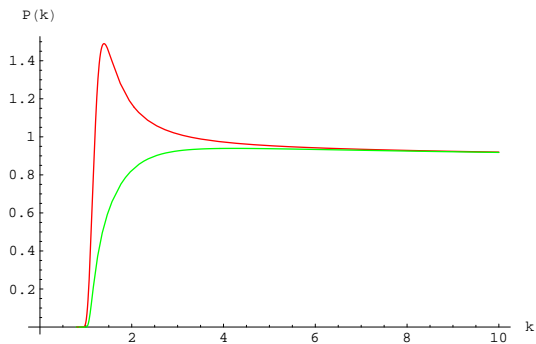}}
\caption[]{Primordial power spectrum $P(k)$ versus $k$, for pre-inflation matter (red/upper line)
and radiation (green/lower line) eras.}
\vspace{0.2cm} \hrule
\label{Pphimr}
\end{figure}

Finally, in Figures (\ref{phimr}), (\ref{Pphimr}), we compare the plots for fields and power
spectra, matter and radiation cases, in the same diagrams.

We see that for both matter and radiation eras there is an exponential suppression of the low $k$-modes.
The matter diagram presents an interesting cusp, just before dropping down, that is absent in the radiation diagram.
In the next section the effects of these features on the CMB spectrum will be further investigated,
via deeper numerical analysis with the help of CMBFAST, a code specialized for cosmological
simulations of the CMB power spectrum.
%
\subsection{Numerical computation of the primordial power spectrum}

\label{PS}

\subsubsection{Matter era}

\begin{figure}[ht]
    \begin{center}
    \includegraphics[width=0.4\textwidth]{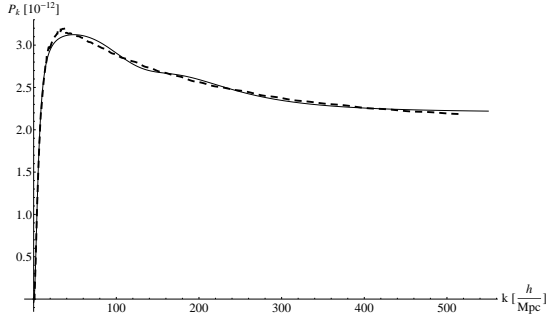}
    {\small \caption{The numerical solution for the primordial power spectrum (dashed line)
    and its fitting function (full line) in the pre-inflation matter era scenario.}
    \label{fig:PowerFitMatterTilt}}
    \end{center}
\end{figure}

For the numerical solution it is most convenient to use the equation of motion for the scalar
field perturbation $\phi$ written as a function of the variable $a$. So we can use equation (\ref{eqphiam})
cast as
\begin{equation}
    [Aa+Ca^4]\,\phi''_k \ + \ \Bigl[ \frac{5}{2} A +
    4 C a^3 \Bigr]\phi'_k \ + \ \frac{k^2}{\kappa}\, \phi_k \ = \ 0.
    \label{eq:EoMphimatter}
\end{equation}
The field perturbation $\phi_k(a)$ is a function of $k$ and $a$. \\
In one of the previous sections we determined the parameter of the matter contribution in the
Friedmann equation, $A$, to be
\begin{equation}
    A = 10^{-1/2} \,\frac{\epsilon_p}{V_p}.
\label{Ame}
\end{equation}
After a radiation-dominance era and a subsequent
period of matter-dominance before inflation, we can, with the help of Eq.(\ref{scalef}) and
the assumption of $a(t_p) = 1$, compute the scale factor at the onset of inflation
\begin{equation*}
    a_{inf} = 10^{7/2}.
\end{equation*}
When inflation starts, the two competing factors of matter and inflation in the Friedmann equation
(\ref{adotm}) must be of the same order of magnitude. This condition, or more rigorously,
condition (\ref{inflcond}), allows us to fix the coefficient $C$ as
\begin{equation}
    C = \frac{A}{2\,a_{inf}^3} = \frac{1}{2} \,10^{-11} \,\frac{\epsilon_p}{V_p}.
\label{Cme}
\end{equation}
For the numerical solution of the equation, we have to use the horizon crossing condition (\ref{kam})
to write $a$ as a function of $k$. The equation for the scalar field
perturbation,~\eqeqref{eq:EoMphimatter}, is evaluated numerically for a fixed $k_i$, repeatedly
for many different choices of $k_i$, and the values of $\phi(a(k_i),k_i)$ are assembled to form
an evolution of the field perturbations in dependence of $k$. \\
The boundary conditions for the numerical solution have been obtained in section \ref{scafieqs}.
Using them to solve the differential equation~\eqref{eq:EoMphimatter} leads to the result shown
in~\figref{fig:PowerFitMatterTilt}. However, the power spectrum as a function of $k$ is only a
collection of data points. In order to be of any use for the CMBFAST code, it has to be given as
an analytical function, which is obtained by fitting the data points with an opportune
mathematical function. The function used for the fitting is
\begin{equation}
    P(k) = a - \frac{b}{1+\frac{k^2}{c}} + \frac{d}{1+\frac{k^4}{e}} - \frac{f}{1+\frac{k^6}{g}},
\end{equation}
where the parameters $a,...,g$ can be found as:

\footnotesize{
\begin{table}[htp]
\centering
\begin{tabular}{|@{\hspace{0.3cm}}c@{\hspace{0.3cm}}|@{\hspace{0.3cm}}c@{\hspace{0.3cm}}|}
\hline
 parameter & value \\
\hline
a & $2.205\cdot10^{-12}$ \\
\hline
b & $3.233\cdot10^{-12}$ \\
\hline
c & $0.03$ \\
\hline
d & $2.578\cdot10^{-12}$ \\
\hline
e & $1.680\cdot10^{-9}$ \\
\hline
f & $1.593\cdot10^{-12}$ \\
\hline
g & $6.584\cdot10^{-14}$ \\
\hline
\end{tabular}
\label{parametermatter}
\end{table}}
\normalsize
~\\

Fig.\,\ref{fig:PowerFitMatterTilt} shows the fitting function (full line) together with the
numerical curve of the power spectrum (dashed line).

With these results, it is then possible to continue with the evaluation of the CMB power spectrum
by putting the fitting function into CMBFAST.

\subsubsection{Radiation era with totally evaporating black holes}
\begin{figure}[ht]
    \begin{center}
    \includegraphics[width=0.4\textwidth]{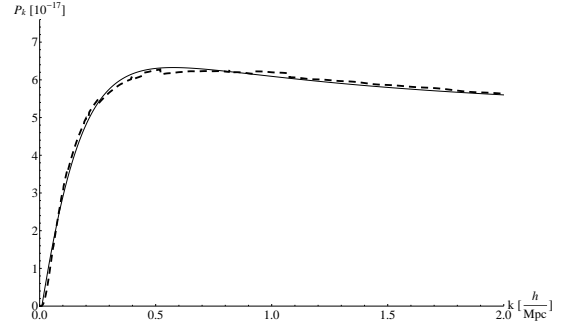}
    {\small \caption{The numerical solution for the primordial power spectrum (dashed line)
    and its fitting function (full line) in the pre-inflation radiation era scenario,
    with completely evaporating black holes (no GUP active).}
    \label{fig:PowerFitRadTilt}}
    \end{center}
\end{figure}

The equation of motion for the scalar field perturbation is Eq.(\ref{eqphiar}), which can be cast
in the form
\begin{equation}
     [B + Ca^4]\,\phi''_k  \ + \ \Bigl[ \frac{2B}{a} + 4Ca^3 \Bigr]\, \phi'_k  \ + \
     \frac{k^2}{\kappa} \, \phi_k \ = \ 0.
    \label{eq:EoMphirad}
\end{equation}
The $k(a)$-relation for the radiation case is given by Eq.(\ref{kar}).
We shall compare the situation of the previous section,
when massive remnants are left by the black hole evaporation,
with the present situation, in which black holes are nucleated but then disappear again completely
into radiation.\\
Following the scenario of radiation in pre-inflation era developed in subsection \ref{nogup},
we can determine the parameters $B$ and $C$ in this case.
When $\tau$ spans the interval between the end of black hole evaporation, $\tau_r \simeq 10^4$,
and the onset of inflation, $\tau_{inf} \simeq 10^8$,
then the scale factor evolves in full radiation era as
\be
a(\tau) = a(t_p)\left(\frac{\tau_r}{\tau_c}\right)^{1/6}\, \tau^{1/2}
\ee
Reminding $a(t_p)=1$ and $\tau_c\simeq10^3$
we get a scale factor at the onset of inflation of
\begin{equation}
    a(\tau_{inf}) = 10^{25/6} \sim 10^{4.16}.
\end{equation}
The radiation content of the Universe is fixed at $\tau_r$, the end of the black hole
evaporation era. Therefore, since by then $R_H=10^4\ell_p$,
\be
\label{Bre}
    B &=& \rho_r(\tau_r) \cdot a^4(\tau_r)\\
    &=& \frac{10^4 \,\sqrt{\zeta}\,\epsilon_p}{(10^4)^3 V_p} \cdot
    [(10^3)^{-1/6} (10^4)^{2/3}]^4 = 10^{2/3}\, \sqrt{\zeta}\,\,\frac{\epsilon_p}{V_p}.\nonumber
\ee
Here, as before, we choose again $\zeta \sim 1$.
Thus, the inflationary parameter $C$ can be fixed by requiring that it is of the same order as
the radiation parameter $B$ at the onset of inflation
\begin{equation}
    C \simeq \frac{B}{a_{inf}^4} \sim 10^{-16}.
\label{Cre}
\end{equation}
Thus we now solve the equation for a scenario without GUP in the same way as before,
and investigate the differences in the primordial power spectrum.
Fig.\,\ref{fig:PowerFitRadTilt} shows the result (dashed line).

The fitting in this case has been done using the function
\begin{equation}
    P(k) = a + b\cdot \mathrm{Arctan}(c\cdot k) + d\cdot \mathrm{Arctan}^2(c\cdot k),
\end{equation}
with the parameters $a,...,d$ to be adjusted by Mathematica. The resulting fitting parameters are
given in the following table, and the curve (full line) together with the numerical result
can be seen in~\figref{fig:PowerFitRadTilt}.

\footnotesize{
\begin{table}[htp]
\centering
\begin{tabular}{|@{\hspace{0.3cm}}c@{\hspace{0.3cm}}|@{\hspace{0.3cm}}c@{\hspace{0.3cm}}|}
\hline
 parameter & value \\
\hline
a & $-3.701\cdot10^{-18}$ \\
\hline
b & $1.261\cdot10^{-16}$ \\
\hline
c & $3.116$ \\
\hline
d & $-5.935\cdot10^{-17}$ \\
\hline
\end{tabular}
\label{parametermatterrad}
\end{table}}
\normalsize

\subsubsection{Radiation era without black holes}
\begin{figure}[ht]
    \begin{center}
    \includegraphics[width=0.4\textwidth]{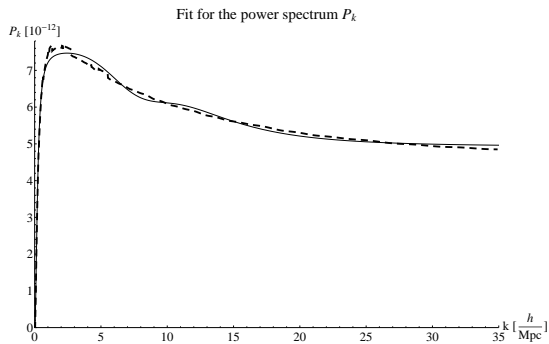}
    {\small \caption{The numerical solution for the primordial power spectrum (dashed line) and its fitting
function (full line) in the pre-inflation radiation era scenario without black holes.}
    \label{fig:PowerFitRadOnlyTilt}}
    \end{center}
\end{figure}

The equation of motion for the scalar field perturbations can be directly taken over from the previous
section - it describes the universe evolving from a state of radiation dominance into the inflationary period.
Simply the parameters $B$ and $C$ in the equation have to be evaluated anew. \\
The scale factor at the onset of inflation is, as previously mentioned,
\begin{equation}
    a(\tau_{inf}) = 10^{7/2}.
\end{equation}
At the Planck time, the radiation had Planck density, and so
\begin{equation}
    B = \rho_p(\tau_p) \cdot a^4(\tau_p) =  1\, \frac{\epsilon_p}{V_p}.
\end{equation}
Thus, the inflationary parameter can be fixed as
\begin{equation}
    C = \frac{B}{a_{inf}^4} \sim 10^{-14}.
\end{equation}
Again the equation for a scenario without GUP is solved in the same way as before, and the resulting primordial
power spectrum is shown in~\figref{fig:PowerFitRadOnlyTilt} (dashed line).

The fitting in this case is done using the same function as in the matter case,
\begin{equation}
    P(k) = a - \frac{b}{1+\frac{k^2}{c}} + \frac{d}{1+\frac{k^4}{e}} - \frac{f}{1+\frac{k^6}{g}},
\end{equation}
with the parameters $a,...,g$ as in the following table:

\footnotesize{
\begin{table}[htp]
\centering
\begin{tabular}{|@{\hspace{0.3cm}}c@{\hspace{0.3cm}}|@{\hspace{0.3cm}}c@{\hspace{0.3cm}}|}
\hline
 parameter & value \\
\hline
a & $-4.566\cdot10^{-15}$ \\
\hline
b & $3.383\cdot10^{-15}$ \\
\hline
c & $0.0089$ \\
\hline
d & $-6.561\cdot10^{-15}$ \\
\hline
e & $152.074$ \\
\hline
f & $3.783\cdot10^{-16}$ \\
\hline
g & $3.564\cdot10^{-7}$ \\
\hline
\end{tabular}
\label{parameterradonly}
\end{table}}
\normalsize

%
%

\section{The CMB power spectrum}

In this section the previously calculated primordial power spectra, in the cases with GUP,
without GUP, and without black holes, as well as the analytical results for the approximated equations, will be
fed into the CMBFAST code~\cite{Selj1996} to obtain the CMB temperature anisotropy spectrum
that can be measured today. We will compare our results to the WMAP seven year data, and to the result for the
CMB spectrum obtained by following
the standard inflationary scenario without any pre-inflation era. \\
The standard inflation (SI) model is in principle a very good fit to the data, considering
all the different features it has to explain. Only the mode with $l=2$ in the CMB spectrum is
very low in comparison to the following data points. If this is not simply a statistical
feature but the indicator of a new physical phenomenon, then the SI model has to be modified
in order to satisfy the drop at the low $l$ modes. For now, only the $l=2$ mode can be used
to construct models with a suppression at low $l$ modes, but in the far future it will be
possible to tell whether the tendency of the power spectrum to drop will continue further,
whether it will stay at a lower but constant level or whether it might even rise up again.

\begin{figure}[ht]
    \begin{center}
    \includegraphics[width=0.4\textwidth]{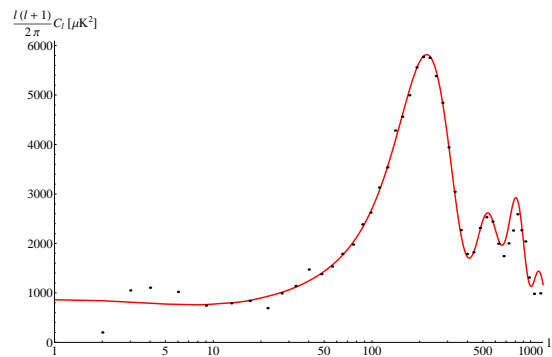}
    {\small \caption{The binned CMB temperature anisotropy spectrum as measured by the WMAP
    satellite, in comparison with the standard inflationary scenario.}
    \label{fig:LCDMbinned}}
    \end{center}
\end{figure}

To produce the CMB power spectrum corresponding to the scenario of standard inflation, we assumed
a scalar spectral index of $n_s = 0.963$ and a running of the index $\alpha_{n_s} = -0.022$, i.e. the
most recent result of the WMAP observations~\cite{Koma2010}. \\
In the calculation of the CMB power spectrum, there is one parameter that can be varied,
the number of e-folds of inflation. The total number of e-folds of inflation (from the start,
when the mode $k_i$ left the horizon, to the end of inflation) can be given (see~\cite{Alab2006}) as
\begin{equation}
    N_{tot} = N(k_0) + ln\Bigl(\frac{k_0}{k_i}\Bigr).
\end{equation}
$k_0$ is the currently largest mode within the horizon,
\begin{equation}
    k_0 = 0.002\, h Mpc^{-1}.
\end{equation}
This is the pivot scale for the wavenumber that the WMAP team has been using in constraining inflation
models from their data~\cite{Koma2009,Koma2010,Peir2006}.
$N(k_0)$ is the number of e-folds from the time during inflation when this mode $k_0$ crossed
outside the horizon,
\begin{equation}
    N(k_0) = ln\Bigl(\frac{k}{k_0}\Bigr)\,.
\end{equation}
Observations can only constrain the number $N(k_0)$, as currently the mode $k_0$ reenters the horizon,
but no information can be given about whether there were more e-folds $\Delta N$ of inflation
\emph{before} $k_0$ exited the horizon during inflation. The constraint on $N(k_0)$ stated
in~\cite{Alab2006} is
\begin{equation}
    N(k_0) = 54 \pm 7.
\end{equation}
Usually in the power spectrum $k$ is normalized to $k_0$, and the power spectrum is taken
at values $P \bigl( \frac{k}{k_0} \bigr)$, where $k_0 = 0.002\, h Mpc^{-1}$. By dividing $k$
in the expression of the power spectrum by numbers smaller than $k_0$, we can add more e-folds
to inflation accounting for the time before $k_0$ exited the horizon:
\begin{equation}
    N_{tot} = ln\Bigl(\frac{k}{k_0}\Bigr) + ln\Bigl(\frac{k_0}{k_i}\Bigr) = ln\Bigl(\frac{k}{k_i}\Bigr).
\end{equation}
Instead of taking the power spectrum as before normalized over $k_0$, we take it as
$P \Bigl( \frac{k}{k_i} \Bigr)$, where any number with $k_i < k_0$ is possible. \\
So, varying $k_i$ is equivalent to adding e-folds $\Delta N$ to the experimentally
constrained number $N(k_0) = 54 \pm 7$.

\subsection{Results from the numerically computed primordial power spectrum}
\begin{figure*}[htp]
\centering
\includegraphics[width=0.4\textwidth]{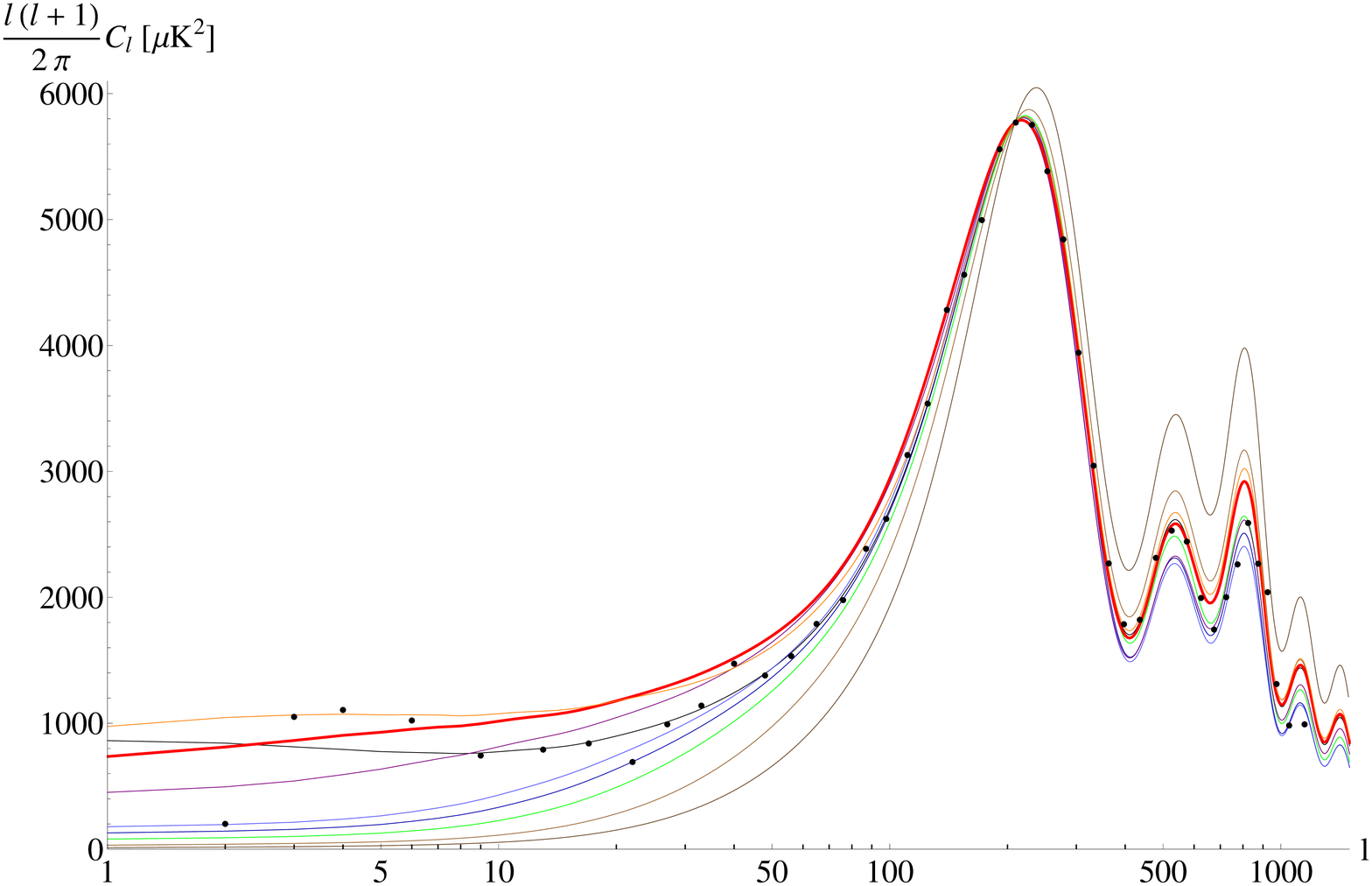}\hspace{5 mm}
\includegraphics[width=0.08\textwidth]{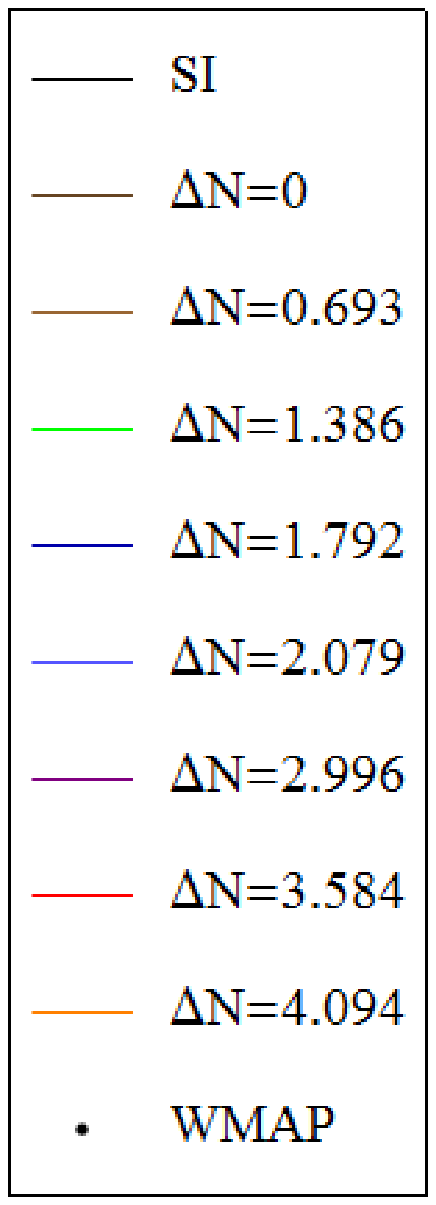}
\includegraphics[width=0.4\textwidth]{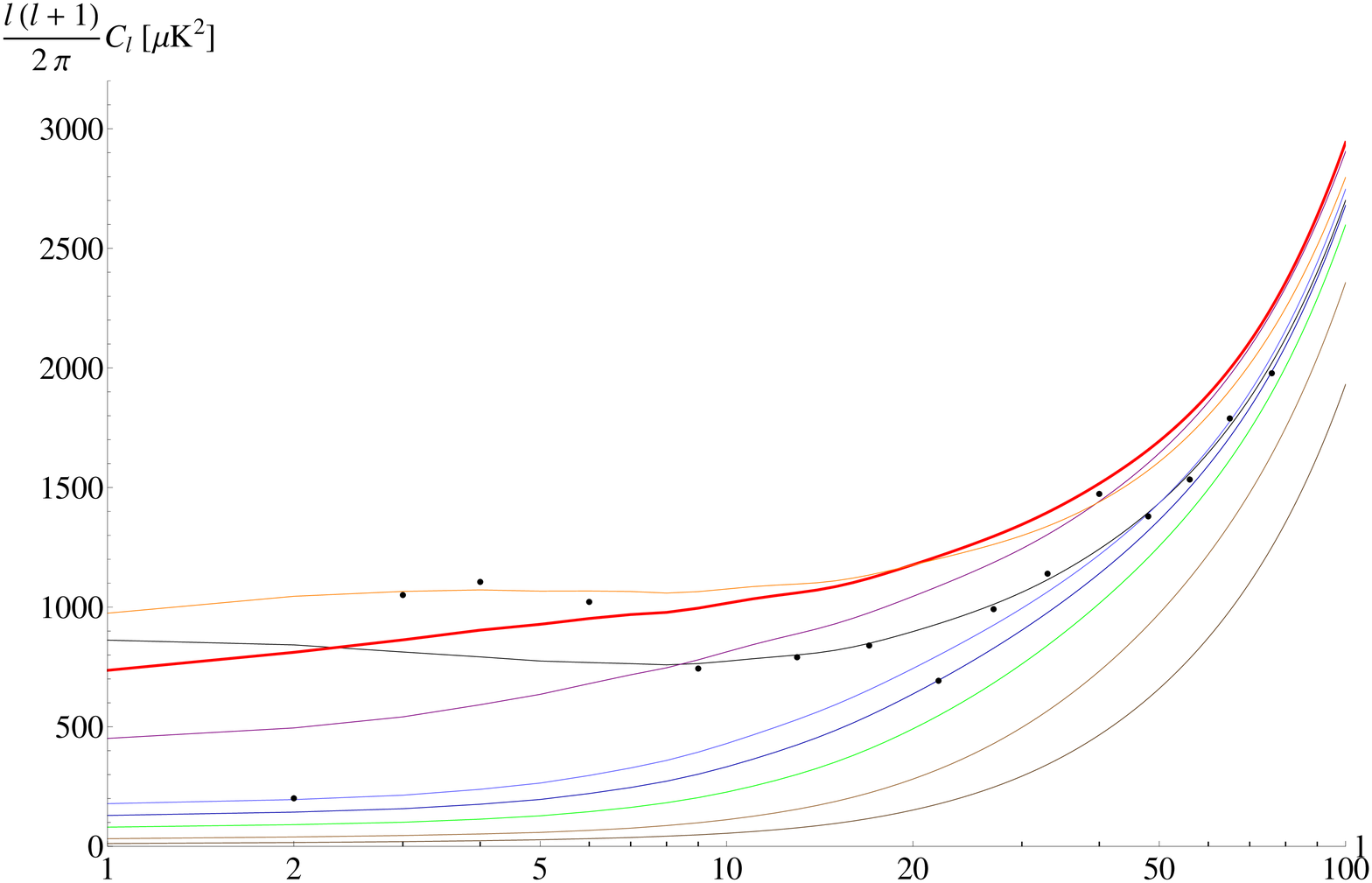}
\caption\small{{The CMB power spectrum for a pre-inflation matter era, for various cases of
$\Delta N$. Overall view and zoom into the lower multipole region.}}
\label{fig:CMBmatter}
\end{figure*}

There are three cases to present from the numerical calculations. \\
For a matter-dominated era before inflation the CMB power spectrum as obtained by
CMBFAST can be seen in~\figref{fig:CMBmatter}.

\normalsize

\begin{figure*}[htp]
\centering
\includegraphics[width=0.4\textwidth]{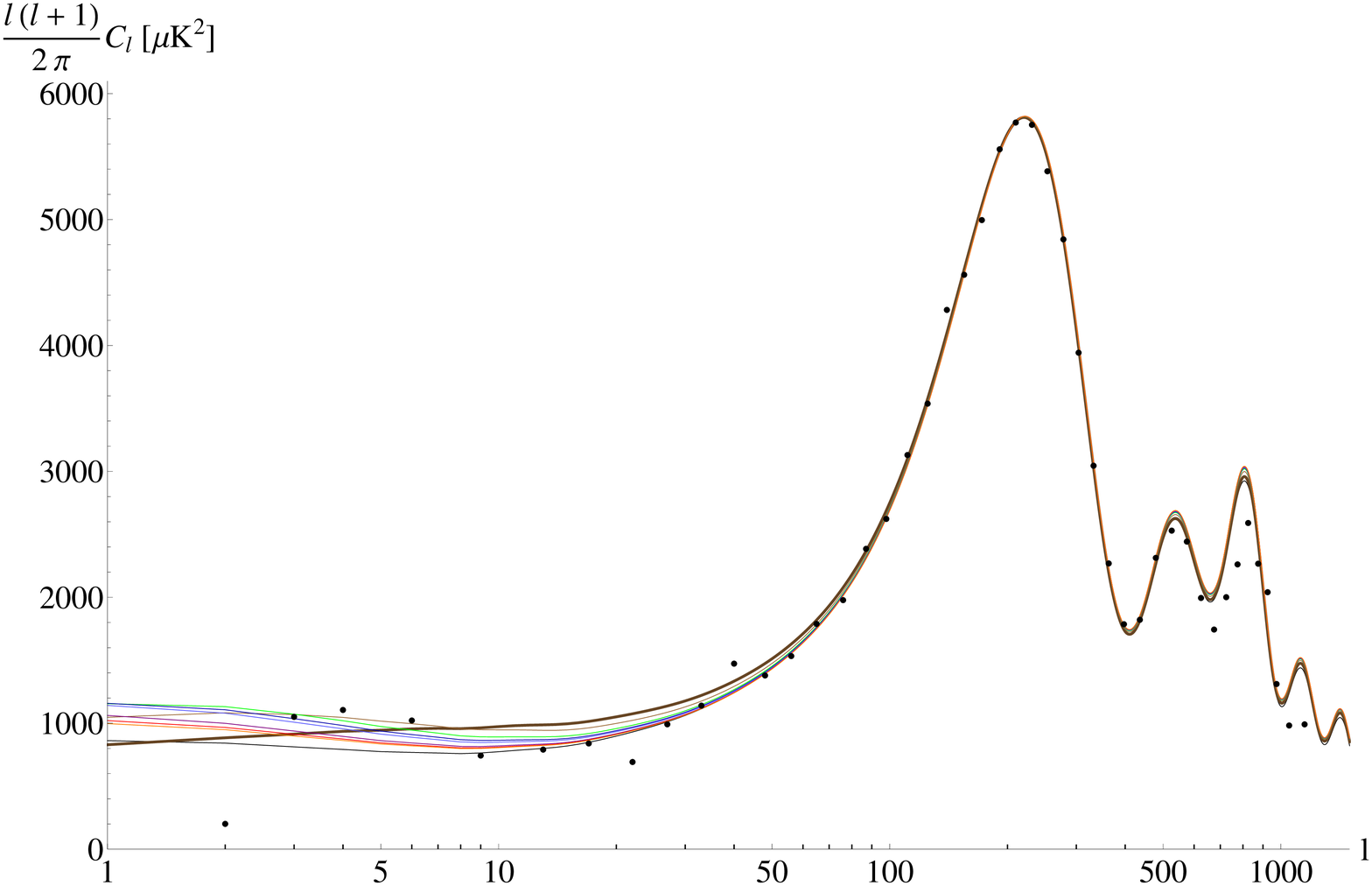} \hspace{5 mm}
\includegraphics[width=0.08\textwidth]{Legend.eps}
\includegraphics[width=0.4\textwidth]{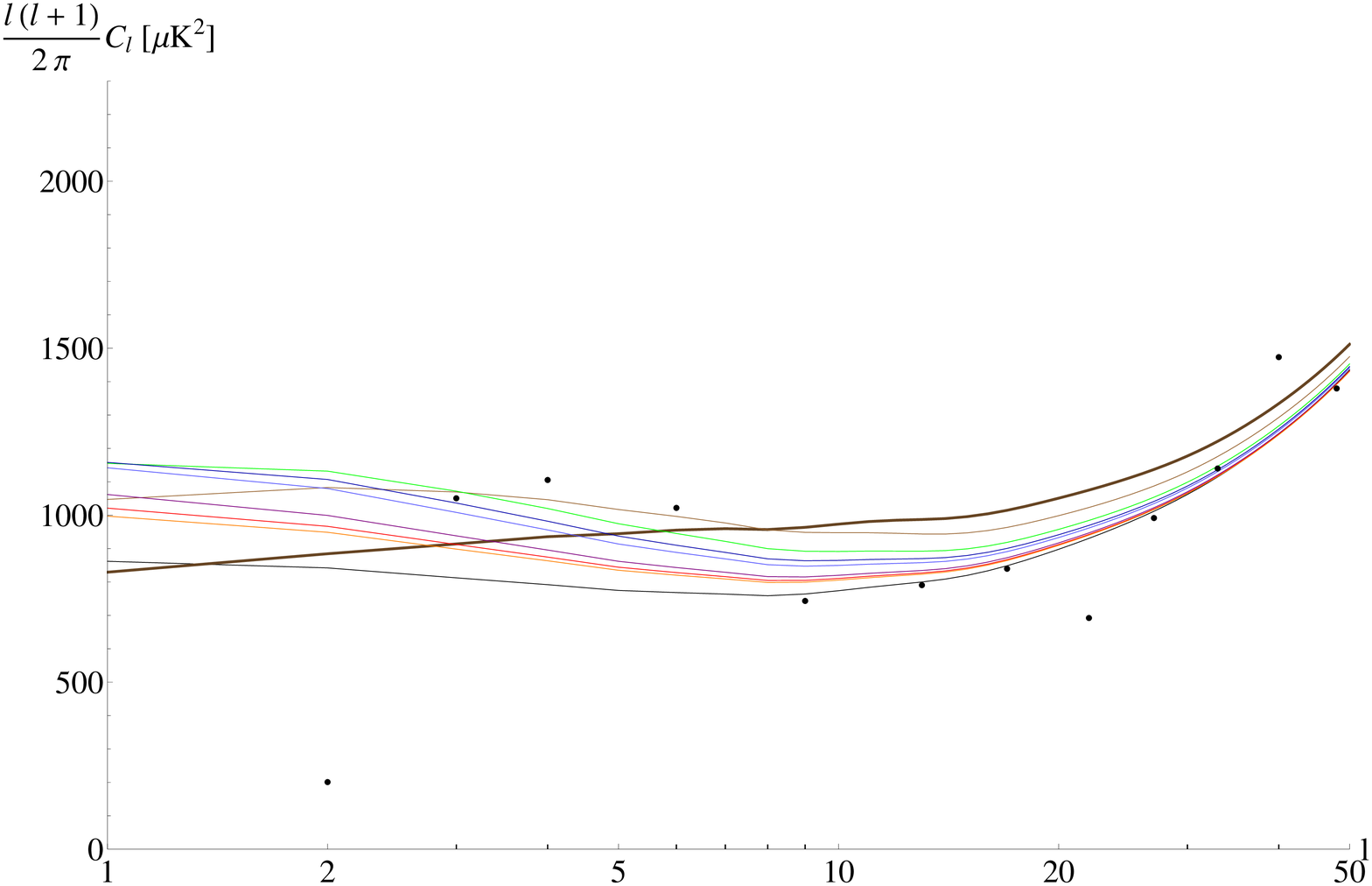}
\caption\small{{The CMB power spectrum for a pre-inflation radiation era, for various cases of
$\Delta N$. Overall view and zoom into the lower multipole region.}}
\label{fig:CMBrad}
\end{figure*}

For the case when the GUP is turned off, the CMB power spectrum can be seen in~\figref{fig:CMBrad}.

\normalsize

\begin{figure*}[htp]
\centering
\includegraphics[width=0.4\textwidth]{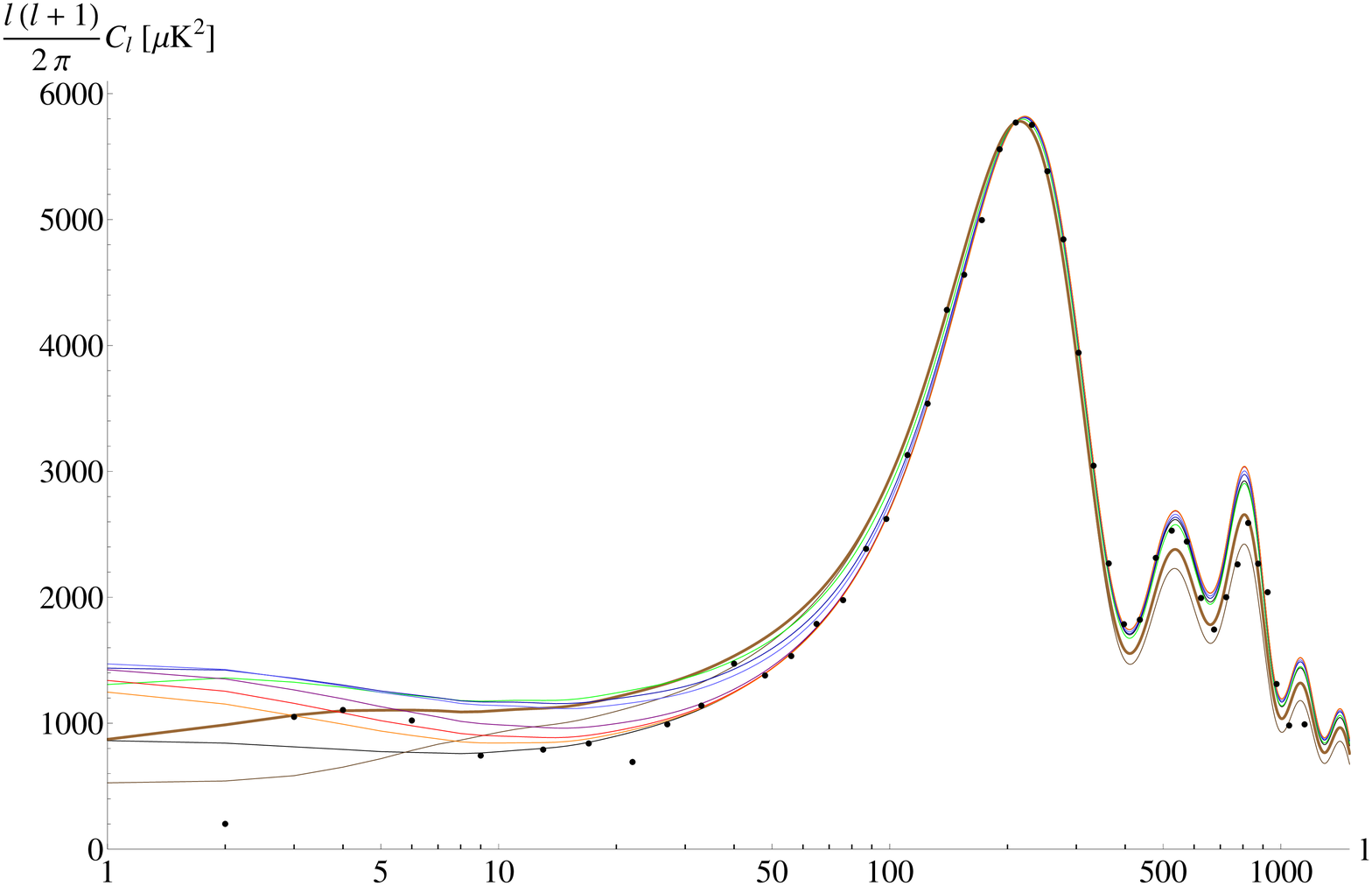} \hspace{5 mm}
\includegraphics[width=0.08\textwidth]{Legend.eps}
\includegraphics[width=0.4\textwidth]{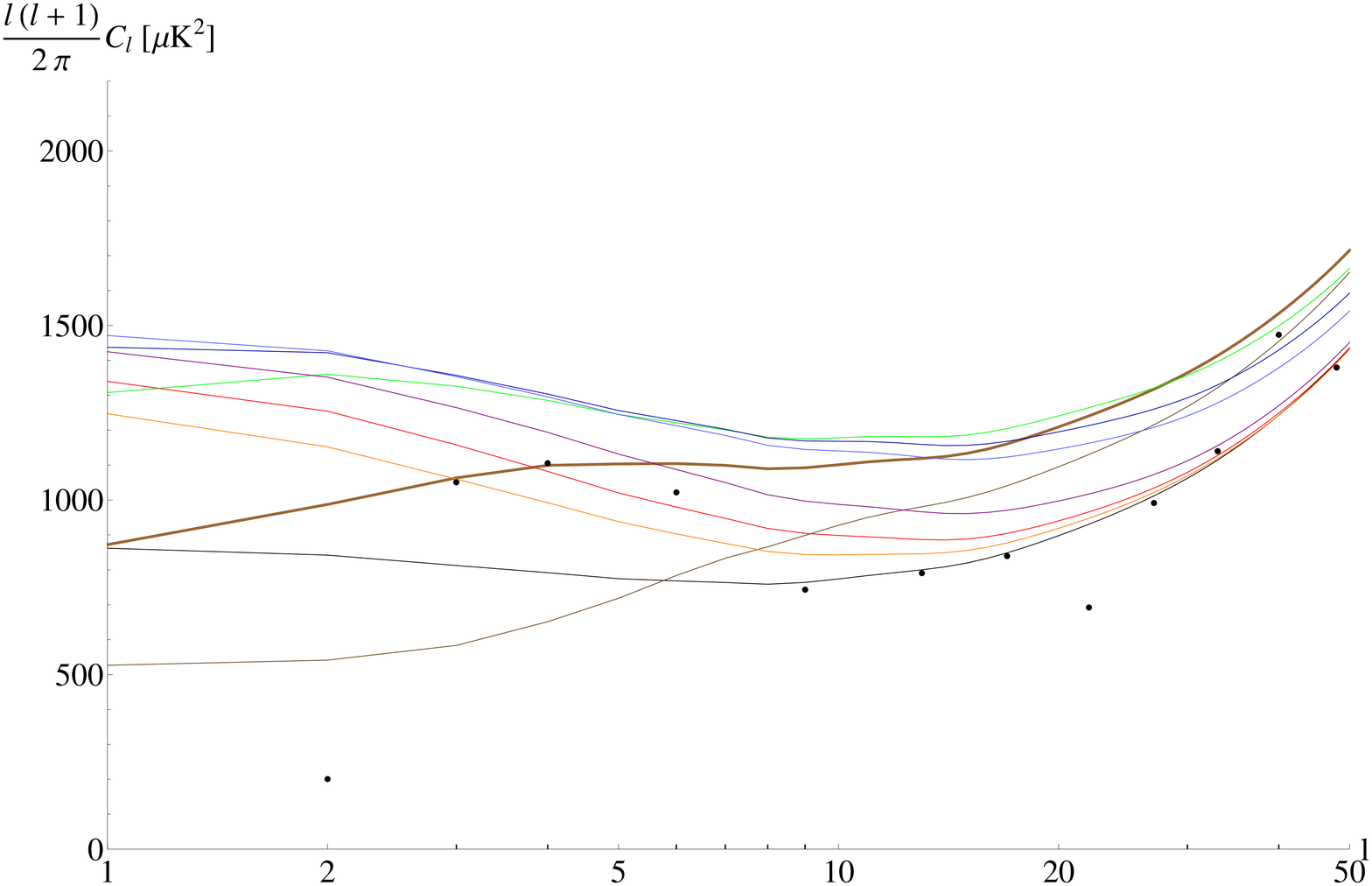}
\caption\small{{The CMB power spectrum for a pre-inflation radiation era without any black holes,
for various cases of $\Delta N$. Overall view and zoom into the lower multipole region.}}
\label{fig:CMBradonly}
\end{figure*}

For the case when there are no black holes at all, only pure radiation,
the CMB power spectrum can be seen in~\figref{fig:CMBradonly}.

\normalsize

\subsection{Results from the analytical solutions for $P(k)$}
\begin{figure*}[htp]
\centering
\includegraphics[width=0.4\textwidth]{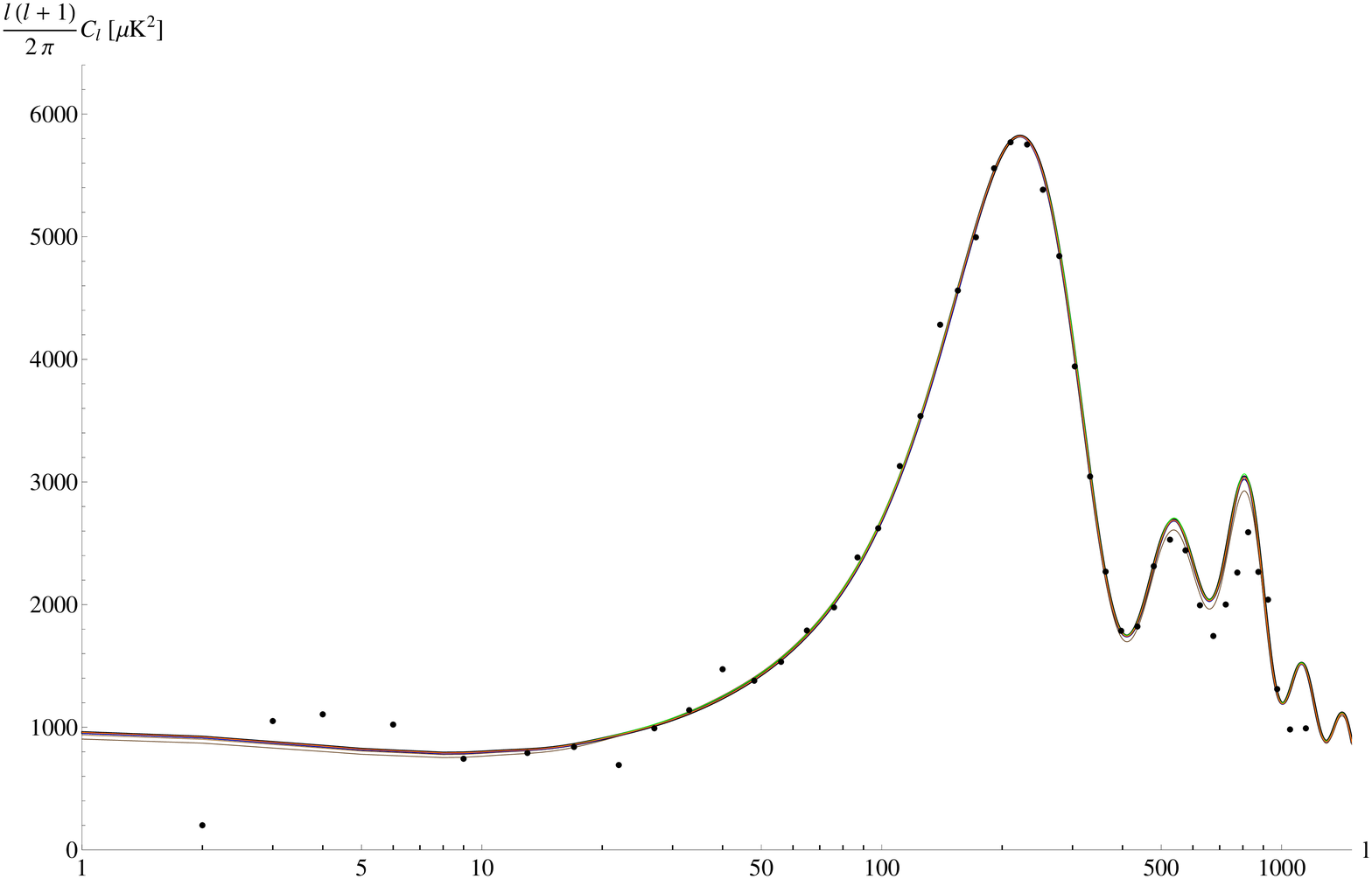} \hspace{5 mm}
\includegraphics[width=0.08\textwidth]{Legend.eps}
\includegraphics[width=0.4\textwidth]{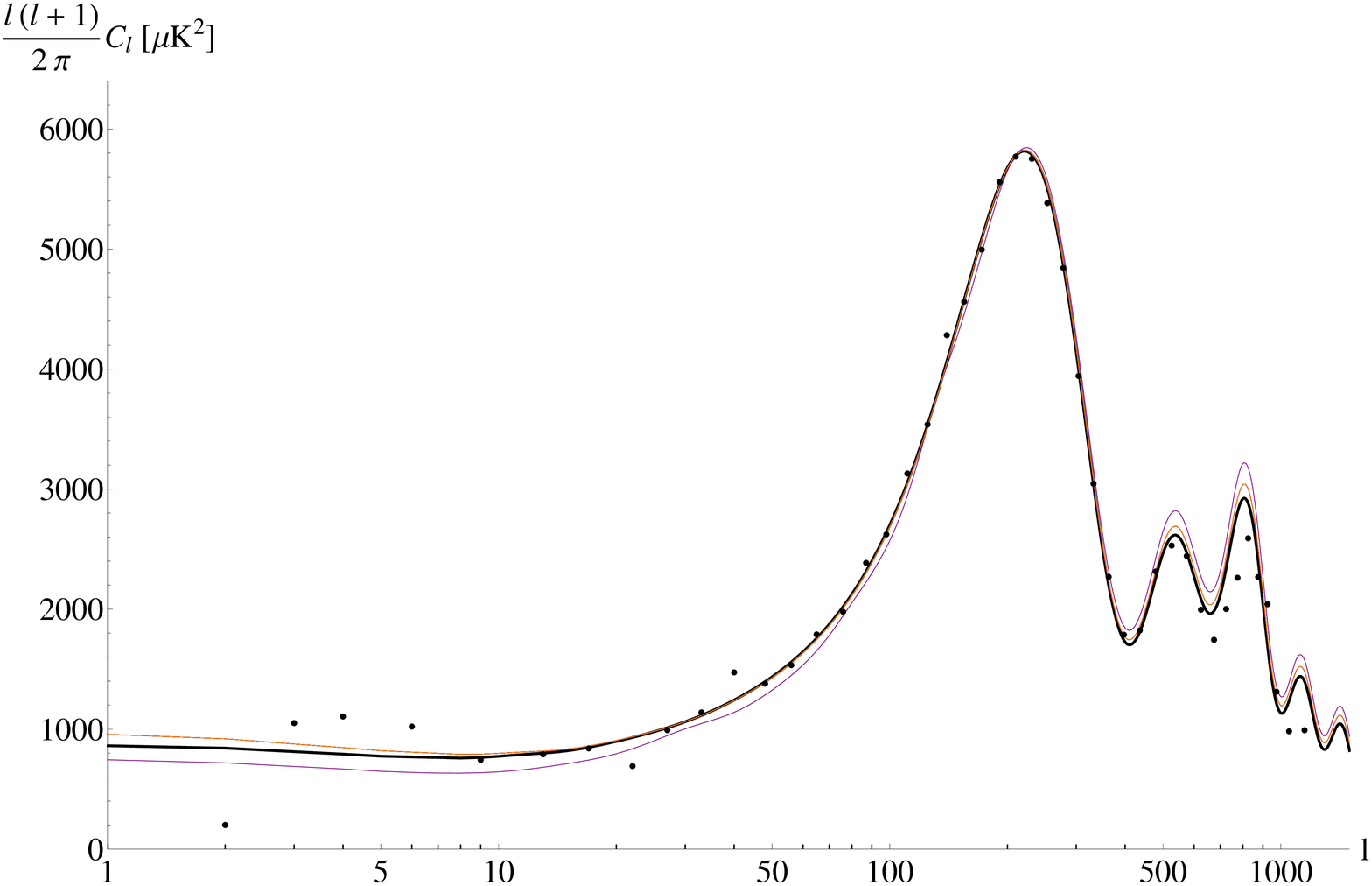}
\caption\small{{The CMB power spectrum from the analytic solution of the approximated differential equation
with matter dominance and radiation dominance, for various cases of $\Delta N$.}}
\label{fig:CMBanal}
\end{figure*}

As regard the first and the second case of the list of the previous paragraph,
the power spectrum obtained by WKB solving the approximated
differential equation for a pre-inflationary matter and radiation era was fed into CMBFAST as well.
The left picture in~\figref{fig:CMBanal} shows the outcome for the CMB temperature anisotropy spectrum
in the case when the GUP is valid and thus the pre-inflation era is matter-dominated
(we use the coefficients $A,C$ from Eqs.(\ref{Ame}, \ref{Cme})), whereas the right
picture in~\figref{fig:CMBanal} shows the results in the case where GUP does not hold, which corresponds
to a pre-inflation radiation era with completely evaporating black holes (the coefficients
$B,C$ from Eqs.(\ref{Bre}, \ref{Cre}) have been used).

\normalsize

\subsection{Comparison}
\begin{figure*}[htp]
\centering
\includegraphics[width=0.3\textwidth]{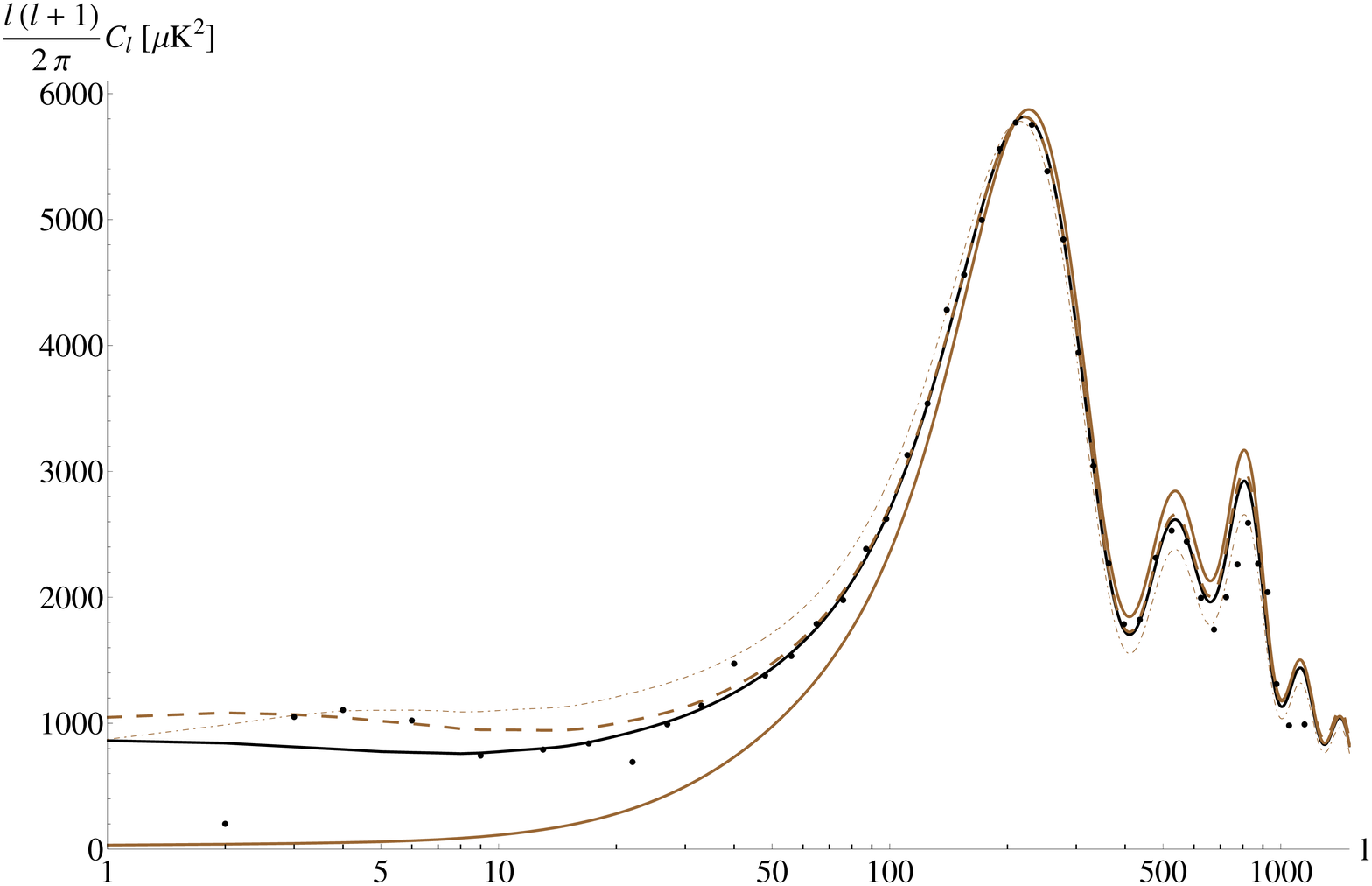} \hspace{10 mm}
\includegraphics[width=0.3\textwidth]{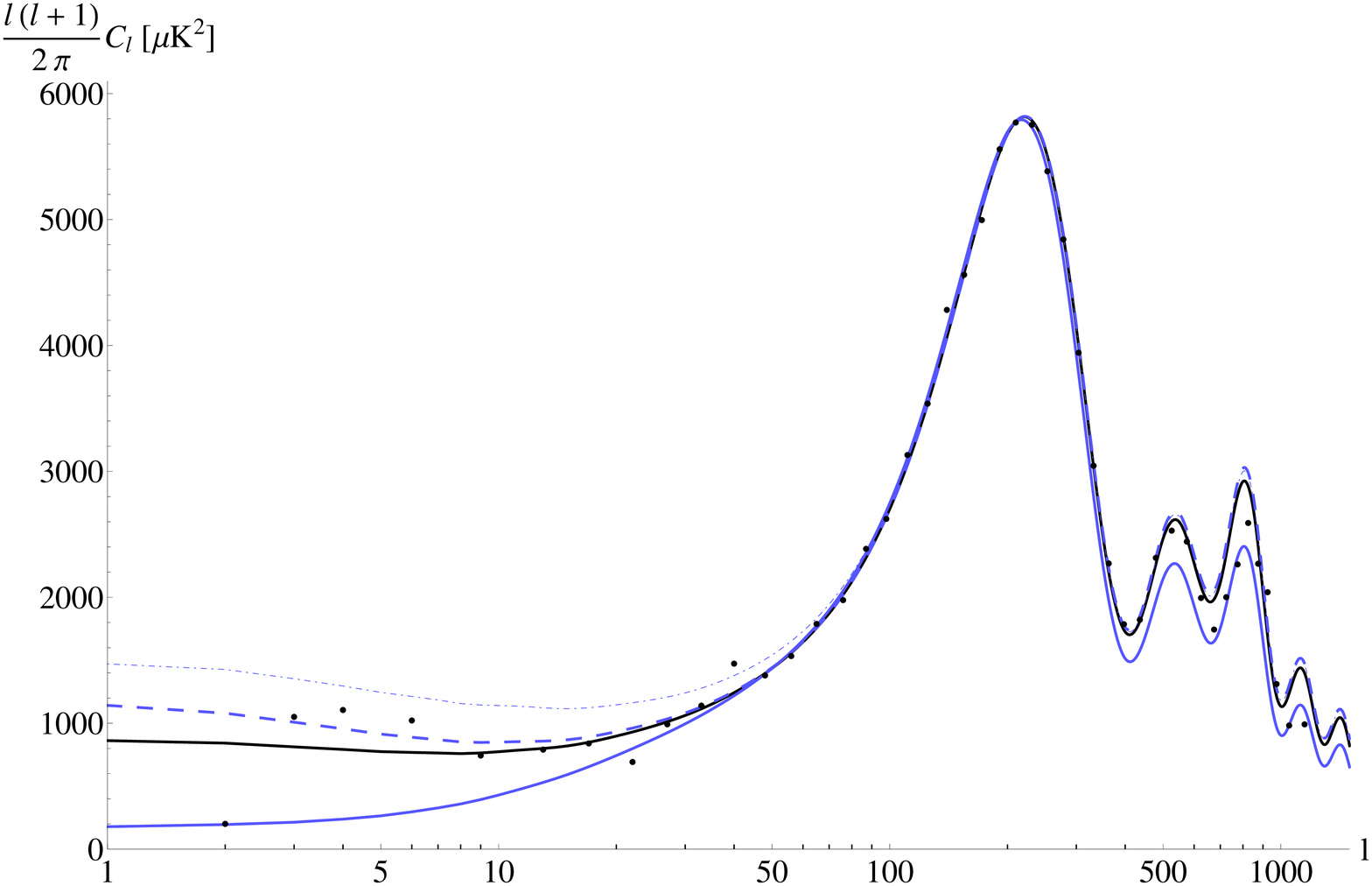} \\
~\\
\includegraphics[width=0.3\textwidth]{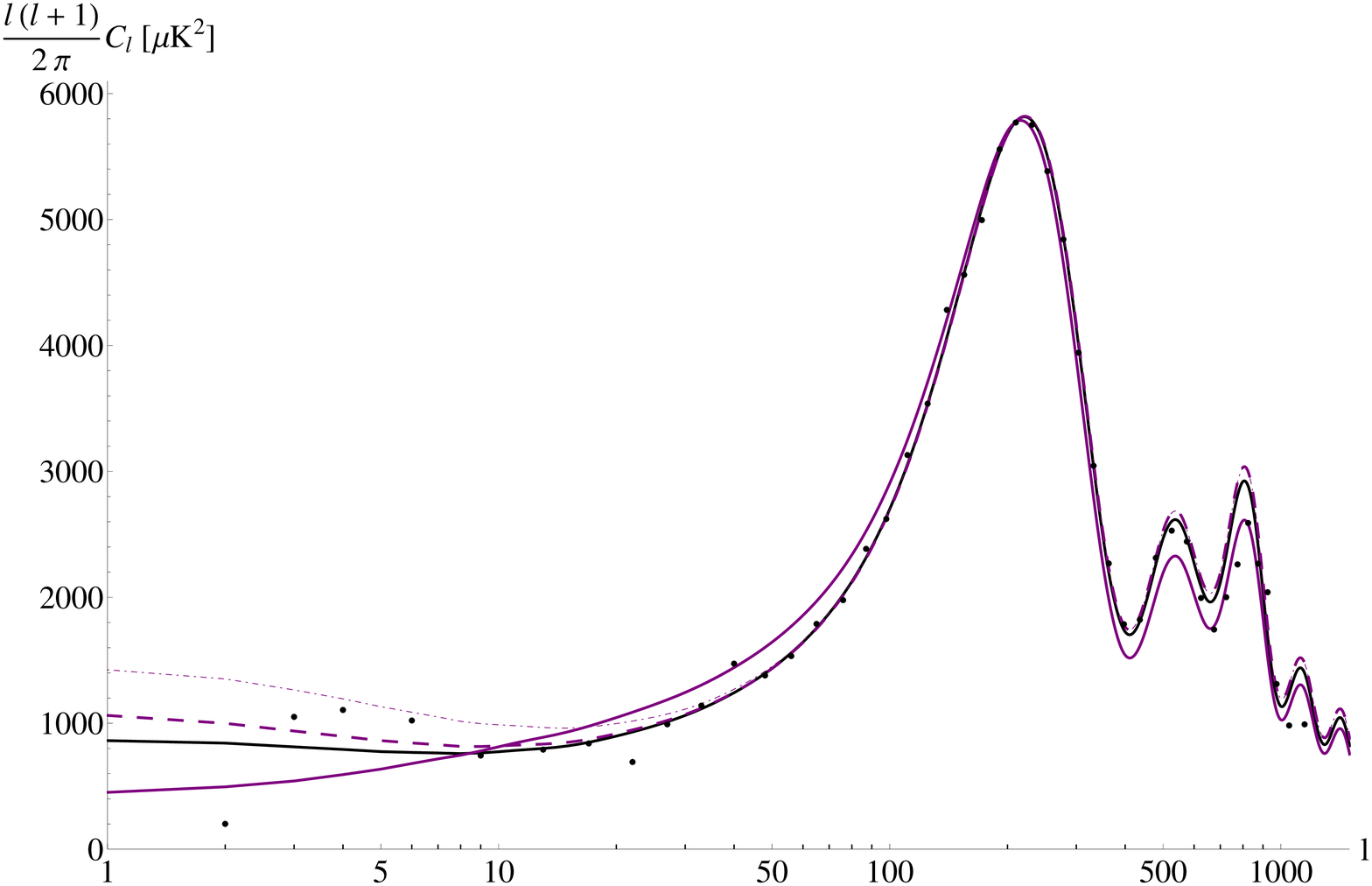} \hspace{10 mm}
\includegraphics[width=0.3\textwidth]{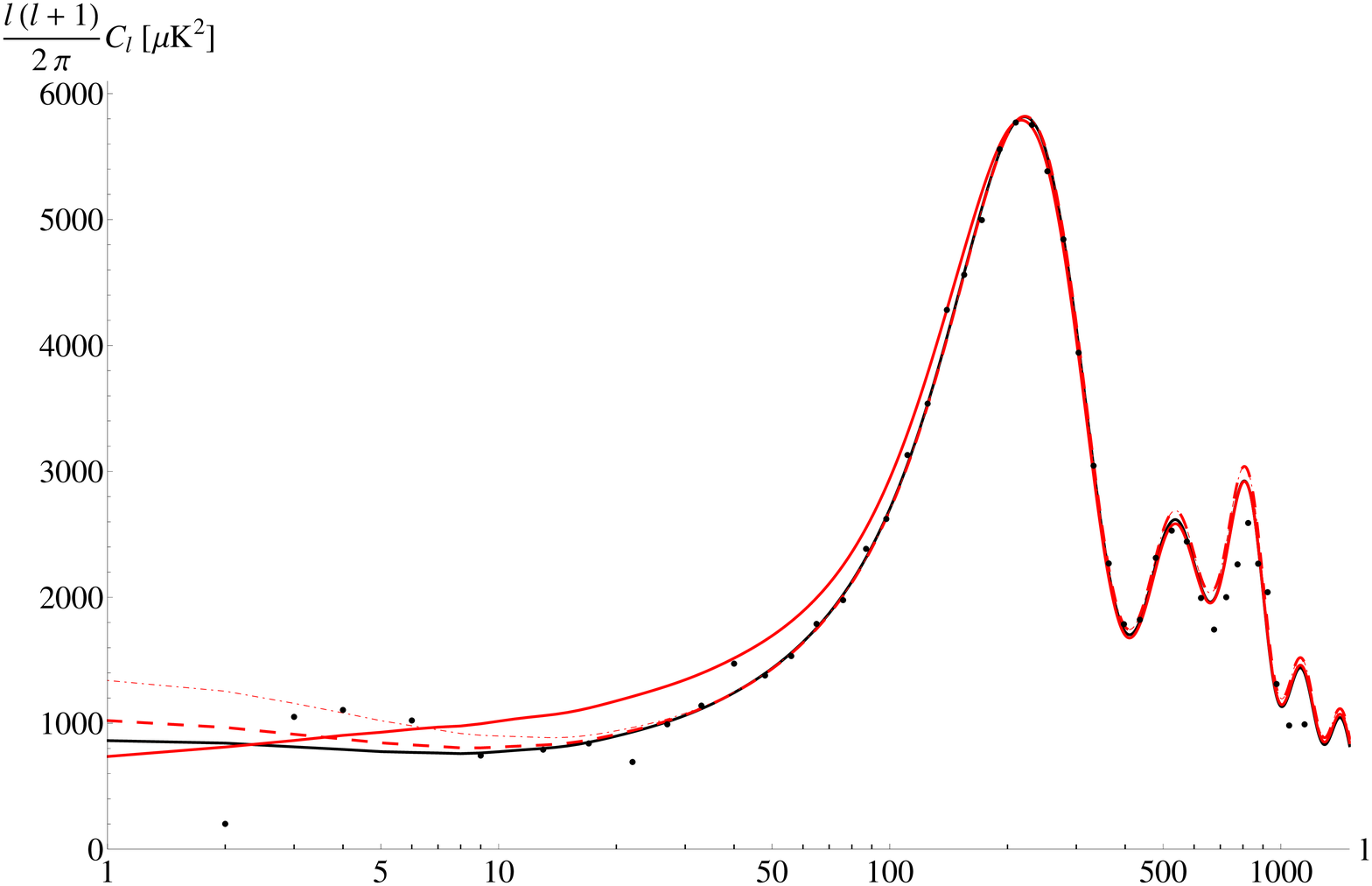} \\
~\\
\includegraphics[width=0.3\textwidth]{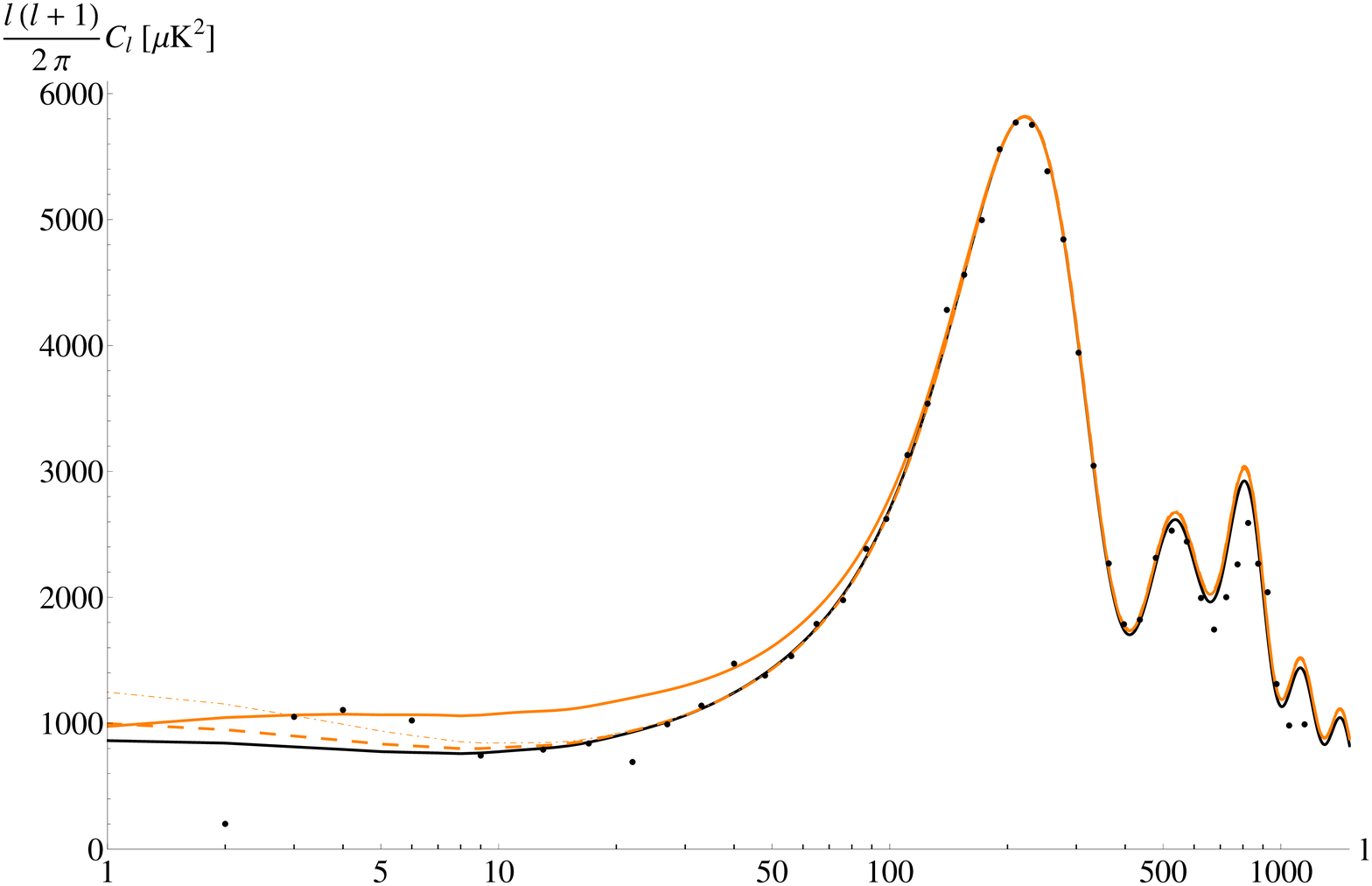}

\caption\small{{Comparing the CMB power spectrum for pre-inflation matter (full lines),
radiation with black holes (dashed lines) and radiation without black holes (dashed-dotted lines)
eras for $\Delta N=0.693147$, $\Delta N=2.07944$, $\Delta N=2.99573$, $\Delta N=3.58352$ and $\Delta N=4.09434$.}}
\label{fig:CMBcomp}
\end{figure*}

To have a better impression of how the three scenarios compare to each other,
in~\figref{fig:CMBcomp} there are several graphs showing three curves for different
values of $\Delta N$. Full lines represent the cases with GUP, dashed lines the cases
without GUP, and the dashed-dotted lines the case of pure radiation, where no black holes existed at all.

\normalsize

\subsection{$\chi^2$ Calculations}
\label{sec:chisquare}
The goodness of the fit of each of the different curves can be quantitatively expressed by
a $\chi^2$-value, calculated from the formula
\begin{equation}
    \chi^2 = \frac{1}{N} \cdot \sum_{i=1}^N{\frac{(D_i - T_i)^2}{C_i^2}},
    \label{eq:chi2}
\end{equation}
where $D_i$ is the $i-th$ data point, $T_i$ is the corresponding value calculated by the model,
and $C_i$ is the error bar of measurement for the $i-th$ data point. The theoretical model,
calculated for different numbers of e-folds are compared with the WMAP 7-year-results for the
unbinned CMB temperature spectrum. The error bar of the measurement
is provided along with the WMAP spectrum data. \\
The $\chi^2$-value for the SI model calculated according to~\eqeqref{eq:chi2} is $\chi^2_{SI} = 1.154$.
For our model of matter-dominance in the pre-inflation era, the $\chi^2$-values for different
number of e-foldings are given in the following table.

\footnotesize{
\begin{table}[htp]
\centering
\begin{tabular}{|@{\hspace{0.1cm}}c@{\hspace{0.1cm}}|@{\hspace{0.1cm}}c@{\hspace{0.1cm}}|@{\hspace{0.1cm}}c@{\hspace{0.1cm}}|@{\hspace{0.1cm}}c@{\hspace{0.1cm}}|@{\hspace{0.1cm}}c@{\hspace{0.1cm}}|@{\hspace{0.1cm}}c@{\hspace{0.1cm}}|@{\hspace{0.1cm}}c@{\hspace{0.1cm}}|@{\hspace{0.1cm}}c@{\hspace{0.1cm}}|@{\hspace{0.1cm}}c@{\hspace{0.1cm}}|}
\hline
$\Delta N$ & $0$ & $0.693$ & $1.386$ & $1.792$ & $2.079$ & $2.996$ & $3.584$ & $4.094$ \\
\hline
$\chi^2$ & $4.801$ & $1.69$ & $1.206$ & $1.353$ & $1.463$ & $1.378$ & $1.171$ & $1.172$ \\
\hline
\end{tabular}
\end{table}}
\normalsize

For the scenario of radiation-dominance in the pre-inflation era, with totally evaporating black holes,
the $\chi^2$-values for different number of e-foldings are given as follows.

\footnotesize{
\begin{table}[htp]
\centering
\begin{tabular}{|@{\hspace{0.1cm}}c@{\hspace{0.1cm}}|@{\hspace{0.1cm}}c@{\hspace{0.1cm}}|@{\hspace{0.1cm}}c@{\hspace{0.1cm}}|@{\hspace{0.1cm}}c@{\hspace{0.1cm}}|@{\hspace{0.1cm}}c@{\hspace{0.1cm}}|@{\hspace{0.1cm}}c@{\hspace{0.1cm}}|@{\hspace{0.1cm}}c@{\hspace{0.1cm}}|@{\hspace{0.1cm}}c@{\hspace{0.1cm}}|@{\hspace{0.1cm}}c@{\hspace{0.1cm}}|}
\hline
$\Delta N$ & $0$ & $0.693$ & $1.386$ & $1.792$ & $2.079$ & $2.996$ & $3.584$ & $4.094$ \\
\hline
$\chi^2$ & $1.121$ & $1.132$ & $1.141$ & $1.144$ & $1.147$ & $1.151$ & $1.152$ & $1.153$ \\
\hline
\end{tabular}
\end{table}}
\normalsize

For the scenario of radiation-dominance without the presence of black holes, the $\chi^2$-values for
different number of e-foldings are given as

\footnotesize{
\begin{table}[htp]
\centering
\begin{tabular}{|@{\hspace{0.1cm}}c@{\hspace{0.1cm}}|@{\hspace{0.1cm}}c@{\hspace{0.1cm}}|@{\hspace{0.1cm}}c@{\hspace{0.1cm}}|@{\hspace{0.1cm}}c@{\hspace{0.1cm}}|@{\hspace{0.1cm}}c@{\hspace{0.1cm}}|@{\hspace{0.1cm}}c@{\hspace{0.1cm}}|@{\hspace{0.1cm}}c@{\hspace{0.1cm}}|@{\hspace{0.1cm}}c@{\hspace{0.1cm}}|@{\hspace{0.1cm}}c@{\hspace{0.1cm}}|}
\hline
$\Delta N$ & $0$ & $0.693$ & $1.386$ & $1.792$ & $2.079$ & $2.996$ & $3.584$ & $4.094$ \\
\hline
$\chi^2$ & $1.562$ & $1.33$ & $1.17$ & $1.157$ & $1.156$ & $1.155$ & $1.154$ & $1.154$ \\
\hline
\end{tabular}
\end{table}}
\normalsize

\begin{figure}[ht]
    \begin{center}
    \includegraphics[width=0.4\textwidth]{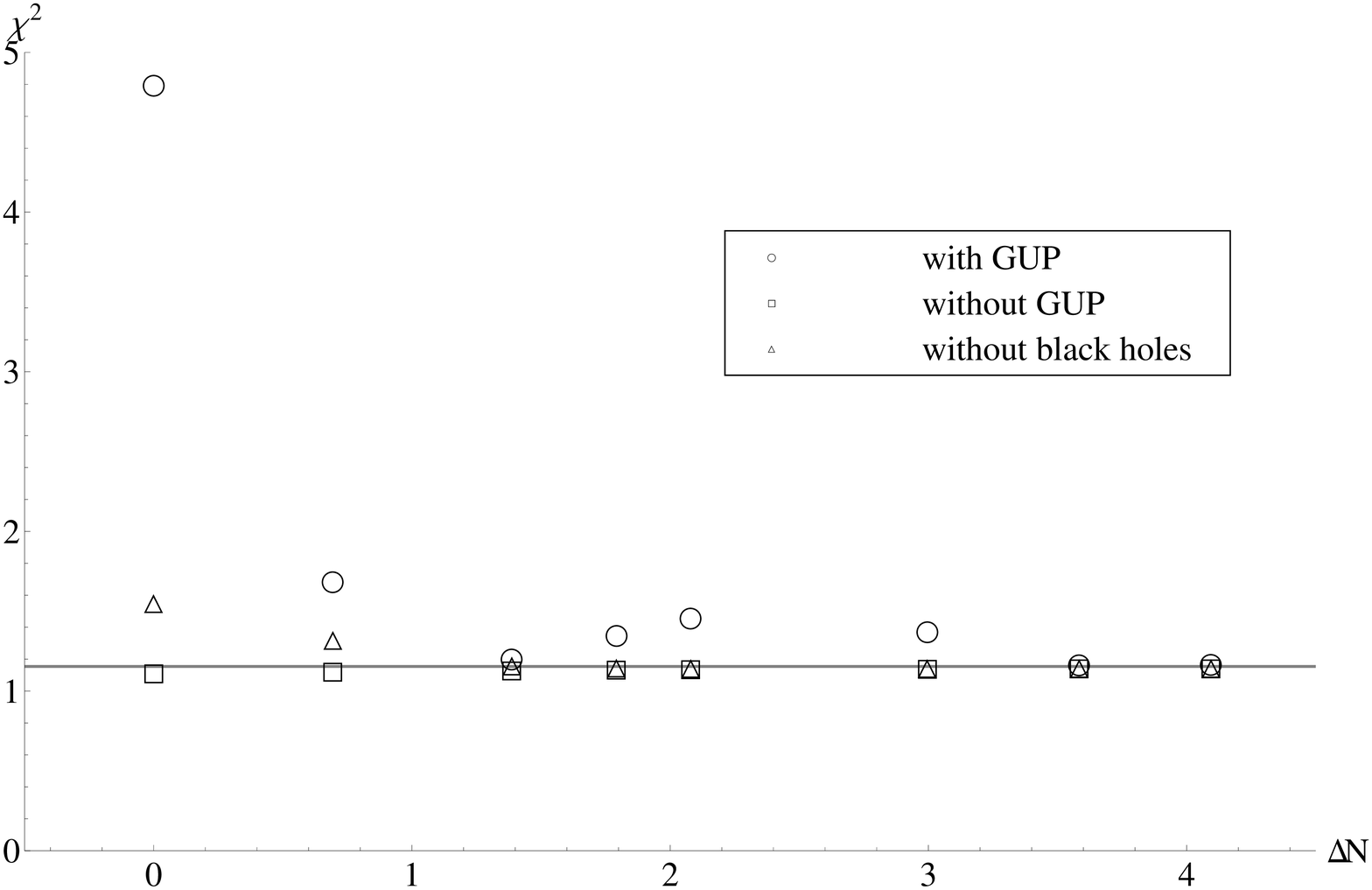}
    {\small \caption[$\chi^2$-values for a pre-inflation matter and radiation era.]{The distribution of $\chi^2$-values
for a pre-inflation matter and radiation era, for various cases of $\Delta N$. Circles represent the matter model,
whereas squares stand for the radiation model and rotated squares denote the case with radiation only.
The horizontal line denotes the value of the SI scenario.}
    \label{fig:chi2matter}}
    \end{center}
\end{figure}

Fig.\,\ref{fig:chi2matter} should help to make the numbers more understandable.
It shows the $\chi^2$-values for the three models over the number of e-foldings $\Delta N$
added to the standard number of $54$ e-folds. The horizontal line represents the value for the SI model.

\normalsize

\section{Conclusions and Outlook}

In this work, we investigated the effects of an era before inflation on the CMB power spectrum
measured today. We utilized the phenomenon of black hole nucleation from quantum fluctuations
of the metric in very early times to argue for the existence of several thousand micro
black holes in the pre-inflation era, which cause the universe to be matter-dominated
from about $\tau \simeq 10^3\,\,t_p$, until the onset of inflation ($\tau \simeq 10^6\, t_p$).
By setting up the Friedmann
equation of the universe evolving from matter dominance to an inflationary phase it is possible
to calculate the power spectrum of primordial fluctuations of a scalar field $\Phi$ living in
this scenario, and then process this primordial power spectrum by CMBFAST code to yield the CMB
temperature anisotropy spectrum measured today.
As an alternative to the matter dominance scenario, we also investigated
the implications of a radiation-dominated era before inflation by relinquishing the claim of
the Generalized Uncertainty Principle, stating that black holes can only evaporate down to
Planck size. These two cases have been calculated both numerically and in analytical approximations.
A third case is presented, in which black holes never existed and the pre-inflation era is purely radiation-dominated.

From the overall analysis that has been done on the three different scenarios, only the model
with matter-dominance in pre-inflation era is really successful in the suppression of the $l=2$ mode,
and incorporates the desired effect on the power spectrum very well.
It asymptotes to the standard inflationary
model for higher $\Delta N$, which is expected as with increasing duration of inflation the
effects of a pre-inflationary era are shifted to larger scales and thus would be expected to
influence the scales that are still to enter the horizon.

For the case of radiation in pre-inflation era, with black holes evaporating until zero mass (no GUP),
the result is rather poor - only one of the curves
shows a suppression in the lower modes, all of them actually lie above the result
given by a standard inflationary scenario without a pre-inflation era. The model shows a
very good accordance for the high $l$ region.\\
In contrast to that, the model with only radiation in the pre-inflationary era turns out to be a
little better; two of the curves show a suppression, while the others lie higher than the curve
obtained by the standard inflationary scenario. This can be traced back to the shape of the primordial
power spectrum, which in this case rather resembles the matter case than the radiation case with black holes.

The CMB power spectrum, produced with
the analytical solutions obtained by approximating the differential equation,
looks very similar to the standard inflation picture with only
slight deviations. But on the other hand, we know that the WKB analytical solution has been mainly
used as a qualitative guide to choose the correct numerical solution among several possibilities,
and so we cannot expect from it a perfect fitting of the data. Especially because we pushed
the analytical approximation to the second order only.
Of course, it is important to investigate the differential equation analytically as to obtain
the correct boundary conditions for the numerical simulation and to firmly support the correctness
of the numerical result.

Although the suppression of the lower modes in the case of pre-inflation matter dominance is there,
the overall fit to the current data turns out to be not so successful. A drawback of the matter
model definitely is the bad fit for large $l$ when only a few e-folds are added; the behavior
for large $l$ becomes better with increasing $\Delta N$, but it is not very good for
small $\Delta N$. The reason why the model without GUP is so much better in the overall fit than the other two cases
can be found in the fitting process for the primordial power spectrum. The shape of $P_k$ in
the matter model and in the pure radiation is more eccentric and harder to be fitted than in the case of radiation without GUP,
where the power spectrum is rather smooth and the $\arctan$-fit matches the curve quite well.
In the matter case, the fitting function is more complicated and the quality of the fit is
definitely worse.
This of course influences the $\chi^2$-value of the model, which is quite large for the
smaller $\Delta N$. A stronger suppression on several of the low $l$ modes also leads to a
larger deviation from the modes with $l\geqslant3$. The SI model is without doubt the best
fit in general; however, it doesn't capture the drop at the $l=2$ mode. The model with matter
in pre-inflation might not be the best fit on all scales, but in the future, with further
modes on larger scales than $l=2$ being suppressed, the model might become more successful
than the standard inflation scenario. For sure, if the suppression of the $l=2$ mode is to
be continued with a suppression on even larger scales, the extension of standard inflation
to having a pre-inflationary era is required, otherwise the drop for the lower modes will
remain unexplained.

The quality of the results has been put into numbers by a $\chi^2$ analysis, which is the best way in
the current situation to give a solid statement about the success of the three models today.

In terms of this
analysis, the pre-inflationary matter era is disfavored compared to the SI model. We could think
that the roles might be really different in the far future, and the modes that are
still to enter the horizon might be more successfully described by the matter era scenario.
However, even without looking too far into the future, but keeping our mind on the present,
the pre-inflation matter era model seems to be the only one, among those here studied, able
to capture and describe the low $l$ modes suppression. Further refinement of the model are obviously in order.
And in principle, if a better matching with the observations will be achieved, the model
can serve to directly check the validity of the Generalized Uncertainty Principle,
as relation (\ref{eq:Afinal}) explicitly suggests, and more indirectly, of the Holographic Principle.
These are surely intriguing avenues for future research.
Only with time it will become clear whether the model with
matter-dominance in pre-inflation era is superior to standard inflation
in its success to explain the CMB power spectrum.
%
%
%

%
%
%
%
%
%
%
%
%
%
%
%
%
%
%
%
%
%
\section*{Acknowledgements}
%
%
%
The authors would like to thank Kin W. Ng and Ron J. Adler for
enlightening conversations. F.S. would like to thank Mariam Bouhmadi
Lopez for conversations and for having drawn his attention on the
wonderful world of elliptic functions. This research is supported by
Taiwan National Science Council under Project No. NSC
97-2112-M-002-026-MY3 and by US Department of Energy under Contract
No. DE-AC03-76SF00515. We also acknowledge the support of the
National Center for Theoretical Sciences of Taiwan.
%
%

%

\section*{Appendix 1}
\label{nucrate}
Here we compute the nucleation rate $n_*$ for micro black holes,
i.e. the number of micro black holes
of critical mass $M$ created in a thermal bath of gravitons, via
gravitational instabilities of hot flat space,
per unit volume per unit time.
Essentially, we follow the procedure detailed in Ref. \cite{kapusta}.
Our discussion, as in that reference, will be based on the standard Heisenberg principle.
As said, the GUP will be implemented in our argument only by considering the cutoff imposed
on minimum masses and maximum temperatures.
Critical mass and temperature are linked by the relation (\ref{H}), which in standard units reads
\be
M = \frac{\hbar c^3}{8 \pi G k_B T}\,.
\label{CM}
\ee
The probability for a quantum vacuum fluctuation to produce a black hole of critical mass $M$ is
$\exp(-\Delta F / (k_B T))$, where $\Delta F$ is the change in the free energy of the system with
$T$ and $V$ held fixed. Now $\Delta F = F - F_g$, where $F$ is the free energy of the black hole
and $F_g$ is the free energy of the thermal gravitons displaced by the black hole. $F$ is related to
the rest energy of the black hole $E = M c^2$ by
\be
E = F - T\frac{dF}{dT}
\ee
This gives us a differential equation for $F$
\be
\frac{dF}{dT}= \frac{F}{T} - \frac{E(T)}{T}
\label{dF}
\ee
where, from (\ref{CM})
\be
E(T)=Mc^2=\frac{\hbar c^5}{8 \pi G k_B T}\,.
\ee
Integrating $F'(T)$ we find
\be
F(T) = \frac{\hbar c^5}{16 \pi G k_B T}\,.
\ee
To compute the thermal free energy $F_g$ of displaced gravitons we need, as it is clear from Eq.(\ref{dF}),
an expression for the total energy $E$ of such thermal gravitons. This can be obtained from Eq.(\ref{sb1}),
explicitly rewritten as
\be
E^{g}_{\rm TOT}(V) \ = \ \frac{\pi^2\, k_B^4}{15\, c^3\, \hbar^3}\, V \, T^4
\ee
where we chose the greybody factor for gravitons $\Gamma_{g} = 1$,
and we dropped the correction function $A(\beta,T)$, as the effects of GUP are
considered only through the cutoff on masses and temperatures.
Since the volume of the displaced gravitons coincides with that of the black hole ($4\pi R_S^3/3$, with
$R_S=2GM/c^2$), we have
\be
E^{g}_{\rm TOT}(T) = \frac{k_B T}{720}\,.
\ee
Therefore Eq.(\ref{dF}) yields
\be
F_g(T) = -\frac{k_B T}{720}\log\frac{T}{T_p}
\ee
So, finally, expressing things in Planck units we have
\be
\frac{\Delta F}{k_B T} = \frac{1}{16 \pi \Theta^2} + \frac{1}{720} \log\Theta
\label{F/T}
\ee
Note that, in the range of interest, namely for $0<\Theta<1$, we have
$|\log\Theta/720| \ll 1/(16\pi\Theta^2)$, so the second term in (\ref{F/T}) can be neglected.

Knowing the probability of one statistical fluctuation, $\exp(-\Delta F / (k_B T))$, we have to
estimate the density for such fluctuations. Consider the fluctuations on the smallest scale possible,
namely with a wavelength $\lambda_{\rm min} = \alpha \ell_p$, $\alpha$ of order $1$.
Then, imagining a cubic lattice, the number of statistical fluctuations in the unit volume is
\be
n_0 = \left(\frac{2}{\lambda_{\rm min}}\right)^3
\ee
and the number of fluctuations per unit volume able to produce a black hole of critical mass $M$ will be
\be
n_* &=& n_0 \exp\left(-\frac{\Delta F}{k_B T}\right) \nonumber \\
&=& \left(\frac{2}{\lambda_{\rm min}}\right)^3 \exp\left(-\frac{1}{16 \pi \Theta^2}\right)
\ee
The number of micro holes with critical mass $M$, created per unit time per unit volume,
can be therefore computed as
\be
\frac{dn_*}{d\tau} = n_* \left(\frac{1}{8 \pi \Theta^3}\right)\frac{d\Theta}{d\tau}
\label{nstarev}
\ee
The value of $d\Theta/d\tau$ can be obtained using Eqs.(\ref{H}) and (\ref{ereqpl}),
with the correction function $A(\beta,T)=1$, and $\Gamma_\gamma=1$. We have
\be
\frac{d\Theta}{d\tau} \,=\, \frac{2\pi^2}{15}\,\,
\,\,\Theta^4
\ee
So finally
\be
\frac{dn_*}{d\tau} = n_* \,\frac{\pi}{60}\,\,\Theta
\ee
In the statistical probability of one fluctuation, we should also include a term for the quantum
correction of the free energy of the black hole.
It can be shown \cite{Hawking} that for a Schwarzschild metric we have
\be
\frac{F^{quantum}}{k_B T} = -\frac{212}{45}\log\left(\frac{\mu c^2}{k_B T}\right)
\ee
where $\mu$ is a regulator mass of the order of the Planck mass. The final formula for
the number of micro holes with critical mass $M$, created per unit time per unit volume,
reads, in Planck units,
\be
\frac{dn_*}{d\tau} &=& n_* \,\frac{\pi}{60}\,\,
\Theta \left(\frac{\tilde{\mu}}{2\Theta}\right)^\frac{212}{45} \\
&=& \left(\frac{2}{\lambda_{\rm min}}\right)^3\frac{\pi\,\,\Theta}{60}
\left(\frac{\tilde{\mu}}{2\Theta}\right)^\frac{212}{45}
\exp\left(-\frac{1}{16\pi\Theta^2}\right)\nonumber
\ee
Since in Planck units $\ell_p = \tilde{\mu} = 1$ the previous formula can be usefully rewritten as
\be
\frac{dn_*}{d\tau} \ = \ \frac{8\pi}{15 \cdot 64\,\pi^3}\,\, \Theta^{-\frac{167}{45}}\,\,
\exp\left(-\frac{1}{16\pi\Theta^2}\right)
\label{nucformula}
\ee
which agrees with the numerical pre-factor of Ref.\cite{kapusta} since we chose
$\alpha = \pi\cdot2^{-32/135} \simeq 2.66$.
%
%
%
%
%
%
\section*{Appendix 2}
%
%
%
In this Appendix we construct the cosmic scale factor $a$ as a function of the conformal time $\eta$.
Once we have $a(\eta)$, we shall be able to write equation (\ref{eqv}) for $v_k(\eta)$.

In order to arrive to the function $a(\eta)$, two different, but equivalent, procedures can be specified.\\

I) When the scale factor $a(t)$ is a known function of the cosmic time $t$, then we can compute
(in principle, explicitly) the function $\eta(t)$
\be
\eta = \int d\eta = \int \frac{dt}{a(t)} \quad \Rightarrow \quad \eta = \eta(t).
\label{eta}
\ee
Hence, inverting the last relation we get
\be
t=t(\eta) \quad \Rightarrow \quad a=a(t(\eta))=a(\eta)\,.
\ee
There exists however also another alternative procedure.\\

II) Sometimes the integral in (\ref{eta}) is not easily doable, and moreover the object we are
usually interested in is the function $a(\eta)$, and not $\eta(t)$. Such function can be directly computed
by re-writing the equation of motion for $a$ in terms of the conformal time $\eta$, instead of
the cosmic time $t$. The equation of motion (\ref{adot}) reads, in conformal time,
\be
\left(\frac{da}{d\eta}\right)^2 \ = \ \kappa\left(Ca^4 + Aa + B\right)
\ee
or
\be
\frac{da}{\sqrt{Ca^4 + Aa + B}} \ = \ \sqrt{\kappa}\,d\eta
\label{aeta}
\ee
The function $a(\eta)$ can be obtained directly from the integration of the previous equation.\\
In the following, we shall integrate equation (\ref{aeta}) in the two cases of our interest: pre-inflation
radiation era, and pre-inflation matter era.\\
\noindent \textbf{Pre-inflation radiation era}: In this case there is no matter, therefore $A=0$.
Equation (\ref{aeta}) reads
\be
\ \sqrt{\kappa}\,d\eta \ = \ \frac{da}{\sqrt{Ca^4 + B}}
\ee
With the substitution
\be
a \ = \ x\left(\frac{B}{C}\right)^{1/4}
\label{ax}
\ee
the equation becomes
\be
(\kappa^2 BC)^{1/4}\,d\eta \ = \ \frac{dx}{\sqrt{1+x^4}}\,.
\ee
We can now make use of the formula (see \cite{GrR})
\be
\int\frac{dx}{\sqrt{1+x^4}} \ = \ \frac{1}{2} \,\,F(\alpha, k)
\label{xF}
\ee
 where $F(\alpha, k)$ is the {\em elliptic integral of the first kind}
\be
F(\alpha, k) = \int_0^\alpha\frac{d\mu}{\sqrt{1-k^2\sin^2\mu}}\,.
\label{eli}
\ee
In our specific case, Eq.(\ref{xF}), we have
\be
k&=&\frac{1}{\sqrt{2}}\nonumber\\
\alpha&=&\arccos\left(\frac{1-x^2}{1+x^2}\right)
\label{alfax}
\ee
Then
\be
2\,(\kappa^2 BC)^{1/4}\,\eta \ = \ \int_0^\alpha\frac{d\mu}{\sqrt{1-k^2\sin^2\mu}}\,.
\ee
Inverting the last integral, we get the Jacobi \emph{amplitude}
\be
\alpha \ = \ {\rm am}[2\,(\kappa^2 BC)^{1/4}\,\eta]
\ee
and, because of relation (\ref{alfax}),
\be
\frac{1-x^2}{1+x^2} &=& \cos {\rm am}[2\,(\kappa^2 BC)^{1/4}\,\eta] \nonumber\\
&=&{\rm cn}\left[2\,(\kappa^2 BC)^{1/4}\,\eta, \,\,\frac{1}{\sqrt{2}}\right]
\ee
where cn is the Jacobi \emph{cosine-amplitude} (see again \cite{GrR} for definitions and properties).
Reminding the relation (\ref{ax}) between $a$ and $x$, finally we can write
\be
a^2(\eta) = \left(\frac{B}{C}\right)^{\frac{1}{2}}\,
\frac{1 - {\rm cn}\left[2\,(\kappa^2 BC)^{1/4}\,\eta, \,\,\frac{1}{\sqrt{2}}\right]}
{1 + {\rm cn}\left[2\,(\kappa^2 BC)^{1/4}\,\eta, \,\,\frac{1}{\sqrt{2}}\right]}\,.
\ee
It is interesting to check the small $\eta$ limit (or, equivalently, the small $C$ limit)
of the previous relation. Considering the MacLaurin
expansion for cn
\be
{\rm cn}(u,k) = 1 - \frac{1}{2} u^2 + O(u^4)
\label{ML}
\ee
we have
\be
a^2(\eta) \ = \ \left(\frac{B}{C}\right)^{\frac{1}{2}}\,\,
\frac{2\,(\kappa^2 BC)^{\frac{1}{2}}\eta^2 + \dots}{1 + 1 + \dots} \ = \ \kappa B \eta^2
\ee
which is the well known expression for the scale factor $a(\eta)$ in pure radiation era.\\
To build equation (\ref{eqv}) for $v_k$ we need to compute $a''(\eta)/a(\eta)$. It is a bit laborious,
but however the result, for the parameter $k=1/\sqrt{2}$, is
\be
\frac{a''(\eta)}{a(\eta)} \ = \ \frac{\beta^2}{2}\,\cdot\,
\frac{1 - {\rm cn}\left(\beta \eta, \, \frac{1}{\sqrt{2}}\right)}
{1 + {\rm cn}\left(\beta \eta, \, \frac{1}{\sqrt{2}}\right)}\,,
\ee
with $\beta=2(\kappa^2BC)^{1/4}$.\\
Therefore the equation for $v_k$ reads
\be
v''_k \ + \ \left(K^2 \ - \ \frac{\beta^2}{2}\,\cdot\,\frac{1 - {\rm cn}(\beta \eta)}
{1 + {\rm cn}(\beta \eta)}\right)v_k \ = \ 0
\ee
where capital $K$ is the cosmological perturbation wave number and has nothing to do with the
elliptic functions parameter $k$. This equation has only a resemblance with the Lame' equation, but
unfortunately, differs from it in a fundamental way.\\
\noindent \textbf{Pre-inflation matter era}: In this case we don't have radiation, i.e. $B=0$,
and the equation for the conformal scale factor $a(\eta)$ reads
\be
\sqrt{\kappa}\,d\eta \ = \ \frac{da}{\sqrt{Ca^4 + Aa}}
\ee
Using the substitution
\be
a \ = \ x\left(\frac{A}{C}\right)^{1/3}
\label{axm}
\ee
we have
\be
(\kappa^3 A^2 C)^{1/6} d\eta \ = \ \frac{dx}{\sqrt{x(1 + x^3)}}
\ee
Again with the help of \cite{GrR2} we can make use of the formula
\be
\int \frac{dx}{\sqrt{x(1+x^3)}} \ = \ \frac{1}{\sqrt[4]3}\,\,F(\alpha,k)
\ee
where $F(\alpha,k)$ is the usual elliptic integral of the first kind (\ref{eli}),
but now
\be
k&=&\frac{\sqrt{2+\sqrt{3}}}{2}\nonumber\\
\alpha&=&\arccos\left(\frac{1+(1-\sqrt{3})x}{1+(1+\sqrt{3})x}\right)\,.
\label{alfaxm}
\ee
Then
\be
\sqrt[4]3\,\, (\kappa^3 A^2 C)^{1/6}\,\eta \ = \ \int_0^\alpha\frac{d\mu}{\sqrt{1-k^2\sin^2\mu}}\,,
\ee
and the Jacobi amplitude reads
\be
\alpha = {\rm am}[\sqrt[4]3\,\, (\kappa^3 A^2 C)^{1/6}\,\eta]
\ee
\vspace{0.5cm}
Reminding relations (\ref{alfaxm}) and (\ref{axm}), finally we have the solution $a(\eta)$ expressed
in terms of Jacobi cosine-amplitude
\be
a(\eta) \ = \ \left(\frac{A}{C}\right)^{\frac{1}{3}}
\frac{1 - {\rm cn}[\beta\eta,k]}{(\sqrt{3}-1)+(\sqrt{3}+1)\,{\rm cn}[\beta\eta, k]}
\ee
where
\be
\beta \ = \ \sqrt[4]3\,\, (\kappa^3 A^2 C)^{1/6}\,\eta \quad {\rm ;} \quad
k \ = \ \frac{\sqrt{2+\sqrt{3}}}{2}\,.
\ee
The limit for small $\eta$, or small $C$, can be computed with the help of Eq.(\ref{ML}), and reads
\be
a(\eta) \ =  \ \frac{\kappa A}{4}\,\, \eta^2
\ee
which is the known expression of the conformal scale factor in pure matter era.\\
As for the construction of the equation for $v_k$, we have, after a somehow long calculation,
\be
&&\frac{1}{\sqrt{3}\beta^2}\,\,\frac{a''(\eta)}{a(\eta)} \ = \\
&&\frac{(1-\sqrt{3})(\sqrt{3}{\rm cn}(\beta\eta) -1) + (1+\sqrt{3})(\sqrt{3} +
{\rm cn}(\beta\eta)){\rm cn}^2(\beta\eta)}{[(\sqrt{3}-1) +
(\sqrt{3} + 1)]^2 (1 - {\rm cn}(\beta\eta))}\nonumber
\ee
The equation for $v_k$ can be henceforth explicitly written down, although it still results to be only
"similar" to the Lame' equation, but not exactly of that known kind. Progress on the analytical exact
solutions of such equations will be reported in future work.

\end{document}